\documentclass[10pt,journal,letterpaper]{IEEEtran}

\usepackage[utf8]{inputenc}
\usepackage{amsmath}
\usepackage{amsthm}
\usepackage{times}
\usepackage{algorithmic}
\usepackage{graphicx}
\usepackage{verbatim}
\usepackage{amssymb}
\usepackage{cite}
\usepackage{epsfig}
\usepackage[usenames]{color}
\usepackage{url}
\usepackage{multirow}
\usepackage{grffile}
\usepackage{epstopdf}
\usepackage{soul}
\usepackage{float}

\newcommand{\tabitem}{~~\llap{\textbullet}~~}

\newcommand\ignore[1]{}

\begin{document}
\title{Deep Learning for IoT Big Data and Streaming Analytics: A Survey}

\author{Mehdi Mohammadi,~\IEEEmembership{Graduate Student Member,~IEEE,} Ala Al-Fuqaha,~\IEEEmembership{Senior Member,~IEEE,} Sameh~Sorour,~\IEEEmembership{Senior Member,~IEEE,} Mohsen Guizani,~\IEEEmembership{Fellow,~IEEE}
\IEEEcompsocitemizethanks{\IEEEcompsocthanksitem Manuscript received September 19, 2017; revised March 30, 2018; accepted
May 23, 2018.

Mehdi Mohammadi and Ala Al-Fuqaha are with the Department of Computer Science, Western Michigan University, Kalamazoo, MI 49008 USA (E-mail: \{mehdi.mohammadi,ala.al-fuqaha\}@wmich.edu.). Sameh Sorour (E-mail: samehsorour@uidaho.edu) and Mohsen Guizani (E-mail: mguizani@ieee.org) are with the Department of Electrical and Computer Engineering, University of Idaho, Moscow, ID 83844 USA. 
\protect
}
}

\markboth{IEEE Communications Surveys \& Tutorials ,~Vol.~x, No.~x, xxxxx~201x}%
{Shell \MakeLowercase{\textit{et al.}}: Bare Demo of IEEEtran.cls for Computer Society Journals}

\IEEEcompsoctitleabstractindextext{
\begin{abstract}

In the era of the Internet of Things (IoT), an enormous amount of sensing devices collect and/or generate various sensory data over time for a wide range of fields and applications. Based on the nature of the application, these devices will result in big or fast/real-time data streams. Applying analytics over such data streams to discover new information, predict future insights, and make control decisions is a crucial process that makes IoT a worthy paradigm for businesses and a quality-of-life improving technology. 
In this paper, we provide a thorough overview on using a class of advanced machine learning techniques, namely Deep Learning (DL), to facilitate the analytics and learning in the IoT domain. We start by articulating IoT data characteristics and identifying two major treatments for IoT data from a machine learning perspective, namely IoT big data analytics and IoT streaming data analytics. We also discuss why DL is a promising approach to achieve the desired analytics in these types of data and applications. The potential of using emerging DL techniques for IoT data analytics are then discussed, and its promises and challenges are introduced. We present a comprehensive background on different DL architectures and algorithms. We also analyze and summarize major reported research attempts that leveraged DL in the IoT domain. The smart IoT devices that have incorporated DL in their intelligence background are also discussed. DL implementation approaches on the fog and cloud centers in support of IoT applications are also surveyed. Finally, we shed light on some challenges and potential directions for future research. At the end of each section, we highlight the lessons learned based on our experiments and review of the recent literature.

\end{abstract}

}

\maketitle

\IEEEdisplaynotcompsoctitleabstractindextext
\IEEEpeerreviewmaketitle

\section{Introduction}\label{sec:Introduction}

The vision of the Internet of Things (IoT) is to transform traditional objects to being smart by exploiting a wide range of advanced technologies, from embedded devices and communication technologies to Internet protocols, data analytics, and so forth \cite{al2015internet}. The potential economic impact of IoT is expected to bring many business opportunities and to accelerate the economic growth of IoT-based services. Based on McKinsey's report on the global economic impact of IoT \cite{manyika2013disruptive}, the annual economic impact of IoT in $2025$ would be in the range of \$$2.7$ to \$$6.2$ trillion. Healthcare constitutes the major part, about $41$\% of this market, followed by industry and energy with $33$\% and $7$\% of the IoT market, respectively. Other domains such as transportation, agriculture, urban infrastructure, security, and retail have about $15$\% of the IoT market totally. These expectations imply the tremendous and steep growth of the IoT services, their generated data and consequently their related market in the years ahead.

\begin{figure}
	\begin{center}		
		\includegraphics[width=.5\textwidth]{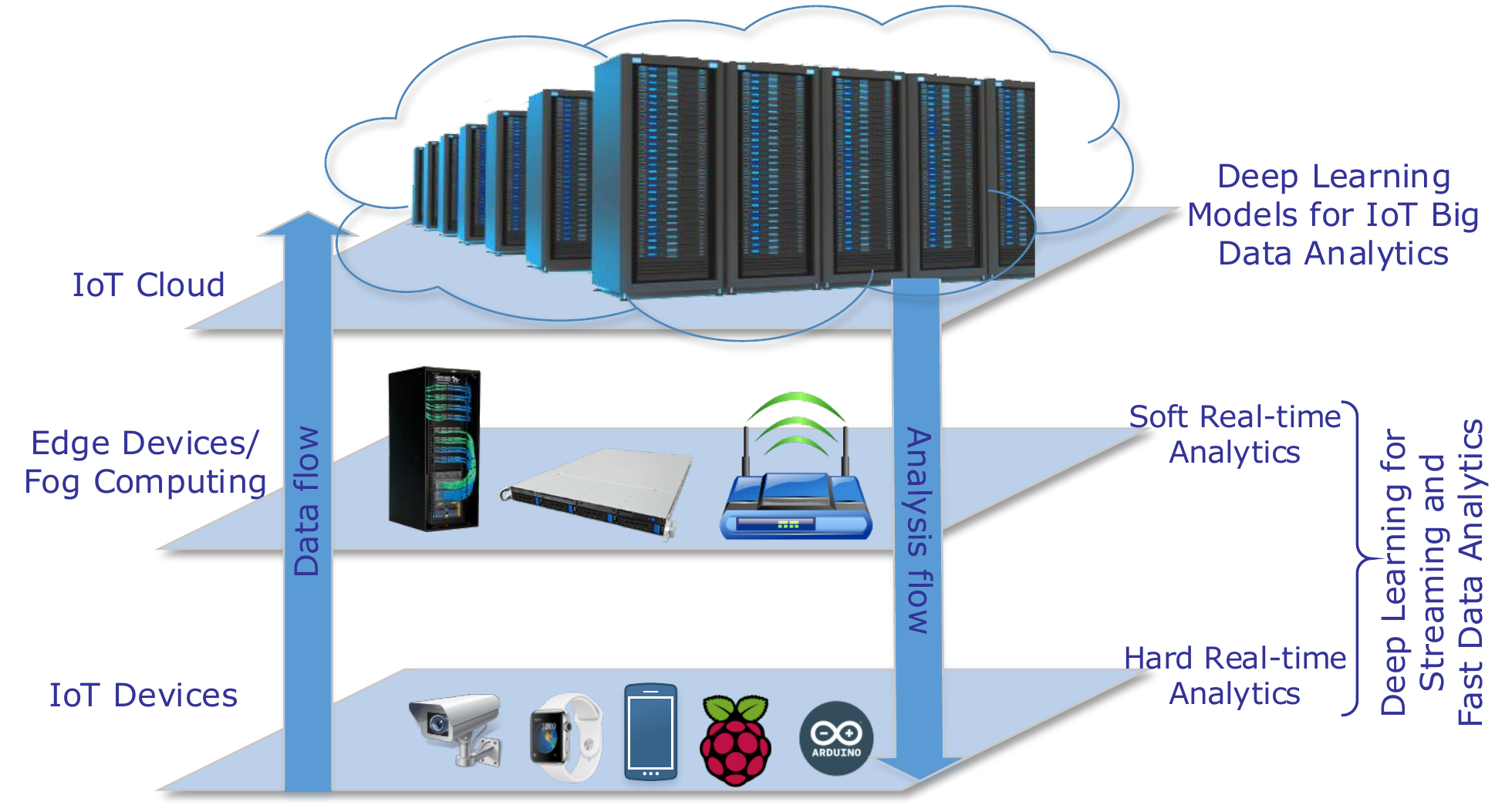}		
	\end{center}
	
	\caption{IoT data generation at different levels and deep learning models to address their knowledge abstraction.}\label{fig:fig-intro}	
\end{figure}

Indeed, machine learning (ML) will have effects on jobs and the workforce, since parts of many jobs may be ``suitable for ML applications" \cite{brynjolfsson2017can}. This will lead to increase in demand for some ML products and the derived demand for the tasks, platforms, and experts needed to produce such products. The economic impact of machine learning in McKinsey's report \cite{manyika2013disruptive} is defined under knowledge work automation; ``the use of computers to perform tasks that rely on complex analyses, subtle judgments, and creative problem solving". The report mentions that advances in ML techniques, such as deep learning and neural networks, are the main enablers of knowledge work automation. Natural user interfaces, such as speech and gesture recognition are other enablers that are highly benefiting from ML technologies. The estimated potential economic impact of knowledge work automation could reach $\$5.2$ trillion to $\$6.7$ trillion per year by $2025$. Figure  shows the break down of this estimate in different occupations. Compared to the economic impact of IoT, this estimation asserts the more attention toward the extraction of value out of data and the potential impacts of ML on the economic situation of individuals and societies. These economic impacts have serious consequences on individuals and countries, since people need to adapt to new means of earning income suitable for them to maintain their desired living standard.

\begin{figure}
	\begin{center}		
		\includegraphics[width=.45\textwidth]{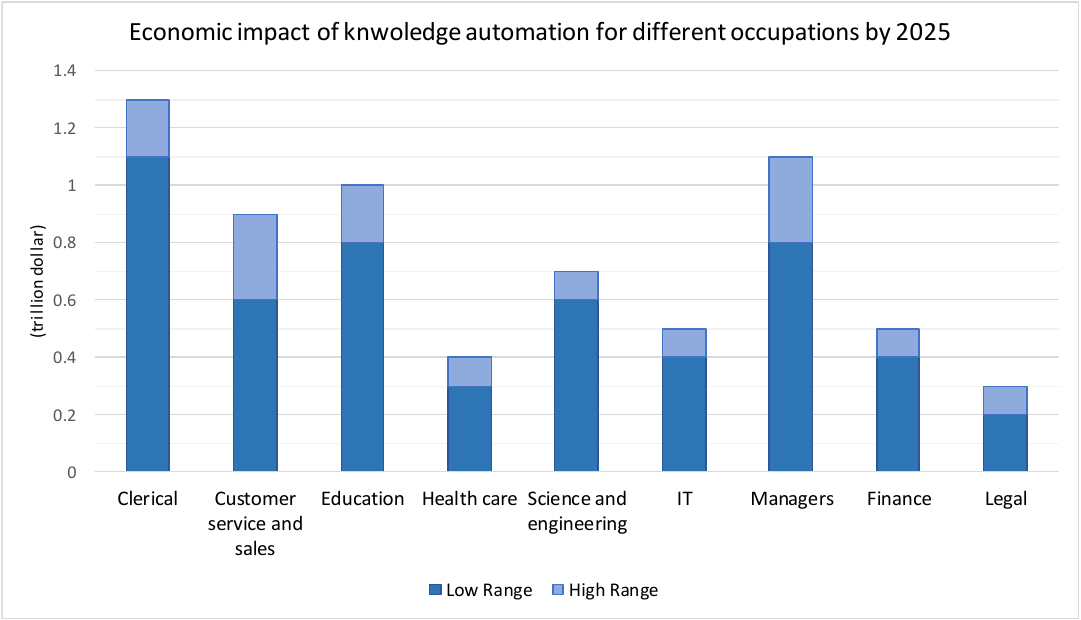}		
	\end{center}
	\caption{The break down of estimated economic impact of \$5.2 trillion to \$6.7 trillion per year for machine learning in 2025.}\label{fig:ml_economic}	
\end{figure}

In recent years, many IoT applications arose in different vertical domains, i.e., health, transportation, smart home, smart city, agriculture, education, etc. The main element of most of these applications is an intelligent learning mechanism for prediction (i.e., regression, classification, and clustering), data mining and pattern recognition or data analytics in general. Among the many machine learning approaches, Deep Learning (DL) has been actively utilized in many IoT applications in recent years. These two technologies (i.e., DL and IoT) are among the top three strategic technology trends for 2017 that were announced at Gartner Symposium/ITxpo 2016 \cite{gartner05_09_2017}. The cause of this intensive publicity for DL refers to the fact that traditional machine learning approaches do not address the emerging analytic needs of IoT systems. Instead, IoT systems need different modern data analytic approaches and artificial intelligence (AI) methods according to the hierarchy of IoT data generation and management as illustrated in Figure~\ref{fig:fig-intro}.

The growing interest in the Internet of Things (IoT) and its derivative big data need stakeholders to clearly understand their definition, building blocks, potentials and challenges. IoT and big data have a two way relationship. On one hand, IoT is a main producer of big data, and on the other hand, it is an important target for big data analytics to improve the processes and services of IoT \cite{mohammadi2018enabling}. Moreover, IoT big data analytics have proven to bring value to the society. For example, it is reported that, by detecting damaged pipes and fixing them, the Department of Park Management in Miami has saved about one million USD on their water bills \cite{chen2014big}.

IoT data are different than the general big data. To better understand the requirements for IoT data analytics, we need to explore the properties of IoT data and how they are different from those of general big data. 
IoT data exhibits the following characteristics \cite{chen2014big}:

\begin{itemize}

\item Large-Scale Streaming Data: A myriad of data capturing devices are distributed and deployed for IoT applications, and generate streams of data continuously. This leads to a huge volume of continuous data. 
\item Heterogeneity: Various IoT data acquisition devices gather different information resulting in data heterogeneity.
\item Time and space correlation: In most of IoT applications, sensor devices are attached to a specific location, and thus have a location and time-stamp for each of the data items. 
\item High noise data: Due to tiny pieces of data in IoT applications, many of such data may be subject to errors and noise during acquisition and transmission.

\end{itemize}

Although obtaining hidden knowledge and information out of big data is promising to enhance the quality of our lives, it is not an easy and straightforward task. For such a complex and challenging task that goes beyond the capabilities of the traditional inference and learning approaches, new technologies, algorithms, and infrastructures are needed \cite{chen2014big-deep}. Luckily, the recent progresses in both fast computing and advanced machine learning techniques are opening the doors for big data analytics and knowledge extraction that is suitable for IoT applications.

Beyond the big data analytics, IoT data calls for another new class of analytics, namely fast and streaming data analytics, to support applications with high-speed data streams and requiring time-sensitive (i.e., real-time or near real-time) actions. Indeed, applications such as autonomous driving, fire prediction, driver/elderly posture (and thus consciousness and/or health condition) recognition demands for fast processing of incoming data and quick actions to achieve their target. Several researchers have proposed approaches and frameworks for fast streaming data analytics that leverage the capabilities of cloud infrastructures and services~\cite{zaharia2012fast,engle2012shark}. However, for the aforementioned IoT applications among others, we need fast analytics in smaller scale platforms (i.e., at the system edge) or even on the IoT devices themselves. For example, autonomous cars need to make fast decisions on driving actions such as lane or speed change. Indeed, this kind of decisions should be supported by fast analytics of possibly multi-modal data streaming from several sources, including the multiple vehicle sensors (e.g., cameras, radars, LIDARs, speedometer, left/right signals, etc.), communications from other vehicles, and traffic entities (e.g., traffic light, traffic signs). In this case, transferring data to a cloud server for analysis and returning back the response is subject to latency that could cause traffic violations or accidents. A more critical scenario would be detecting pedestrians by such vehicles. Accurate recognition should be performed in strict real-time to prevent fatal accidents. These scenarios imply that fast data analytics for IoT have to be close to or at the source of data to remove unnecessary and prohibitive communication delays. 

\subsection{Survey Scope}
DL models in general bring two important improvements over the traditional machine learning approaches in the two phases of training and prediction. First, they reduce the need for hand crafted and engineered feature sets to be used for the training \cite{lecun2015deep}. Consequently, some features that might not be apparent to a human view can be extracted easily by DL models.  In addition, DL models improve the accuracy\footnote{Accuracy in this work in general refers to the degree to which the result of the prediction conforms to the ground truth values. Readers may also face top-2 or top-3 accuracy in the text. In general, top-N accuracies refers to considering the N highest-probability answers of the prediction model and checking whether that set contains the expected value or not. Therefore, top-1 accuracy refers to the output with the highest probability. Likewise, top-3 accuracy refers to the three most probable predictions. For example, if we feed a picture of a tiger to a model that recognizes animal images, and it returns the list of possible outputs as dog:$0.72$, tiger:$0.69$, and cat:$0.58$, the top-1 accuracy will output the answer set containing only ``dog'', which is counted as wrong. On the other hand, the top-2 and top-3 accuracies will result in output sets containing ``tiger'' as an answer, and are thus counted as correct.}.

In this paper, we review a wide range of deep neural network (DNN) architectures and explore the IoT applications that have benefited from DL algorithms. The paper identifies five main foundational IoT services that can be used in different vertical domains beyond the specific services in each domain. It will also discuss the characteristics of IoT applications and the guide to matching them with the most appropriate DL model. This survey focuses on the confluence of two emerging technologies, one in communication networks, i.e., IoT and the other in artificial intelligence, i.e., DL, detailing their potential applications and open issues. The survey does not cover traditional machine learning algorithms for IoT data analytics as there are some other attempts, mentioned in section~\ref{sec:relatedWork}, that have covered such approaches. Moreover, this survey also does not go into the details of the IoT infrastructure from a communications and networking perspective.

\begin{figure*}
	\begin{center}		
		\includegraphics[width=.95\textwidth]{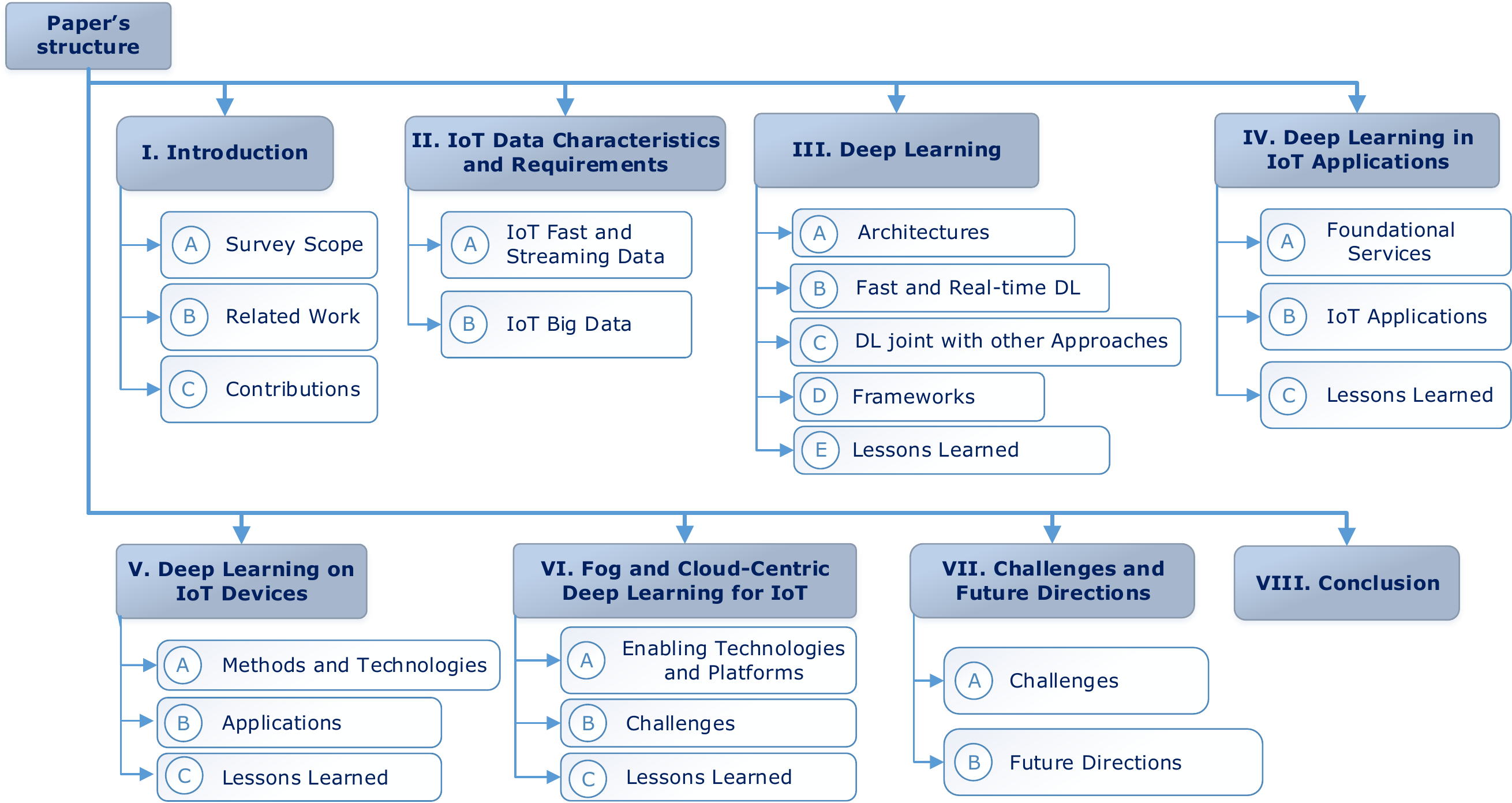}		
	\end{center}	
	\caption{Structure of the survey.}\label{fig:paper_struct}	
\end{figure*}

\subsection{Related Work} \label{sec:relatedWork}

To the best of our knowledge, there does not exist an article in the literature that is dedicated to surveying the specific relation between IoT data and DL as well as applications of DL methods in IoT. There are few works presenting common data mining and machine learning methods that have been used in IoT environments. The work presented in \cite{tsai2014data} by Tsai \textit{et al.} focused on data mining approaches in IoT. It addressed different classification, clustering, and frequent pattern mining algorithms for the IoT infrastructure and services. However, that work did not consider DL approaches, which is the focus of our survey. Moreover, their focus is mainly on offline data mining, while we also consider learning and mining for both real-time (i.e., fast) and big data analytics. 

In \cite{perera2014context}, Perera \textit{et al.} have reviewed different classes of machine learning approaches (supervised and unsupervised, rules, fuzzy logic, etc.) in the reasoning phase of a context-aware computing system, and have discussed the potentials of applying those methods in IoT systems. Nonetheless, they also did not study the role of DL on the context reasoning.  

The work in \cite{alsheikh2014machine} by Alsheikh \textit{et al.} provides a survey of machine learning methods for wireless sensor networks (WSNs). In that work, the authors studied machine learning methods in the functional aspects of WSNs, such as routing, localization, and clustering, as well as non-functional requirements, such as security and quality of service. They reviewed several algorithms in supervised, unsupervised, and reinforcement learning approaches. This work focuses on the infrastructure of WSN (which is one potential infrastructure for implementing IoT applications), while our work is not dependent on the sources of data (i.e., IoT infrastructures) and covers a wide range of IoT applications and services. Moreover, the focus of \cite{alsheikh2014machine} was on traditional machine learning methods, whereas this article focuses on advanced and DL techniques.

Finally, Fadlullah \textit{et al.} \cite{fadlullah2017state} addressed DL approaches in network traffic control systems. While this work primarily focuses on the infrastructure of network, it differs from our work that focuses on the usage of DL in IoT applications. 

Beyond the specific works on the IoT, Qiu \textit{et al.}~\cite{qiu2016survey} reviewed several traditional machine learning techniques along with several advanced techniques including DL for processing general big data. In specific, they highlighted the connection of different machine learning techniques with signal processing technologies to process and analyze timely big data applications.

\subsection{Contributions}\label{sec:contributions}

This paper is intended for IoT researchers and developers who want to build analytics, AI systems, and learning solutions on top of their IoT infrastructure, using the emerging DL machine learning approaches. The contributions of this paper can be summarized as follows: 
\begin{itemize}
\item In order to adopt DL approaches in the IoT ecosystems, we identify the key characteristics and issues of IoT data.
\item Compared to some related work in the literature that have addressed machine learning for IoT, we review the state-of-the-art DL methods and their applicability in the IoT domain both for big data and streaming data analytics.
\item We review a wide range of IoT applications that have used DL in their context. We also provide a comparison and a guideline for using different types of DNN in the various IoT domains and applications.
\item We review the recent approaches and technologies for deploying DL on all levels of IoT hierarchy from resource constrained devices to the fog and the cloud.
\item We highlight the challenges and future research directions for the successful and fruitful merging of DL and IoT applications. 

\end{itemize}

The rest of this paper is organized as follows. In section \ref{sec:iotData}, we highlight the IoT data characteristics and describe what IoT big data as well as fast and streaming data are, and how they are different from the general big data. Section \ref{sec:deeplearning} presents several common and successful architectures of DNNs. It also includes a brief description of advancements toward real-time and fast DL architectures as well as state-of-the-art algorithms that are joint with DL. A succinct review of several frameworks and tools with different capabilities and algorithms that support DNNs is also presented. IoT applications in different domains (e.g., healthcare, agriculture, ITS, etc.) that have used DL will be surveyed in section \ref{sec:applications}. Section \ref{sec:DL_on_devices} reviews the attempts to bring DNN to the resource constraint devices. Section \ref{sec:cloud_centric_DL} explains the works that investigated bringing the DNN models to the scale of fog and cloud computing. Future research direction and open challenges are presented in section~\ref{sec:challenges}. The paper is concluded in Section \ref{sec:Conclusion} with a summary of its main take-away messages. Figure~\ref{fig:paper_struct} depicts the structure of the paper.

\section{IoT Data Characteristics and Requirements for Analytics}\label{sec:iotData}

IoT data can be streamed continuously or accumulated as a source of big data. Streaming data refers to the data generated or captured within tiny intervals of time and need to be promptly analyzed to extract immediate insights and/or make fast decisions. Big data refers to huge datasets that the commonly used hardware and software platforms are not able to store, manage, process, and analyze. These two approaches should be treated differently since their requirements for analytic response are not the same. Insight from big data analytics can be delivered after several days of data generation, but insight from streaming data analytics should be ready in a range of few hundreds of milliseconds to few seconds. 

Data fusion and sharing play a critical role in developing ubiquitous environments based on IoT data. This role is more critical for time-sensitive IoT applications where a timely fusion of data is needed to bring all pieces of data together for analysis and consequently providing reliable and accurate actionable insights. Alam \textit{et al.}~\cite{alam2017data} presented a survey paper in which data fusion techniques for IoT environments are reviewed followed by several opportunities and challenges.

\subsection{IoT fast and streaming data} \label{sec:IoT_fast_data}

Many research attempts suggested streaming data analytics that can be mainly deployed on high-performance computing systems or cloud platforms. The streaming data analytics on such frameworks is based on data parallelism and incremental processing~\cite{li2015supporting}. By data parallelism, a large dataset is partitioned into several smaller datasets, on which parallel analytics are performed simultaneously. Incremental processing refers to fetching a small batch of data to be processed quickly in a pipeline of computation tasks. Although these techniques reduce time latency to return a response from the streaming data analytic framework, they are not the best possible solution for time-stringent IoT applications. By bringing streaming data analytics closer to the source of data (i.e., IoT devices or edge devices) the need for data parallelism and incremental processing is less sensible as the size of the data in the source allows it to be processed rapidly. However, bringing fast analytics on IoT devices introduces its own challenges such as limitation of computing, storage, and power resources at the source of data.

\subsection{IoT Big data} \label{sec:IoT_big_data}
IoT is well-known to be one of the major sources of big data, as it is based on connecting a huge number of smart devices to the Internet to report their frequently captured status of their environments. Recognizing and extracting meaningful patterns from enormous raw input data is the core utility of big data analytics as it results in higher levels of insights for decision-making and trend prediction. Therefore, extracting these insights and knowledge from the big data is of extreme importance to many businesses, since it enables them to gain competitive advantages. In social sciences, Hilbert \cite{hilbert2016big} compares the impact of big data analytics to that of the invention of the telescope and microscope for astronomy and biology, respectively. 

Several works have described the general features of big data from different aspects \cite{hilbert2016big,fan2013mining,hu2014toward,demchenko2013addressing} in terms of volume, velocity, and variety. However, we adopt the general definition of big data to characterize the IoT big data through the following “6V's” features: 

\begin{itemize}

\item Volume: Data volume is a determining factor to consider a dataset as big data or traditional massive/ very large data. The quantity of generated data using IoT devices is much more than before and clearly fits this feature. 
\item Velocity: The rate of IoT big data production and processing is high enough to support the availability of big data in real-time. This justifies the needs for advanced tools and technologies for analytics to efficiently operate given this high rate of data production.
\item Variety: Generally, big data comes in different forms and types. It may consist of structured, semi-structured, and unstructured data. A wide variety of data types may be produced by IoT such as text, audio, video, sensory data and so on.
\item Veracity: Veracity refers to the quality, consistency, and trustworthiness of the data, which in turn leads to accurate analytics. This property needs special attention to hold for IoT applications, especially those with crowd-sensing data. 
\item Variability: This property refers to the different rates of data flow. Depending on the nature of IoT applications, different data generating components may have inconsistent data flows. Moreover, it is possible for a data source to have different rates of data load based on specific times. For example, a parking service application that utilizes IoT sensors may have a peak data load in rush hours. 
\item Value: Value is the transformation of big data to useful information and insights that bring competitive advantage to organizations. A data value highly depends on both the underlying processes/services and the way that data is treated. For example, a certain application (e.g., medical vital sign monitoring) may need to capture all sensor data, while a weather forecast service may need just random samples of data from its sensors. As another example, a credit card provider may need to keep data for a specific period of time and discard them thereafter. 

\end{itemize}

Beyond the aforementioned properties, researchers \cite{hilbert2016big}\cite{hu2014toward} have identified other characteristics such as: 

\begin{itemize}

\item Big data can be a byproduct or footprint of a digital activity or IoT interplay. The use of Google’s most common search terms to predict seasonal flu is a good example of such digital byproduct \cite{ginsberg2009detecting}. 
\item	Big data systems should be horizontally scalable, that is, big data sources should be able to be expanded to multiple datasets. This attribute also leads to the complexity attribute of big data, which in turn imposes other challenges like transferring and cleansing data.

\end{itemize}

Performing analytics over continuous data flows are typically referred to as stream processing or sometimes complex event processing (CEP) in the literature. Strohbach \textit{et al.} \cite{strohbach2015towards} proposed a big data analytics framework for IoT to support the volume and velocity attributes of IoT data analytics. The integration of IoT big data and streaming data analytics, an open issue that needs more investigation, has been also studied as part of that work. However, their proposed framework is designed to be deployed on cloud infrastructures. Moreover, their focus is on the data management aspect of the framework and did not use advanced machine learning models such as DL. Other off-the-shelf products such as Apache Storm are also available for real-time analytics on the cloud. A big gap in this area is the lack of frameworks and algorithms that can be deployed on the fog (i.e., system edge) or even on the IoT devices. When DL comes to play in such cases, a trade-off between the depth and performance of the DNN should be considered.

\section{Deep Learning}\label{sec:deeplearning}

\begin{figure}	
	\begin{center}		
		\includegraphics[width=0.5\textwidth]{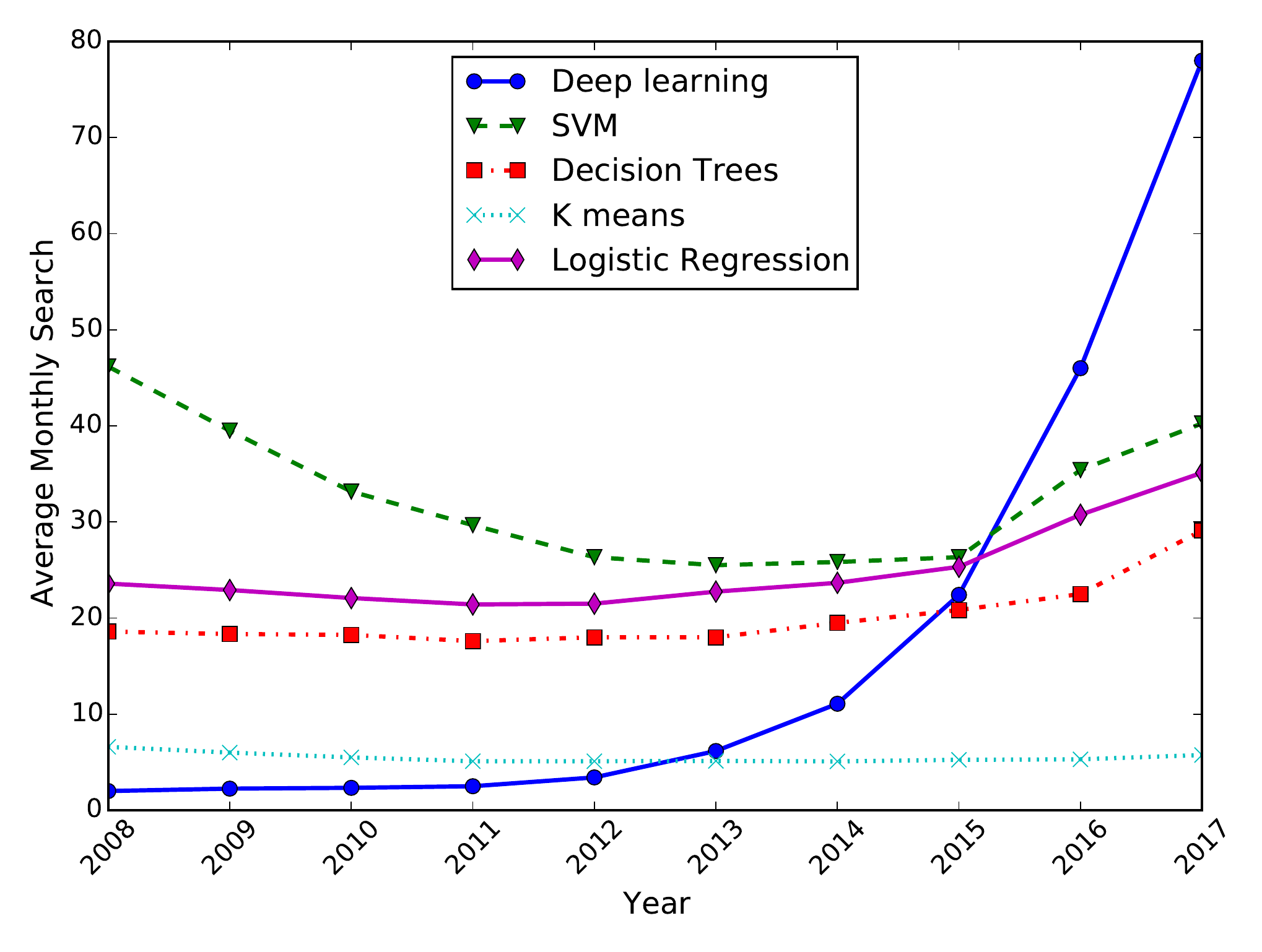}		
	\end{center}
	\caption{Google Trend showing more attention toward deep learning in recent years.}\label{fig:googleTrend}	
\end{figure}

DL consists of supervised or unsupervised learning techniques based on many layers of Artificial Neural Networks (ANNs) that are able to learn hierarchical representations in deep architectures. DL architectures consist of multiple processing layers. Each layer is able to produce non-linear responses based on the data from its input layer. The functionality of DL is imitated from the mechanisms of human brain and neurons for processing of signals. 

DL architectures have gained more attention in recent years compared to the other traditional machine learning approaches. Such approaches are considered as being shallow-structured learning architectures versions  (i.e., a limited subset) of DL. Figure~\ref{fig:googleTrend} shows the searching trend of five popular machine learning algorithms in Google trends, in which DL is becoming more popular among the others. Although ANNs have been introduced in the past decades, the growing trend for DNNs started in 2006 when G. Hinton \textit{et al.} presented the concept of deep belief networks \cite{hinton2006reducing}. Thereafter, the state-of-the-art performance of this technology has been observed in different fields of AI including image recognition, image retrieval, search engines and information retrieval, and natural language processing.

DL techniques have been developed on top of traditional ANNs. Feed-forward Neural Networks (FNNs) \cite{svozil1997introduction} (a.k.a Multilayer Perceptrons - MLPs) have been used in the past decades to train systems, but when the number of layers is increased, they become difficult to train \cite{schmidhuber2015deep}. The small size of training data was another factor that results in overfitted models. Moreover, the limitation in computational capabilities in those days prohibited the implementation of efficient deeper FNNs. These computational limitations have been resolved lately due to hardware advances in general and the development of Graphics Processing Units (GPUs) and hardware accelerators specifically. Beyond the structural aspects and significance of depth of DL architectures, as well as hardware advances, DL techniques have benefited from advancements in effective training algorithms of deep networks including: 
\begin{itemize}
\item Using Rectified Linear Units (ReLUs) as activation function \cite{glorot2011deep},
\item Introducing dropout methods \cite{hinton2012improving},
\item Random initialization for the weights of the network\cite{sutskever2013importance},
\item Addressing the degradation of training accuracy by residual learning networks \cite{he2016deep},
\item Solving vanishing gradient problem as well as exploding gradient problem by introducing and enhancing Long Short-Term Memory networks \cite{hochreiter1997long,mikolov2014learning}.
\end{itemize}

One advantage of DL architectures, compared to the traditional ANNs, is that DL techniques can learn hidden features from the raw data \cite{lecun2015deep}. Each layer trains on a set of features based on the previous layer’s outputs. The inner-most layers can recognize more complex features, since they aggregate and recombine features from the previous layers. This is called the hierarchy of features. For example, in case of a face recognition model, raw image data of portraits as vector of pixels are fed to a model in its input layer. Each hidden layer can then learn more abstract features from the previous layer's outputs, e.g., the first hidden layer identifies the lines and edges, the second layer identifies face parts such as nose, eyes, etc., and the third layer combines all the previous features to generate a face.

However, the reported improvements of DL models are based on empirical evaluations, and there is still no concrete analytical foundation to answer why DL techniques outperform their shallow counterparts. Moreover, there is no clear boundary between deep and shallow networks based on the number of hidden layers. Generally, neural networks with two or more hidden layers that incorporate the recent advanced training algorithms are considered as deep models. Also, recurrent neural networks with one hidden layer are considered as deep since they have a cycle on the units of the hidden layer, which can be unrolled to an equivalent deep network.

\subsection{Architectures}
In this section, we present a brief overview of several common DL models as well as the most cutting-edge architectures that have been introduced in recent years. Interested readers can refer to other literature that surveyed the models and architectures of DL in more details, such as \cite{deng2014tutorial}. Table~\ref{tbl:modelsComparison} summarizes these models, their attributes, characteristics, and some sample applications.

\begin{figure}	
	\begin{center}		
		\includegraphics[width=.45\textwidth]{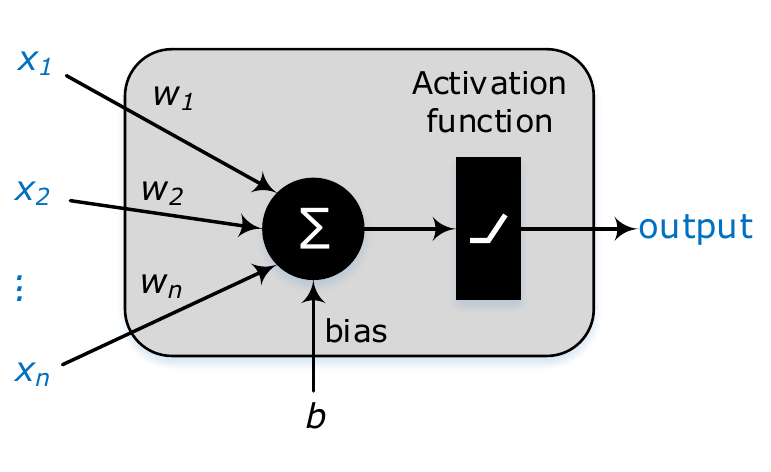}		
	\end{center}
	
	\caption{A neuron is a unit of artificial neural networks, with several inputs and trainable weights and bias.}\label{fig:neuron}	
\end{figure}

A DNN consists of an input layer, several hidden layers, and an output layer. Each layer includes several units called neurons. A neuron receives several inputs, performs a weighted summation over its inputs, then the resulting sum goes through an activation function to produce an output. Each neuron has a vector of weights associated to its input size as well as a bias that should be optimized during the training process. Figure~\ref{fig:neuron} depicts the structure of a neuron.  

\begin{figure}	
	\begin{center}		
		\includegraphics[width=.45\textwidth]{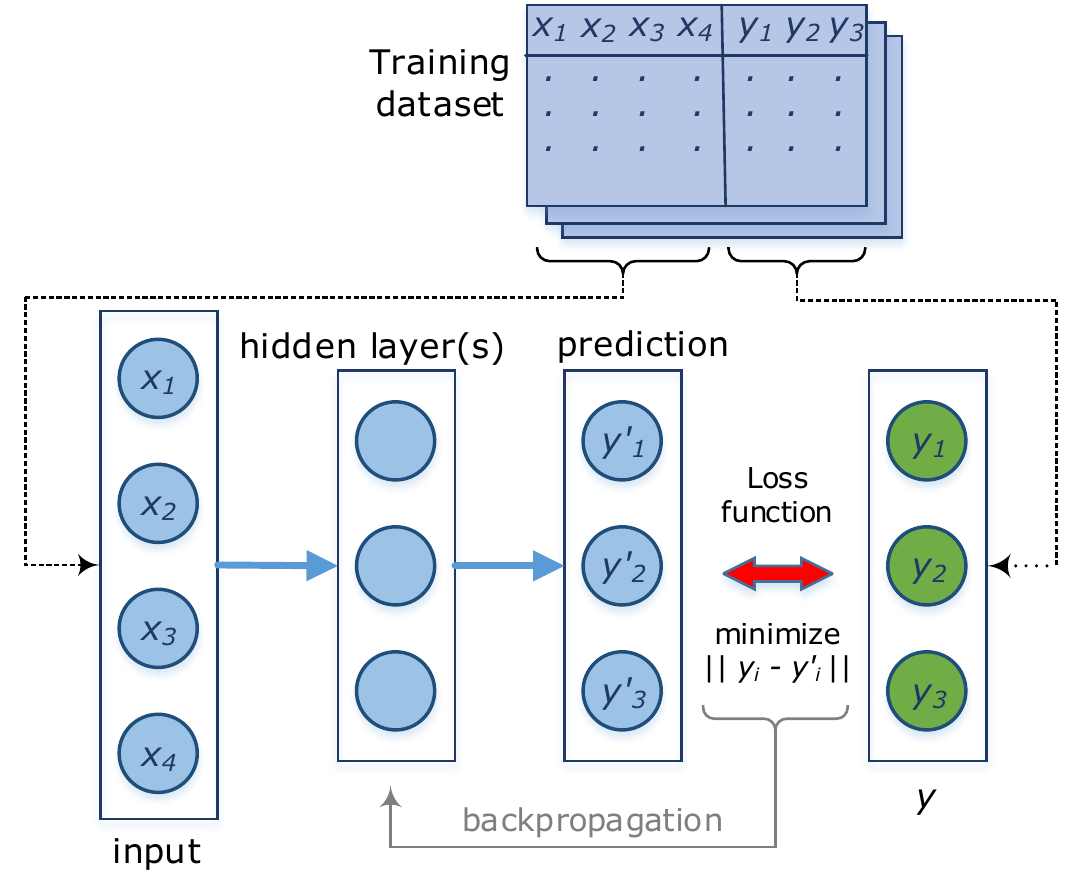}		
	\end{center}
	
	\caption{The overall mechanism of training of a DL model.}\label{fig:DL_mechanism}	
\end{figure}

In the training process, the input layer assigns (usually randomly) weights to the input training data and passes it to the next layer. Each subsequent layer also assigns weights to their input and produces their output, which serves as the input for the following layer. At the last layer, the final output representing the model prediction is produced. A loss function determines the correctness of this prediction by computing the error rate between the predicted and true values. An optimization algorithm such as Stochastic Gradient Descent (SGD) \cite{bottou2010large} is used to adjust the weight of neurons by calculating the gradient of the loss function. The error rate is propagated back across the network to the input layer (known as backpropagation algorithm \cite{rumelhart1986learning,chauvin1995backpropagation}). The network then repeats this training cycle, after balancing the weights on each neuron in each cycle, until the error rate falls below a desired threshold. At this point, the DNN is trained and is ready for inference. In Figure~\ref{fig:DL_mechanism}, the high level mechanism of training for DL models is illustrated.

In a broad categorization, DL models fall into three categories, namely generative, discriminative, and hybrid models. Though not being a firm boundary, discriminative models usually provide supervised learning approaches, while generative models are used for unsupervised learning. Hybrid models incorporate the benefits of both discriminative and generative models.
\\

\begin{table*}
\renewcommand{\arraystretch}{1.25}
\centering
\caption{Summary of deep learning models.}
\label{tbl:modelsComparison}
\begin{tabular}{|l|l|l|l|l|l|}
\hline
\multicolumn{1}{|c|}{\textbf{Model}} & \multicolumn{1}{c|}{\textbf{Category}} & \multicolumn{1}{c|}{\textbf{Learning model}}                                 & \multicolumn{1}{c|}{\textbf{Typical input data}} & \multicolumn{1}{c|}{\textbf{Characteristics}}  &  \multicolumn{1}{c|}{\textbf{Sample IoT Applications}}                                                                                  \\ \hline \hline
AE                          & Generative                    & Unsupervised                                                        & Various                                 & \begin{tabular}[c]{@{}l@{}} \tabitem Suitable for feature extraction,\\ dimensionality reduction \\ \tabitem Same number of input and\\ output units \\\tabitem The output reconstructs\\ input data \\ \tabitem Works with unlabeled data\end{tabular}     & \begin{tabular}[c]{@{}l@{}}\tabitem Machinery fault diagnosis\\ \tabitem Emotion recognition \end{tabular}              \\ \hline
RNN                         & Discriminative                & Supervised                                                          & Serial, time-series                     & \begin{tabular}[c]{@{}l@{}}\tabitem Processes sequences of data \\ through internal memory \\ \tabitem Useful in IoT applications \\with time-dependent data\end{tabular}  & \begin{tabular}[c]{@{}l@{}}\tabitem Identify movement pattern\\ \tabitem Behavior detection \end{tabular}                         \\ \hline
RBM                         & Generative                    & \begin{tabular}[c]{@{}l@{}}Unsupervised, \\ Supervised\end{tabular} & Various                                 & \begin{tabular}[c]{@{}l@{}}\tabitem Suitable for feature \\extraction, dimensionality\\ reduction, and classification \\ \tabitem Expensive training procedure\end{tabular} & \begin{tabular}[c]{@{}l@{}}\tabitem Indoor localization\\ \tabitem Energy consumption \\prediction \end{tabular} \\ \hline
DBN                         & Generative                    & \begin{tabular}[c]{@{}l@{}}Unsupervised, \\ Supervised\end{tabular} & Various                                 & \begin{tabular}[c]{@{}l@{}}\tabitem Suitable for hierarchical \\features discovery\\ \tabitem Greedy training of the \\network layer by layer \end{tabular}  & \begin{tabular}[c]{@{}l@{}}\tabitem Fault detection classification\\ \tabitem Security threat identification \end{tabular}                              \\ \hline
LSTM                        & Discriminative                & Supervised                                                          & \begin{tabular}[c]{@{}l@{}} Serial, time-series, \\long time dependent data\end{tabular}                    & \begin{tabular}[c]{@{}l@{}}\tabitem Good performance with data \\of long time lag\\ \tabitem Access to memory cell is \\protected by gates\end{tabular}     &  \begin{tabular}[c]{@{}l@{}}\tabitem Human activity recognition\\ \tabitem Mobility prediction \end{tabular}                           \\ \hline
CNN                         & Discriminative                & Supervised                                                          & 2-D (image, sound, etc.)                & \begin{tabular}[c]{@{}l@{}}\tabitem Convolution layers take\\ biggest part of computations\\ \tabitem Less connection compared \\to DNNs. \\ \tabitem Needs a large training \\dataset for visual tasks.\end{tabular}  &  \begin{tabular}[c]{@{}l@{}}\tabitem Plant disease detection\\ \tabitem Traffic sign detection \end{tabular}                     \\ \hline
VAE                         & Generative                    & Semi-supervised                                                     & Various                                 &  \begin{tabular}[c]{@{}l@{}} \tabitem A class of Auto-encoders\\ \tabitem Suitable for scarcity of \\labeled data  \end{tabular}  &  \begin{tabular}[c]{@{}l@{}}\tabitem Intrusion detection\\ \tabitem Failure detection \end{tabular}                                                                               \\ \hline
GAN                         & Hybrid                        & Semi-supervised                                                     & Various                                 & \begin{tabular}[c]{@{}l@{}}\tabitem Suitable for noisy data \\ \tabitem Composed of two networks: \\a generator and a discriminator\end{tabular}  & \begin{tabular}[c]{@{}l@{}}\tabitem Localization and wayfinding\\ \tabitem Image to text \end{tabular}                                                   \\ \hline
Ladder Net                  & Hybrid                        & Semi-supervised                                                     & Various                                 & \begin{tabular}[c]{@{}l@{}} \tabitem Suitable for noisy data \\ \tabitem Composed of three networks: \\two encoders and one decoder \end{tabular} & \begin{tabular}[c]{@{}l@{}}\tabitem Face recognition\\ \tabitem Authentication \end{tabular}
\\ \hline
\end{tabular}
\end{table*}

\subsubsection{Convolutional Neural Networks (CNNs)}$\quad$\\
For vision-based tasks, DNNs with a dense connection between layers are hard to train and do not scale well. One important reason is the translation-invariance property of such models. They thus do not learn the features that might transform in the image (e.g., rotation of hand in pose detection). CNNs have solved this problem by supporting translation-equivariance computations. A CNN receives a 2-D input (e.g., an image or speech signal) and extracts high level features through a series of hidden layers. The hidden layers consist of convolution layers as well as fully connected layers at the end. The convolution layer is at the core of a CNN and consists of a set of learnable parameters, called filters, that have the same shape as the input's shape but with smaller dimensions. In the training process, the filter of each convolutional layer goes through the whole input volume (e.g., in case of an image, it goes across the width and length of the image) and calculates an inner product of the input and the filter. This computation over the whole input leads to a feature map of the filter.

Another building block of a CNN is the pooling layers, which operate on the feature maps. The objective of having pooling layers is to reduce the spatial size of the representation, in order to both cut down the number of parameters and computation times and to reduce the chance of overfitting. \textit{Max pooling} is a common approach that partitions the input space into non-overlapping regions and picks the maximum value for each region. 

\begin{figure}	
	\begin{center}		
		\includegraphics[width=.5\textwidth]{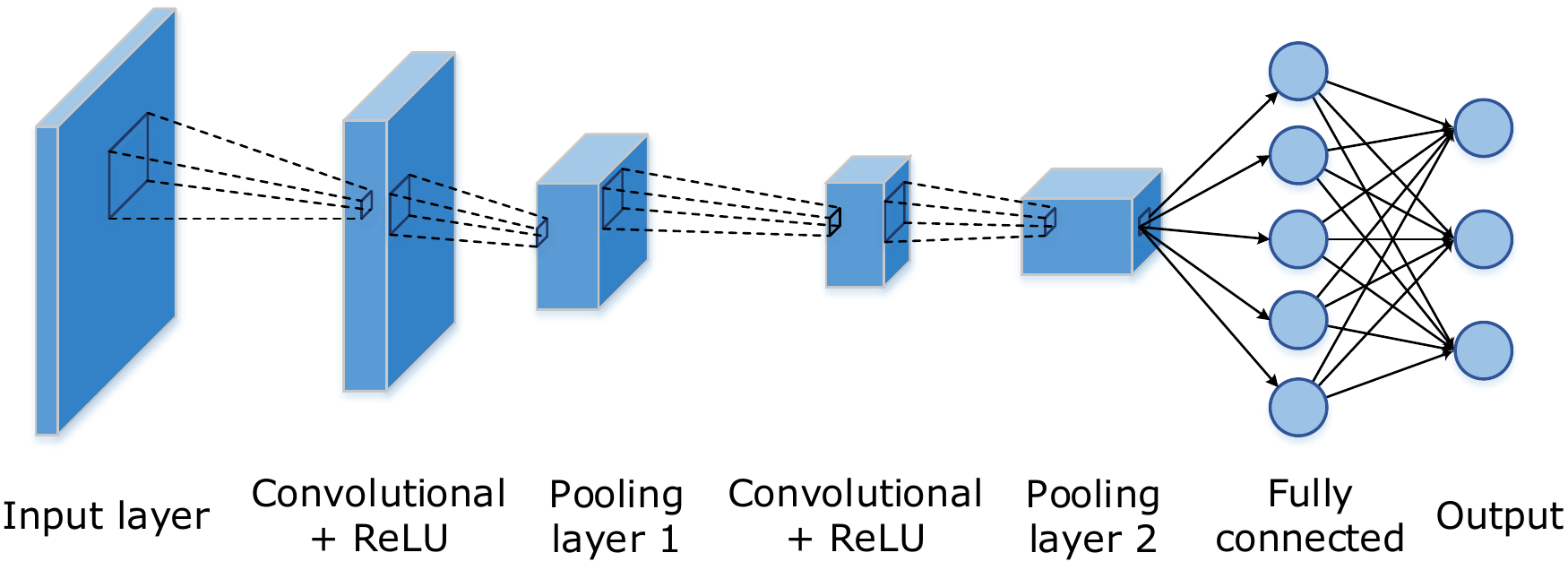}		
	\end{center}
	\caption{Architecture of a CNN.}\label{fig:cnn_model}
\end{figure}

The last important component in CNN is ReLU, which consist of neurons with activation function in the form of $f(x) = \max (0, x)$. The introduction of this activation function in CNN results in a faster training time without affecting the generalization of the network in a sensible negative way \cite{krizhevsky2012imagenet}. Figure~\ref{fig:cnn_model} depicts the structure of a CNN.

A main difference between CNNs and fully connected networks is that each neuron in CNNs is connected only to a small subset of the input. This decreases the total number of parameters in the network and enhances the time complexity of the training process. This property is called local connectivity. 

Many IoT devices, such as drones, smart phones, and smart connected cars, are equipped with cameras. The CNN architecture and its variations have been investigated for a variety of application scenarios that involve these devices. Some typical applications include flood or landslide prediction through drone images, plant disease detection using plant pictures on smart phones, and traffic sign detection using vehicles' cameras.\\

\subsubsection{Recurrent Neural Networks (RNNs)}$\quad$\\
In many tasks, prediction is dependent on several previous samples such that, in addition to classifying individual samples, we also need to analyze the sequences of inputs. In such applications, a feed-forward neural network is not applicable since it assumes no dependency between input and output layers. RNNs have been developed to address this issue in sequential (e.g., speech or text) or time-series problems (sensor's data) with various length. Detecting drivers' behaviors in smart vehicles, identifying individual's movement patterns, and estimating energy consumption of a household are some examples where RNNs can be applied. The input to an RNN consists of both the current sample and the previous observed sample. In other words, the output of an RNN at time step $t-1$ affects the output at time step $t$. Each neuron is equipped with a feedback loop that returns the current output as an input for the next step. This structure can be expresses such that each neuron in an RNN has an internal memory that keeps the information of the computations from the previous input. 

\begin{figure}
	\begin{center}		
		\includegraphics[width=.45\textwidth]{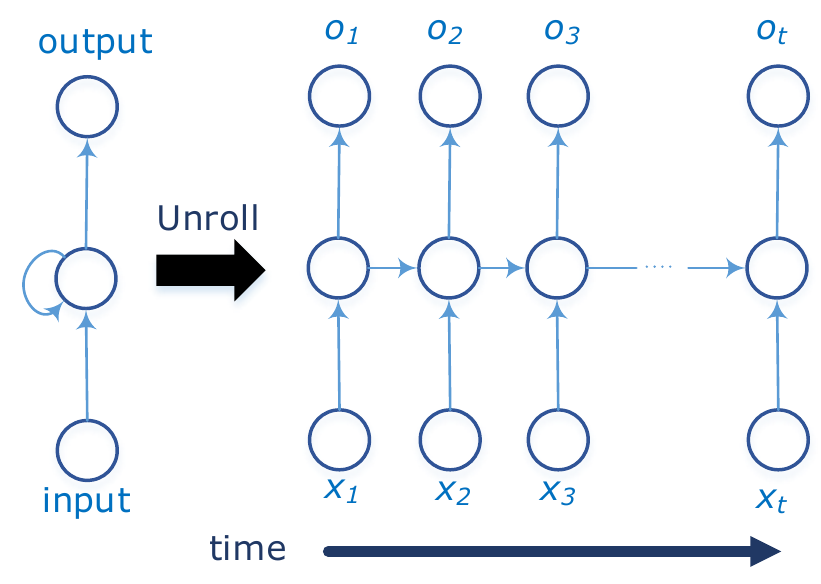}		
	\end{center}
	
	\caption{Structure of a recurrent neural network.}\label{fig:rnn}	
\end{figure}

To train the network, an extension of the backpropagation algorithm, called Backpropagation Through Time (BPTT) \cite{werbos1990backpropagation}, is used. Due to the existence of cycles on the neurons, we cannot use the original backpropagation here, since it works based on error derivation with respect to the weight in their upper layer, while we do not have a stacked layer model in RNNs. The core of BPTT algorithm is a technique called unrolling the RNN, such that we come up with a feed-forward network over time spans. Figure~\ref{fig:rnn} depicts the structure of an RNN and unrolled concept. 

Traditional RNNs can be considered as deep models since they can be seen as several non-linear layers of neurons between the input layer and the output layer when they are unfolded in time~\cite{pascanu2013construct}. However, considering the architecture and the functionality of RNNs, the hidden layers in RNNs are supposed to provide a memory instead of a hierarchical representation of features~\cite{hermans2013training}. There are several approaches to make RNNs deeper, including adding more layers between the input and hidden layers, stacking more hidden layers, and adding more layers between hidden layers and the output layer~\cite{pascanu2013construct}.\\ 

\subsubsection{Long Short Term Memory (LSTM)} 
\begin{figure}
	\begin{center}		
		\includegraphics[width=.45\textwidth]{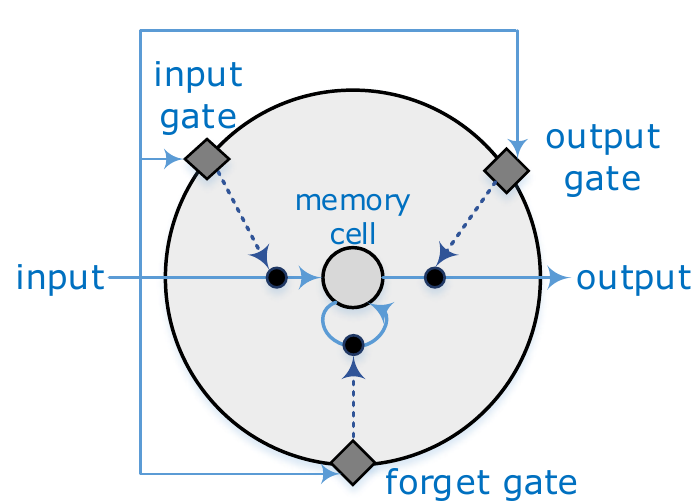}		
	\end{center}
	
	\caption{Structure of a LSTM memory cell. Solid arrow lines show the flow of data and dashed arrow lines show the signals coming from gates.}\label{fig:lstm_cell}	
\end{figure}

$\quad$\\
LSTM is an extension of RNNs. Different variations of LSTM have been proposed, though most of them have followed the same design of the original network \cite{hochreiter1997long}. LSTM uses the concept of gates for its units, each computing a value between 0 and 1 based on their input. In addition to a feedback loop to store the information, each neuron in LSTM (also called a memory cell) has a multiplicative forget gate, read gate, and write gate. These gates are introduced to control the access to memory cells and to prevent them from perturbation by irrelevant inputs. When the forget gate is active, the neuron writes its data into itself. When the forget gate is turned off by sending a $0$, the neuron forgets its last content. When the write gate is set to $1$, other connected neurons can write to that neuron. If the read gate is set to $1$, the connected neurons can read the content of the neuron. Figure~\ref{fig:lstm_cell} depicts this structure.

An important difference of LSTMs compared to RNNs is that LSTM units utilize forget gates to actively control the cell states and ensure they do not degrade. The gates can use \textit{sigmoid} or \textit{tanh} as their activation function. In fact, these activation functions cause the problem of vanishing gradient during backpropagation in the training phase of other models using them. By learning what data to remember in LSTMs, stored computations in the memory cells are not distorted over time. BPTT is a common method for training the network to minimize the error.  

When data is characterized by a long dependency in time, LSTM models perform better than RNN models \cite{chung2014empirical}. This long lag of dependency can be observed in IoT applications such as human activity recognition, predicting educational performance in online programs, and disaster prediction based on environmental monitoring, to name a few.\\

\subsubsection{Autoencoders (AEs)} $\quad$\\
AEs consist of an input layer and an output layer that are connected through one or more hidden layers. AEs have the same number of input and output units. This network aims to reconstruct the input by transforming inputs into outputs with the simplest possible way, such that it does not distort the input very much. This kind of neural networks has been used mainly for solving unsupervised learning problems as well as transfer learning \cite{baldi2012autoencoders}. Due to their behavior of constructing the input at the output layer, AEs are mainly used for diagnosis and fault detection tasks. This is of great interest for industrial IoT to serve many applications such as fault diagnosis in hardware devices and machines, and anomaly detection in the performance of assembly lines.

\begin{figure}[]
	\begin{center}		
		\includegraphics[width=.45\textwidth]{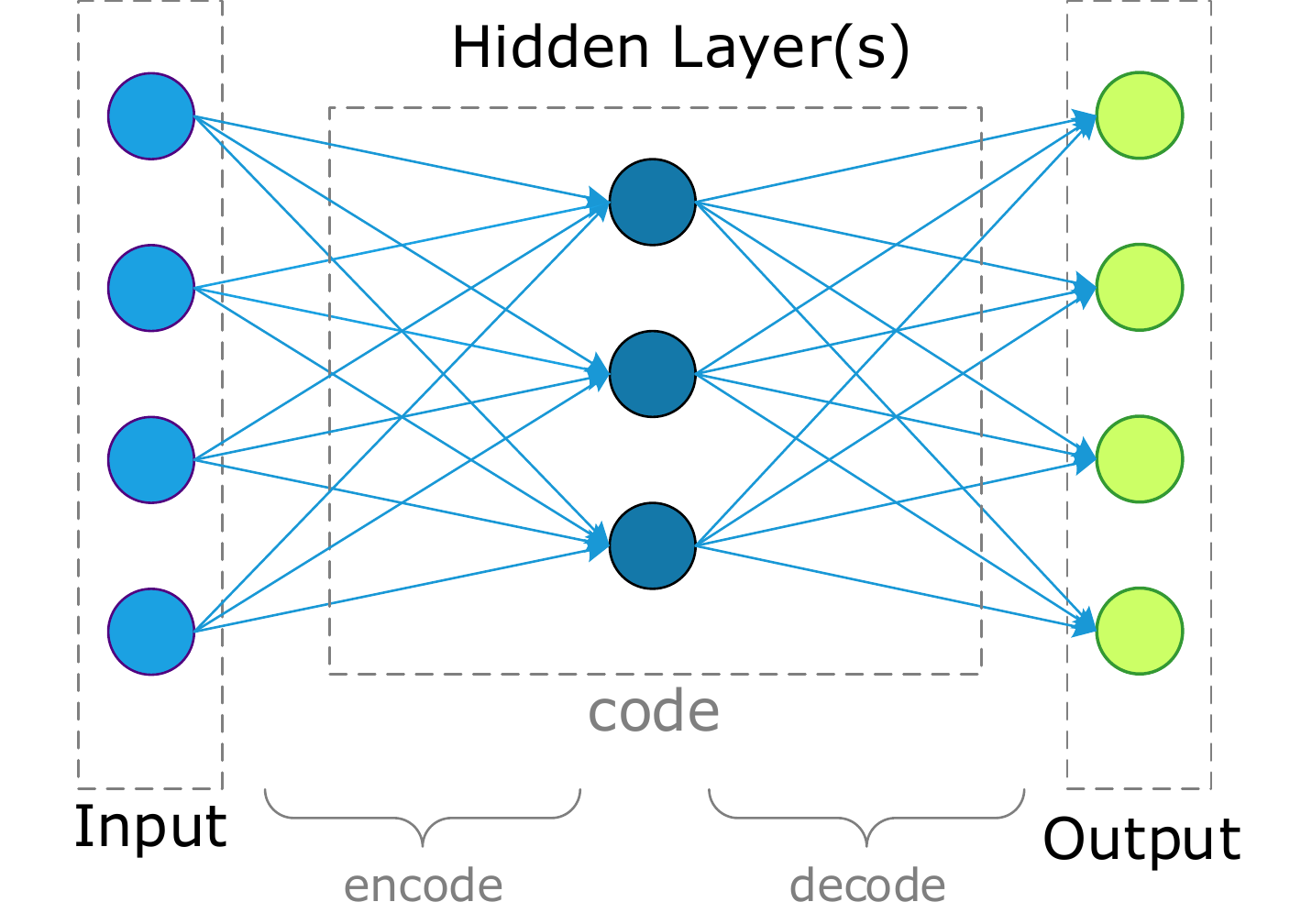}		
	\end{center}
	\caption{Structure of an autoencoder network.}\label{fig:autoencoder}	
\end{figure}

AEs have two main components: An encoder and a decoder. The encoder receives the input and transforms it to a new representation, which is usually called a code or latent variable. The decoder receives the generated code at the encoder, and transforms it to a reconstruction of the original input. The training procedure in AEs involves minimizing reconstruction error, i.e., the output and input showing minimal difference. Figure~\ref{fig:autoencoder} illustrates the structure of a typical AE. There are several variations and extensions of AEs like denoising AE, contractive AE, stacked AE, sparse AE, and variational AE. \\

\subsubsection{Variational Autoencoders (VAEs)}
\begin{figure}[t]
	\begin{center}		
		\includegraphics[width=.45\textwidth]{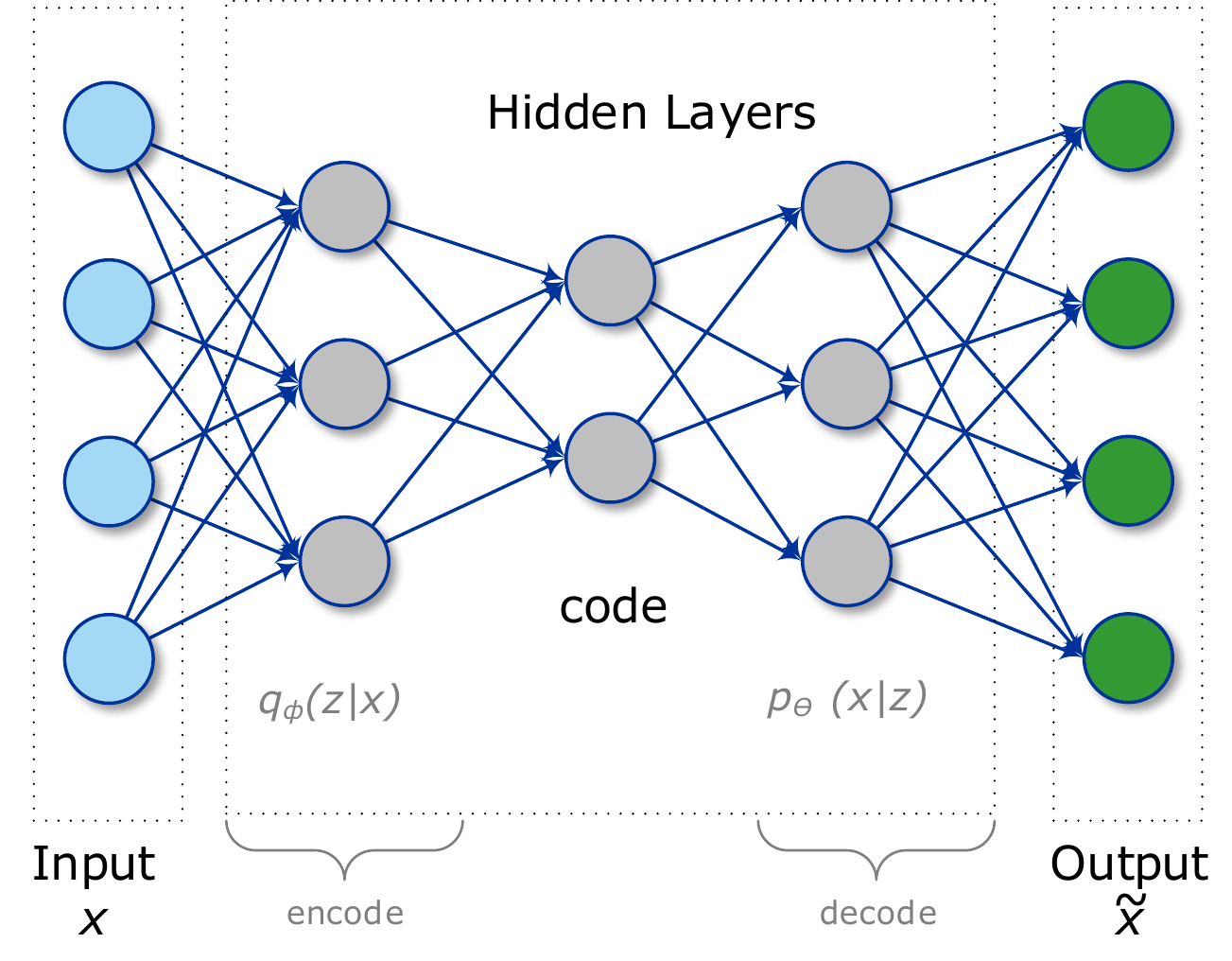}		
	\end{center}
	\caption{Structure of a variational autoencoder network.}\label{fig:vae}	
\end{figure}

$\quad$\\
VAEs, introduced in 2013, are a popular generative model framework whose assumptions on the structure of the data is not strong, while having a fast training process through backpropagation~\cite{doersch2016tutorial}. Moreover, this model has been used for semi-supervised learning~\cite{kingma2014semi}. Therefore, it is a good fit for IoT solutions that deal with diverse data and the scarcity of labeled data. Such applications include failure detection in sensing or actuating levels and intrusion detection in security systems. For each data point $\mathbf{x}$, there is a vector of corresponding latent variables denoted by $\mathbf{z}$. The training architecture of a VAE consists of an encoder and a decoder with parameters $\phi$ and $\theta$, respectively. A fixed form distribution $q_{\phi}(\mathbf{z}|\mathbf{x})$ helps the encoder in estimating the posterior distribution $p_{\theta}(\mathbf{z}|\mathbf{x})$. The model consists of two networks: One generating samples and the other performing approximate inference.
A schematic of the VAE is depicted in Figure~\ref{fig:vae}. \\

\subsubsection{Generative Adversarial Networks (GANs)}
\begin{figure}[t]	
	\begin{center}		
		\includegraphics[width=.45\textwidth]{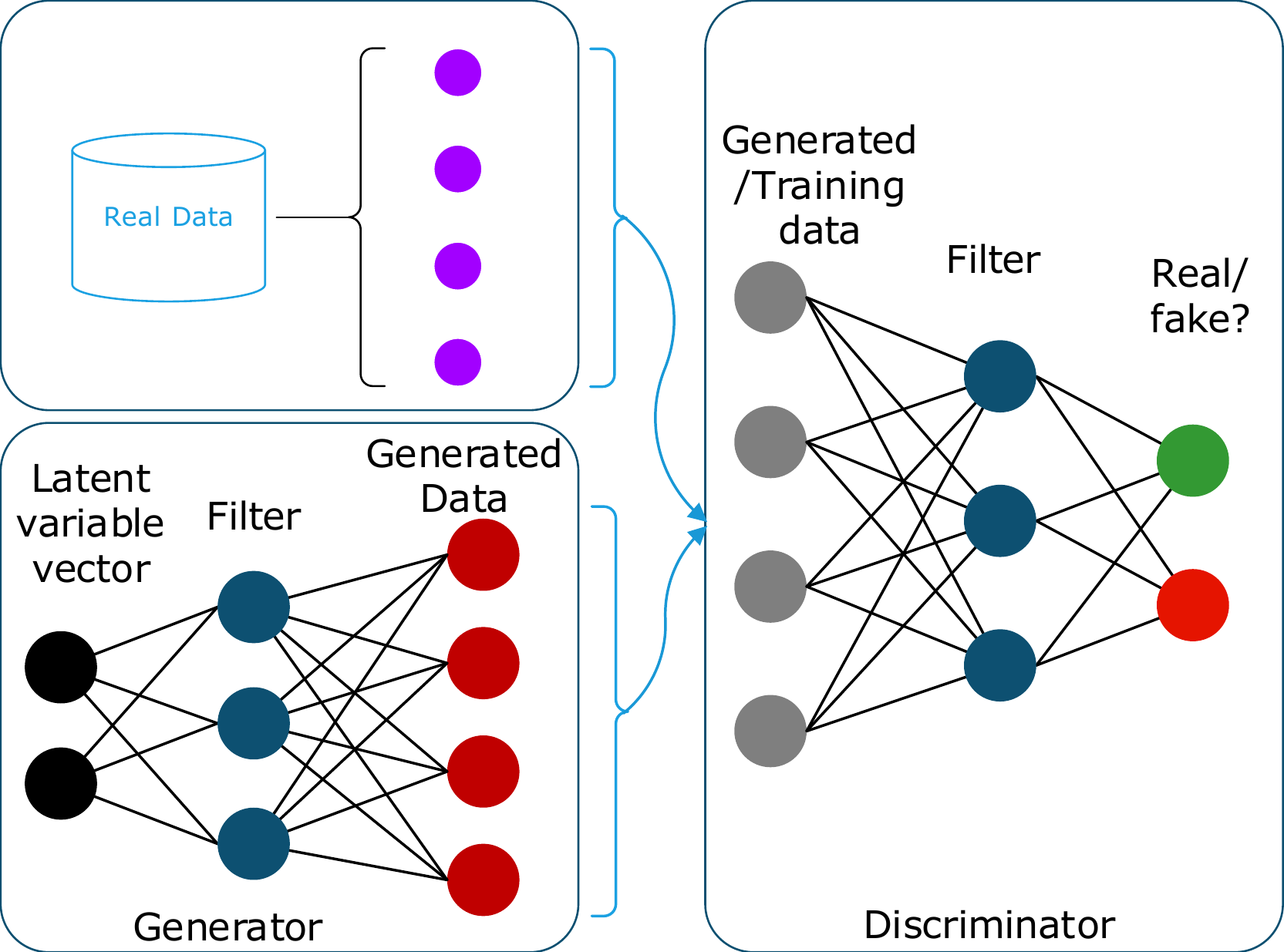}		
	\end{center}
	\caption{Concept of a generative adversarial network.}\label{fig:gan}	
\end{figure}

$\quad$\\
GANs, introduced by Goodfellow \textit{et al.}~\cite{goodfellow2014generative}, consist of two neural networks, namely the generative and discriminative networks, which work together to produce synthetic and high-quality data. The former network (a.k.a. the generator) is in charge of generating new data after it learns the data distribution from a training dataset. The latter network (a.k.a. the discriminator) performs discrimination between real data (coming from training data) and fake input data (coming from the generator). The generative network is optimized to produce input data that is deceiving for the discriminator (i.e., data that the discriminator cannot easily distinguish whether it is fake or real). In other words, the generative network is competing with an adversary discriminative network. Figure~\ref{fig:gan} depicts the concept of GANs. 

The objective function in GANs is based on minimax games, such that one network tries to maximize the value function and the other network wants to minimize it. In each step of this imaginary game, the generator, willing to fool the discriminator, plays by producing a sample data from random noise. On the other hand, the discriminator receives several real data examples from the training set along with the samples from the generator. Its task is then to discriminate real and fake data. The discriminator is considered to perform satisfactorily if its classifications are correct. The generator also is performing well if its examples have fooled the discriminator. Both discriminator and generator parameters then are updated to be ready for the next round of the game. The discriminator's output helps the generator to optimize its generated data for the next round. 

In IoT applications, GANs can be applied for scenarios that require the creation of something new out of the available data. This can include applications in localization and way-finding, where a generator network in GAN produces potential paths between two points, while the discriminator identifies which paths look viable. GANs are also very helpful for developing services for visually impaired people, such as images-to sound-converters using both GANs to generate descriptive texts from a given image \cite{dai2017towards} and another DL model to perform text-to-speech conversion. In an image processing research using GANs, a large number of real celebrity snapshots have been analyzed to create new fake images such that a human cannot identify if they are real images or not \cite{nytimes2018gan}.\\

\subsubsection{Restricted Boltzmann Machine (RBMs)} $\quad$\\
An RBM is a stochastic ANN that consists of two layers: A visible layer that contains the input that we know, and a hidden layer that contains the latent variables. The restriction in RBMs is applied to the connectivity of neurons compared to Boltzmann machine. RBMs should build a bipartite graph, such that each visible neuron should be connected to all hidden neurons and vice versa, but there is no connection between any two units in a same layer. Moreover, the bias unit is connected to all of the visible and hidden neurons. RBMs can be stacked to form DNNs. They are also the building block of deep belief networks. 

\begin{figure}
	\begin{center}		
		\includegraphics[width=.45\textwidth]{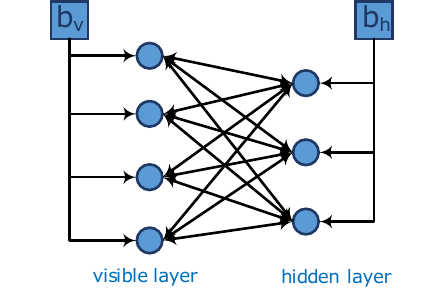}		
	\end{center}
	\caption{Structure of a restricted Boltzmann machine. The visible and hidden layers have separate bias.}\label{fig:rbm}	
\end{figure}

The training data is assigned to visible units. The training procedure can use backpropagation and gradient descent algorithms to optimize the weights of the network. The objective of training RBM is to maximize the product of all probabilities of the visible units. The functionality of RBM is similar to the AEs as it uses forward feeding to compute the latent variables, which are in turn used to reconstruct the input using backward feeding. The structure of an RBM is shown in Figure~\ref{fig:rbm}. 

RBMs can perform feature extraction out of input data. This happens through modeling a probability distribution over a set of inputs that is represented in a set of hidden units. For example, having a set of favorite movies of individuals, an RBM model can have a visible layer consisting of as many neurons as the number of available movies, and a hidden layer consisting of three neurons to represent three different genres such as drama, action and comedy. So, based on the application, the hidden layer can be considered as the output layer. Or it can be complemented with an additional classifier layer to perform classification based on extracted features.

From the types of potential applications where RBMs can be used, we name indoor localization, energy consumption prediction, traffic congestion prediction, posture analysis, and generally any application that benefits from extracting the most important features out of the available ones.\\

\subsubsection{Deep Belief Network (DBNs)} $\quad$\\
DBNs are a type of generative ANNs that consist of a visible layer (corresponding to the inputs) and several hidden layers (corresponding to latent variables). They can extract hierarchical representation of the training data as well as reconstruct their input data. By adding a classifier layer like softmax, it can be used for prediction tasks. 

\begin{figure}
	\begin{center}		
		\includegraphics[width=.45\textwidth]{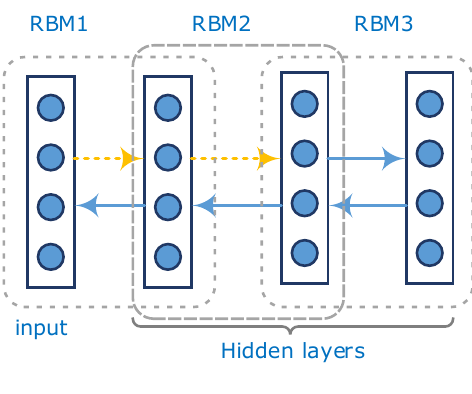}		
	\end{center}
	\caption{Structure of a deep belief network. The dash arrows show the feature extraction path and solid arrows show the generative path.}\label{fig:dbn}
\end{figure}

The training of a DBN is performed layer by layer, such that each layer is treated as an RBM trained on top of the previous trained layer. This mechanism makes a DBN an efficient and fast algorithm in DL~\cite{bengio2009learning}. For a given hidden layer in DBN, the hidden layer of previous RBM acts as the input layer. Figure~\ref{fig:dbn} shows the structure of a typical DBN.

Several applications can benefit from the structure of DBNs, such as fault detection classification in industrial environments, threat identification in security alert systems, and emotional feature extraction out of images.\\

\subsubsection{Ladder Networks} $\quad$\\
Ladder networks were proposed in 2015 by Valpola \textit{et al.}~\cite{valpola2015neural} to support unsupervised learning. Later, they were extended to work in semi-supervised settings~\cite{rasmus2015semi} and have shown state-of-the-art performance for several tasks, such as handwritten digits recognition and image classification. The architecture of a ladder network consists of two encoders and one decoder. The encoders act as the supervised part of the network and the decoder performs unsupervised learning. One of the encoders, called clean encoder, produces the normal computations while the other encoder, called corrupted encoder, adds Gaussian noise to all layers. 

Using a denoising function, the decoder can reconstruct the representations at each layer given the corresponding corrupted data. The difference between the reconstructed and clean data at each layer is used for computing the denoising cost of that layer. In the encoder side, the cost function uses the difference between the corrupted output of encoder layers and the corresponding clean outputs. The training objective is to minimize the sum of cost in the supervised part and unsupervised network. 

\begin{figure}[t]	
	\begin{center}		
		\includegraphics[width=.45\textwidth]{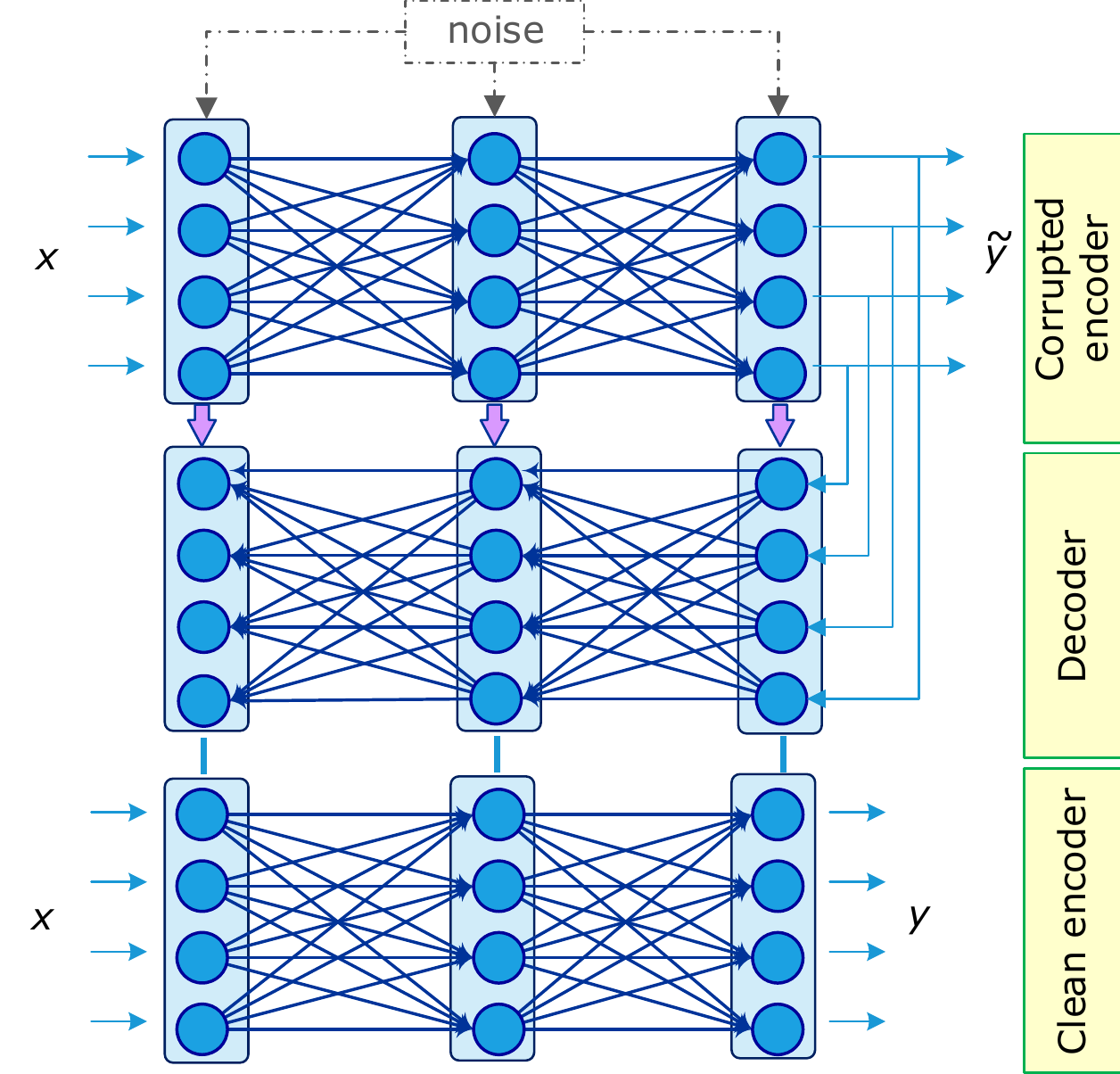}		
	\end{center}
	\caption{Ladder network structure with two layers.}\label{fig:ladder}	
\end{figure}

The initial experimental evaluations of ladder networks \cite{valpola2015neural} are limited to some standard tasks, such as handwritten digits classification over the Modified National Institute of Standards and Technology (MNIST) datasets \cite{lecun2018mnist} or image recognition tasks on the datasets of the Canadian Institute for Advanced Research (CIFAR)-10 \cite{krizhevsky2009learning}. Though it has not been used widely in IoT scenarios, ladder networks have the potential to be used in many vision-based IoT analytics where semi-supervision is a great bonus. Figure~\ref{fig:ladder} shows the structure of a ladder network.

\subsection{Fast and Real-time DL Architectures}

The research works for fast and real-time analytics using DL models over the stream of data are still in their infancy. An initial work in this area that utilizes ANNs is done by Liang \textit{et~al.}~\cite{liang2006fast}. It has extended the extreme learning machine (ELM) networks to apply an online sequential learning algorithm to single hidden layer feed-forward networks. Their framework, called OS-ELM, learns the training data one-by-one as well as chunk-by-chunk, and only newly arrived data go through the training process. This architecture is the base for the real-time manufacturing execution system that is proposed in \cite{yang2016rfid}. In this work, OS-ELM has been used for shop floor object localization using RFID technology. Zou \textit{et al.} \cite{zou2014online} have also reported using this architecture for an indoor localization algorithm based on WiFi fingerprinting, in which the OS-ELM model can bear well the dynamic environmental changes while still showing good accuracy.

For convolutional networks, the architecture proposed by Ren \textit{et al.}, called Faster R-CNN~\cite{ren2017faster} (based on Fast R-CNN~\cite{girshick2015fast}), aims to detect objects in images in real-time. Object detection in images needs more computations and hence consumes more energy compared to the image classification tasks, since the system has a large number of potential object suggestions that need to be evaluated. The proposed architecture is based on applying region proposal algorithms in full CNNs that perform object bounds prediction and objectness score computation at each position at the same time. Their evaluation of the proposed object detection architecture indicates that the run time of the system is between $5$-$17$ frames per second (fps)  given that the original input frames are re-scaled such that the shortest side of the image would be $600$ pixels. Mao \textit{et al.} \cite{mao2017toward} also used Fast R-CNN for embedded platforms reporting a run time of $1.85$ fps with frames scaled to $600$ pixels in the shortest side in embedded CPU+GPU platform, which have been shown to be energy-efficient with a close-to-real-time performance. However, for image processing tasks, we can consider an approach to be truly real-time when it can process and analyze $30$ fps or better. Redmon \textit{et al.} \cite{redmon2016you} developed You Only Look Once (YOLO) that has reached the performance of $45$ fps for input images resized to $448\times448$, and even a smaller version of it, Fast YOLO, achieving $155$ fps, which are suitable for smart cameras.  

\subsection{Joint DL with Other Approaches}\label{sec:deepJoint}

DL architectures also have been used jointly in other machine learning approaches to make them more efficient. The nonlinear function approximation of DL models that can support thousands or even billions of parameters is a strong motivation to use this method in other machine learning approaches in need of such functions. Moreover, the automatic feature extraction in deep models is another motivating reason to exploit these models jointly with other approaches. In the following subsections, a summary of such approaches that are suitable for IoT scenarios is provided.\\

\subsubsection{Deep Reinforcement Learning} 
\begin{figure*}[t]	
	\begin{center}		
		\includegraphics[width=0.9\textwidth]{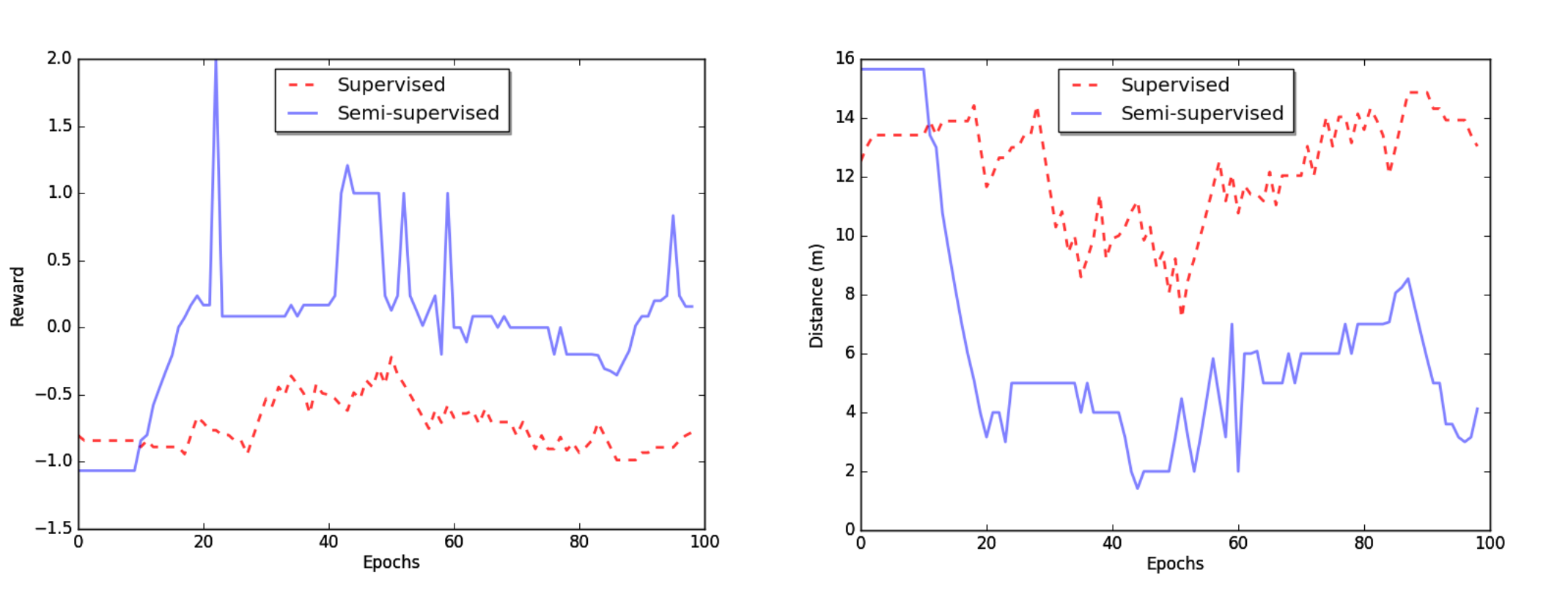}		
	\end{center}
	\caption{Deep reinforcement learning (supervised and semi-supervised): Obtaining rewards (left) and their corresponding accuracy measurement (right) \cite{mohammadi2017semi}.}\label{fig:DRL_sample}	
\end{figure*}

$\quad$\\
Deep Reinforcement Learning (DRL) \cite{mnih2013playing} is a combination of reinforcement learning (RL) with DNNs. It aims to create software agents that can learn by themselves to establish successful policies for gaining maximum long-term rewards. In this approach, RL finds the best policy of actions over the set of states in an environment from a DNN model. The need for a DNN in an RL model becomes evident when the underlying environment can be represented by a large number of states. In such situation, traditional RL is not efficient enough. Instead, a DL model can be used to approximate the action values in order to estimate the quality of an action in a given state. Systems that use DRL in their context are in their infancy, but already have showed very promising results. In the field of IoT, the work presented in \cite{mohammadi2017semi} uses DRL in a semi-supervised setting for localization in smart campus environments. The aim of this work is to localize the users based on received signals from multiple Bluetooth Low Energy (BLE) iBeacons. The learning agent uses DRL to find the best action to perform (i.e., moving in a direction like North, North-West, etc. from a starting point). The reward function is the reciprocal of the distance error to a predefined target, such that the learning agent receives more rewards when it gets closer to its intended target and vice versa. Figure \ref{fig:DRL_sample} shows a sample result of such method when a DNN model helps for gaining more rewards in a semi-supervised setting (left sub-figure in Figure \ref{fig:DRL_sample}) and its reward interpretation to the accuracy (right sub-figure).\\

\subsubsection{Transfer Learning with Deep Models} $\quad$\\
Transfer learning, which falls in the area of domain adaptation and multi-task learning, involves the adaptation and improvement of learning in a new domain by transferring the knowledge representation that has been learned from data of a related domain~\cite{bengio2012deep}. Transfer learning is an interesting potential solution for many IoT applications where gathering training data is not an easy task. For example, considering the training of a localization system through BLE or WiFi fingerprinting using smart phones, the RSSI values at a same time and location for different platforms (e.g., iOS and Android) vary. If we have a trained model for one platform, the model can be transferred to the other platform without re-collecting another set of training data for the new platform.

DL models are well matched to transfer learning due to their ability to learn both low-level and abstract representations from input data. Specifically, Stacked denoising AEs~\cite{bengio2012deep} and other variations of AEs \cite{deng2014introducing} have been shown to perform very well in this area. Transfer learning with DNNs is still an ongoing and active research area in AI community, and we have not seen reported real-world applications in IoT.\\

\subsubsection{Online Learning Algorithms joint with DL}$\quad$\\
As the stream of data generated from IoT applications goes through the cloud platforms for analysis, the role of online machine learning algorithms becomes more highlighted, as the training model needs to be updated by the incremental volume of data. This is opposed to what the current technologies support, which is based on batch learning techniques, where the whole training data set should be available for training and, thereafter, the trained model cannot evolve by new data. Several research works report applying online learning techniques on various DL models, including stacked denoising AEs~\cite{wu2013online}, sum-product networks~\cite{jaini2016online}, and RBMs~\cite{chen2016sequential}.

\subsection{Frameworks}

The rapid growth of interest to use DL architectures in different domains has been supported by introducing several DL frameworks in recent years. Each framework has its own strength based on its supported DL architectures, optimization algorithms, and ease of development and deployment~\cite{bahrampour2015comparative}. Several of these frameworks have been used widely in research for efficient training of DNNs. In this section, we review some of these frameworks.

\textbf{H2O}: H2O is a machine learning framework that provides interfaces for R, Python, Scala, Java, JSON, and CoffeeScript/JavaScript \cite{candel2015deep}. H2O can be run in different modes including standalone mode, on Hadoop, or in a Spark Cluster. In addition to common machine learning algorithms, H2O includes an implementation of a DL algorithm, which is based on feed-forward neural networks that can be trained by SGD with backpropagation. H2O’s DL AE is based on the standard deep (multi-layer) neural net architecture, where the entire network is learned together, instead of being stacked layer-by-layer.

\textbf{Tensorflow}: Initially developed for Google Brain project, Tensorflow is an open source library for machine learning systems using various kinds of DNNs \cite{abadi2016tensorflow}. It is used by many Google products including Google Search, Google Maps and Street View, Google Translate, YouTube and some other products. Tensorflow uses graph representations to build neural network models. Developers can also take advantage of TensorBoard, which is a package to visualize neural network models and observe the learning process including updating parameters. Keras\footnote{https://keras.io/} also provides a high level of programming abstraction for Tensorflow. 

\textbf{Torch}: Torch is an open source framework for machine learning containing a wide range of DL algorithms for easy development of DNN models~\cite{collobert2011torch7}. It has been developed upon Lua programming language to be light-weight and fast for training DL algorithms. It is used by several companies and research labs like Google, Facebook, and Twitter. It supports developing machine learning models for both CPUs and GPUs, and provides powerful parallelization packages for training DNNs.

\textbf{Theano}: Theano is an open source Python-based framework for efficient machine learning algorithms, which supports compiling for CPUs and GPUs~\cite{bastien2012theano}. It uses the CUDA library in optimizing the complicated codes that need to be run on GPUs. It also allows parallelism on CPUs. Theano uses graph representations for symbolic mathematical expressions. Through this representation, symbolic differentiation of mathematical expressions is supported in Theano. Several wrappers including Pylearn2, Keras, and Lasagne provide easier programming experience on top of Theano~\cite{raschka2017python}.

\textbf{Caffe}: Caffe \cite{jia2014caffe} is an open source framework for DL algorithms and a collection of reference models. It is based on C++, supports CUDA for GPU computations, and provides interfaces for Python and Matlab. Caffe separates model representation from its implementation. This has been made possible by defining models by configurations without hard-coding them in the source code. Switching between platforms (e.g., CPU to GPU or mobile devices) is easy by only changing a flag. Its speed on GPU is reported to be $1$ ms/image for prediction and $4$ ms/image for training.

\textbf{Neon}: Neon\footnote{http://neon.nervanasys.com} is another open source DL framework based on Python with high performance for modern DNNs, such as AlexNet \cite{krizhevsky2012imagenet}, Visual Geometry Group (VGG) \cite{parkhi2015deep}, and GoogleNet \cite{szegedy2015going}. It supports developing several commonly used models, such as CNNs, RNNs, LSTMs, and AEs, on both CPUs and GPUs. The list is being extended as they implemented GANs for semi-supervised learning using DL models. It also supports easy changing of the hardware platform back-ends.  

Bahrampour \textit{et al.} in \cite{bahrampour2015comparative} have provided a comparative study for four of the aforementioned tools namely, Caffe, Neon, Theano and Torch. Although the performance of each tool varies in different scenarios, Torch and Theano showed the overall best performance in most of the scenarios. Another benchmarking is provided in \cite{shi2016benchmarking}, comparing the running performance of Caffe, TensorFlow, Torch, CNTK, and MXNet. Table~\ref{tbl:tools} summarizes and compares different DL frameworks. 

\begin{table*}
\centering
\caption{Properties of Frameworks for Developing Deep Learning (As of September 2017).}
\label{tbl:tools}
\begin{tabular}{|l|l|l|l|l|l|}
\hline 
\multicolumn{1}{|c|}{\textbf{Frameworks}} & \multicolumn{1}{c|}{\begin{tabular}[c]{@{}c@{}}\textbf{Core} \\ \textbf{Language}\end{tabular}} & \multicolumn{1}{c|}{\textbf{Interface}}                                       & \multicolumn{1}{c|}{\textbf{Pros}}                                                                                                                                                          & \multicolumn{1}{c|}{\textbf{Cons}}                                                                 & \multicolumn{1}{c|}{\begin{tabular}[c]{@{}c@{}}\textbf{Used in IoT}\\ \textbf{Application}\end{tabular}} \\ \hline \hline
H2O                         & Java                                                                          & \begin{tabular}[c]{@{}l@{}}R, Python,\\ Scala, REST API\end{tabular} & \tabitem Wide range of interfaces                                                                                                                                                         & \begin{tabular}[c]{@{}l@{}}\tabitem Limited number of \\ models are supported \\\tabitem Is not flexible\end{tabular}       &  \cite{mehmood2017utilearn}                                                                                     \\ \hline
Tensorflow                  & C++                                                                           & \begin{tabular}[c]{@{}l@{}}Python, Java,\\ C, C++, Go\end{tabular}   & \begin{tabular}[c]{@{}l@{}}\tabitem Fast on LSTM training\\ \tabitem Support to visualize \\ networks\end{tabular}                                                                               & \begin{tabular}[c]{@{}l@{}}\tabitem Slower training \\ compared to other \\Python-based frameworks\end{tabular}                                                                                          &  \cite{luckow2016deep}                                                                                     \\ \hline
Theano                      & Python                                                                        & Python                                                               & \begin{tabular}[c]{@{}l@{}}\tabitem Supports various models\\ \tabitem Fast on LSTM training \\ on GPU\end{tabular}                                                                               &       \begin{tabular}[c]{@{}l@{}}\tabitem Many low level APIs\end{tabular}                                                                                    &  \cite{ma2015large}                                                                                     \\ \hline
Torch                       & Lua                                                                           & C, C++                                                               & \begin{tabular}[c]{@{}l@{}}\tabitem Supports various models\\ \tabitem Good documentation\\ \tabitem Helpful error debugging\\  messages\end{tabular}                                                    &  \begin{tabular}[c]{@{}l@{}}\tabitem Learning a new language \end{tabular}                                                                                          &  \cite{luckow2016deep}  \cite{lane2015early}                                                                                   \\ \hline
Caffe                       & C++                                                                           & \begin{tabular}[c]{@{}l@{}}Python,\\ Matlab\end{tabular}             & \begin{tabular}[c]{@{}l@{}}\tabitem Provides a collection of \\ reference models\\ \tabitem Easy platform switching\\ \tabitem Very good at convolutional\\ networks\end{tabular}                                                                 & \begin{tabular}[c]{@{}l@{}} \tabitem Not very good for\\ recurrent networks \end{tabular}                                                                                       &  \cite{mittal2016spotgarbage,liu2017new,sladojevic2016deep}                                                                                     \\ \hline
Neon                        & Python                                                                        & Python                                                               & \begin{tabular}[c]{@{}l@{}}\tabitem Fast training time \\ \tabitem Easy platform switching\\ \tabitem Supports modern \\ architectures like GAN\end{tabular}                                                                     & \begin{tabular}[c]{@{}l@{}}\tabitem Not supporting CPU \\multi-threading\end{tabular}                                                                                          &  \cite{liu2016application}                                                                                     \\ \hline
Chainer~~\cite{tokui2015chainer}                     & Python                                                                        & Python                                                               & \begin{tabular}[c]{@{}l@{}}\tabitem Supports modern \\ architectures\\ \tabitem Easier to implement\\ complex architectures\\ \tabitem Dynamic change of model\end{tabular}                              & \begin{tabular}[c]{@{}l@{}}\tabitem Slower forward computation\\  in some scenarios\end{tabular} & \cite{hada2016deep}                                                                                      \\ \hline
Deeplearning4j              & Java                                                                          & \begin{tabular}[c]{@{}l@{}}Python, Scala,\\ Clojure\end{tabular}     & \begin{tabular}[c]{@{}l@{}}\tabitem Distributed training\\ \tabitem Imports models from \\ major frameworks \\ (e.g., TensorFlow, Caffe,\\  Torch and Theano)\\ \tabitem Visualization tools\end{tabular} & \begin{tabular}[c]{@{}l@{}}\tabitem Longer training time \\ compared to other tools\end{tabular} &  \cite{soto2016ceml, borkowski2016predicting}                                                                                     \\ \hline
\end{tabular}
\end{table*}

\subsection{Lessons Learned}
In this section, we reviewed several common DL architectures that can serve in the analytics component of various IoT applications. Most of these architectures work with various types of input data generated by IoT applications. However, to get better performance for serial or time-series data, RNNs and their variations are recommended. In particular, for long term dependencies among data points, LSTM is more favorable due to its concept of gates for memory cells. For cases where the input data is more than one-dimensional, variations of CNNs work better. RBM, DBN, and variations of AE perform well in handling high-dimensionality reduction, and hierarchical feature extraction. Combined with a classification layer, they can be used for a variety of detection and forecasting scenarios. More recent architectures including VAE, GAN, and Ladder Networks are expected to have a great impact on IoT applications since they cover semi-supervised learning. Those are more favorable for IoT applications where a huge amount of data is generated while a small fraction of that can be annotated for machine learning. The role of these architectures can be emphasized by knowing that only about $3$\% of all universe data by $2012$ was annotated, and is hence useful for supervised machine learning~\cite{gantz2012digital}. Table \ref{tbl:modelsComparison} summarizes DL architectures. 

A few attempts toward making DL architectures fast and real-time responsive were also discussed. This avenue needs more exploration and research to be applicable in many time-sensitive IoT applications. Emerging machine learning architectures and techniques that both benefit from DL and address the specific IoT application requirements were also highlighted. Indeed, DRL can support autonomousness of IoT applications, transfer learning can fill the gap of lack of training data sets, and online learning matches the need for stream analysis of IoT data.

We also reviewed several common and powerful frameworks for the development of DL models. For IoT applications, training times, run times, and dynamic update of the trained models are determining factors for a reliable and efficient analytic module. Most of the current frameworks follow the pattern of ``define-and-run'' instead of ``define-by-run''~\cite{tokui2015chainer}. The former does not allow dynamic updates of the model while the latter supports such modifications. Chainer~\cite{tokui2015chainer} is a framework that follows the latter pattern and can handle dynamic changes of the model. 

\section{DL Applications in IoT}\label{sec:applications}

DL methods have been shown promising with state-of-the-art results in several areas, such as signal processing, natural language processing, and image recognition. The trend is going up in IoT verticals. Some neural network models work better in special domains. For example, convolutional networks provide better performance in applications related to vision, while AEs perform very well with anomaly detection, data denoising, and dimensionality reduction for data visualization. It is important to make this link between the kind of neural network model that best fits each of the different application domains.

\begin{figure*}	
	\begin{center}		
		\includegraphics[width=1\textwidth]{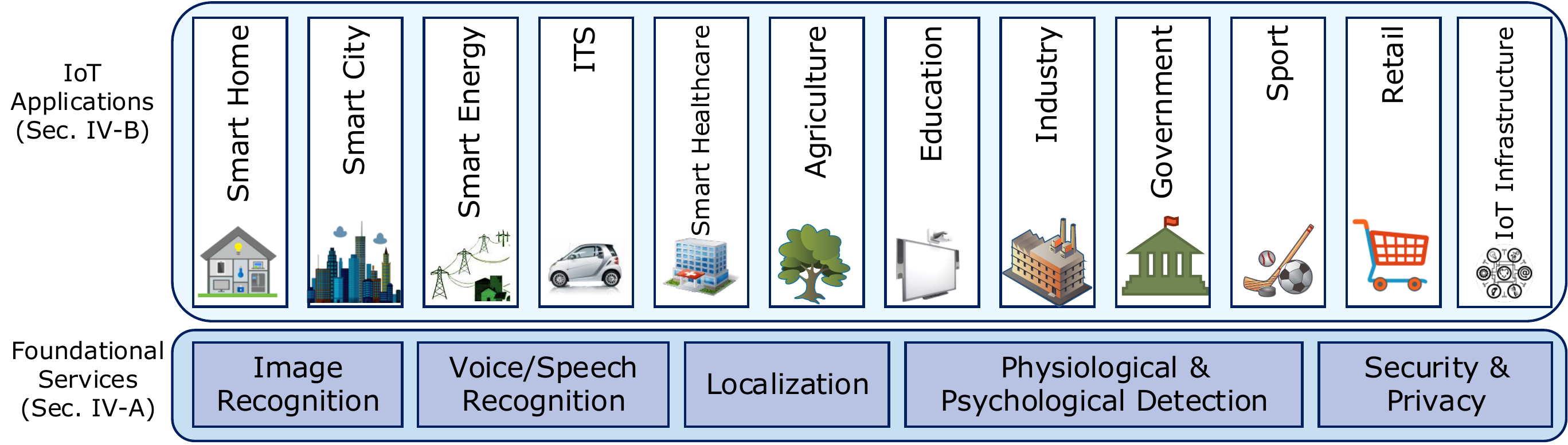}		
	\end{center}
	\caption{IoT applications and the foundational services.} \label{fig:applications}	
\end{figure*}

In this section, we review successful applications of DL in IoT domains. Based on our observation, many IoT related applications utilize vision and image classification (like traffic sign recognition, or plant disease detection that we will discuss in Section \ref{sec:complex_apps}) as their base intelligent service. There are other services, such as human pose detection, which are utilized for smart home applications or intelligent car assistance. We identify several kinds of these services as foundational services on which other IoT applications can be built. The common property of these services is that they should be treated in a fast analytic mode instead of piling their data for later analytics. Indeed, each domain may have specific services beyond these foundational services. Figure~\ref{fig:applications} shows the foundational services and the IoT applications on top of them. 

In the following subsections, we first review foundational services of IoT that use DL as their intelligence engine, then highlight the IoT applications and domains where a combination of foundational services as well as specific ones may be utilized. 

\subsection{Foundational Services}

\subsubsection{Image Recognition} $\quad$\\
A large portion of IoT applications address scenarios in which the input data for DL is in the form of videos or images. Ubiquitous mobile devices equipped with high resolution cameras facilitate generating images and videos by everyone, everywhere. Moreover, intelligent video cameras are used in many places like smart homes, campuses, and manufacturers for different applications. Image recognition/classification and object detection are among the fundamental usages of such devices. 

One issue with the IoT-related systems that have addressed image recognition is the use of specific source datasets for evaluation of their performance. Most of these systems employ the available common image datasets such as the MNIST dataset of handwritten digits \cite{lecun2018mnist}, VGG face dataset \cite{parkhi2015deep}, CIFAR-10 and CIFAR-100 tiny images dataset, etc. Though being good for comparison with other approaches, those datasets do not show the specific characteristics of IoT systems. For example, the input for the task of vehicle detection in smart connected cars would not be always a clear image, and there are cases where the input image is at night, or in a rainy or foggy weather. These cases are not handled through the available datasets and hence the models trained based on these datasets are not comprehensive enough.\\

\subsubsection{Speech/Voice Recognition} $\quad$\\
With the massive proliferation of smart mobile devices and wearables, automatic speech recognition is becoming a more natural and convenient way for people to interact with their devices \cite{mit02_15_2017voice}. Also, the small size of mobile devices and wearables nowadays lower the possibility of having touch screens and keyboards as a means of input and interaction with these devices. However, the main concern for providing speech/voice recognition functionality on resource-constrained devices is its energy-intensiveness, especially when the data is processed through neural networks. In a typical speech recognition neural network model, voice data is represented as the raw input to the network. The data is processed through the hidden layers, and the likelihood of the voice data to a particular speech sound is presented at the output layer.

Price \textit{et al.} \cite{price2017scalable} have reported that they have built a special-purpose low-power DL chip for automatic speech recognition. The new specialized chip consumes a tiny amount of energy between 0.2 and 10 milliwatts, 100 times lesser than the energy consumption for running a speech recognition tool in current mobile phones. In the new chip, DNNs for speech recognition have been implemented. For the sake of energy saving, three levels of voice activity recognition are designed with three separate neural networks, each of which having a different level of complexity. A lowest complexity network, thus consuming the lowest amount of energy, detects voice activity by monitoring the noise in the environment. If this network identifies a voice, the chip runs the next complexity level recognition network whose task is acoustic modeling to identify if the voice looks like speech. If the output of this network is a high likelihood, then the third network, having the highest energy consumption, is triggered to run to identify individual words.\\ 

\subsubsection{Indoor Localization} $\quad$\\
Providing location aware services, such as indoor navigation and location aware marketing in retailers, are becoming prevalent in indoor environments. Indoor localization may also have applications in other sectors of IoT, such as in smart homes, smart campuses, or hospitals. The input data generated from such applications usually comes from different technologies, such as vision, visible light communication (VLC), infrared, ultrasound, WiFi, RFID, ultrawide band, and Bluetooth. For the approaches based on WiFi or Bluetooth, most of the literature have used mobile phones for receiving signals from the fixed transmitters (i.e., access points or iBeacons), which are called fingerprints. Among these fingerprinting approaches, several attempts reported the use of DL models to predict the location \cite{wang2015deepfi, gu2015semi, zhang2016deep}. 

DL has been used successfully to locate indoor positions with high accuracy. In a system called DeepFi \cite{wang2015deepfi}, a DL method over fingerprinting WiFi channel state information data has been utilized to identify user positions. This system consists of offline training and online localization phases. In the offline training phase, DL is exploited to train all the weights based on the previously stored channel state information fingerprints. Other works \cite{gu2015semi}, \cite{zhang2016deep} report using variations of DL models in conjunction with other learning methods to extract features and estimate positions. These experiments assert that the number of hidden layers and units in DL models has a direct effect on the localization accuracy. In \cite{liu2017fusion}, a CNN is used for indoor localization by fusion of both magnetic and visual sensing data. Moreover, a CNN has been trained in \cite{becker2017indoor} to determine the indoor positions of users by analyzing an image from their surrounding scene. 

Lu \textit{et al.} have also used LSTM networks for localizing soccer robots~\cite{lu2017multimodal}. In this application, data collected from several sensors, namely Inertia Navigation System (INS) and vision perceptions, are analyzed to predict the position of the robot. The authors reported improved accuracy and efficiency compared to two baseline methods, namely standard Extended Kalman Filtering (EKF) and the static Particle Filtering.\\

\subsubsection{Physiological and Psychological State Detection} $\quad$\\
IoT combined with DL techniques has been also employed to detect various physiological and psychological states of humans, such as pose, activity, and emotions. Many IoT applications incorporate a module for human pose estimation or activity recognition to deliver their services, e.g., smart homes, smart cars, entertainment (e.g., XBox), education, rehabilitation and health support, sports, and industrial manufacturing. For example, convenient applications in smart homes are built based on the analysis of occupant's poses. The cameras transfer the video of the occupant to a DNN to find out the pose of the person and take the most appropriate action accordingly. Toshev~\textit{et al.}~\cite{toshev2014deeppose} report a system employing a CNN model to achieve this goal. This sort of services can also be used in education to monitor the attention of students, and in retail stores to predict the shopping behavior of customers \cite{liu2016joint}. 

Ordonez \textit{et al.} \cite{ordonez2016deep} have proposed a DL framework that combines the strength of CNN and LSTM neural networks for human activity recognition from wearable sensor data (accelerometer, gyroscope, etc.). Their model consists of four convolutional layers with ReLUs followed by two LSTM layers and a softmax layer. They showed that this combination outperformed a baseline model that is just based on convolutional layers by $4\%$ on average. The work of Tao \textit{et al.} \cite{tao2016multi} also used LSTM architecture for human activity recognition based on mobile phone sensor's data. Li \textit{et al.} \cite{li2016deepRfidActivity} also report the usage of raw data from passive FRID tags for detecting medical activities in a trauma room (e.g., blood pressure measurement, mouth exam, cardiac lead placement, etc.) based on a deep CNN.

In \cite{pigou2015beyond}, a combined model of CNN and RNN was proposed for gesture recognition in video frames. This model showed better results compared to the models without such combination, and asserted the importance of the recurrence component for such task. In \cite{fragkiadaki2015recurrent}, Fragkiadaki \textit{et al.} proposed a DNN model called Encoder-Recurrent-Decoder (ERD) for human body pose recognition and motion prediction in videos and motion capture data sets. The proposed model consisted of an RNN with an encoder before the recurrent layers and a decoder after them. This architecture was shown to outperform Conditional Restricted Boltzmann Machines (CRBMs) for this application. 

Beyond the physical movements, emotion estimation of humans from video frames has been also investigated in \cite{kahou2016emonets} using a model that consists of a CNN, DBN, and AE. Furthermore, the work in \cite{neverova2016learning} used mobile inertial sensor data for motion detection. It confirmed that human motion patterns can be used as a source of user identification and authentication. The employed model in this system is a combination of convolutional layers and clockwork RNN.\\
 
\subsubsection{Security and Privacy} $\quad$\\
Security and privacy is a major concern in all IoT domains and applications. Smart homes, ITS, Industry, smart grid, and many other sectors consider security as a critical requirement. Indeed, the validity of the functionality of the systems depends on protecting their machine learning tools and processes from attackers.

False Data Injection (FDI) is a common type of attack on data-driven systems. In \cite{he2017real}, He \textit{et al.} proposed a Conditional DBN to extract FDI attack features from the historical data of smart grids, and use these features for attack detection in real-time. The work in \cite{kang2016intrusion} is also related to anomaly detection that may occur in vehicle networks.

Smart phones as great contributers to IoT data and applications are also under serious threats of hacker attacks. Consequently, protecting these devices from a variety of security issues is necessary for IoT perspectives beyond the users' concerns. Yuan \textit{et al.} \cite{yuan2014droid} proposed a DL framework to identify malwares in Android apps. Their architecture is based on a DBN by which they reported accuracy of 96.5\% to detect malware apps.  

The security and privacy preservation of deep machine learning approaches are the most important factors for the acceptance of using these methods in IoT sectors. Shokri \textit{et al.} \cite{shokri2015privacy} proposed a method to address the privacy preservation issues in DL models when they are subject to distributed learning. Their approach was able to preserve both the privacy of participants’ training data and the accuracy of the models at the same time. The core of their approach is based on the fact that stochastic gradient descent optimization algorithms, used in many DL architectures, can be performed in a parallel and asynchronous way. Individual participants can thus independently train the model on their own data and share a portion of their model parameters with other participants. Abadi \textit{et al.} \cite{abadi2016deep} also proposed a method for privacy guarantee in DL models using differentially private stochastic gradient descent algorithm.

\subsection{Applications}\label{sec:complex_apps}

\subsubsection{Smart Homes} $\quad$\\
The concept of smart homes involve a broad range of applications based on IoT, which can contribute to enhancing homes' energy usage and efficiency, as well as the convenience, productivity, and life-quality of their occupants. Nowadays, home appliances can connect to the Internet and provide intelligent services. For example, Microsoft and Liebherr in a collaborative project are applying Cortana DL to the information gathered from inside the refrigerator \cite{vcloudnews09_05_2016}. These analytics and predictions can help the household to have a better control on their home supplies and expenses, and, in conjunction with other external data, can be used for monitoring and predicting health trends.

Over one third of the generated electricity in the U.S. is consumed by the residential sector \cite{manic2016intelligent}, with HVAC and lighting devices consisting the largest source of such consumption in buildings. This demand is expected to grow in a slower pace by smart management of energy as well as the efficiency improvements in appliances. Hence, the ability to control and improve energy efficiency and predict the future need is a must for smart home systems. In the smart home applications, electricity load prediction are the most common applications that employ different DL networks to figure out the task. Manic \textit{et al.} \cite{manic2016intelligent}  performed a comparison analysis of load  forecasting for home energy consumption using three DL architectures, including LSTM, LSTM Sequence-to-Sequence (S2S) and CNN. Their results show that LSTM S2S predicts the future usage better than the other architectures, followed by CNN, and then LSTM. They also compared the same dataset over a conventional ANN, and all of the aforementioned models outperformed the ANN model.  
\\

\subsubsection{Smart City} $\quad$\\
Smart city services span over several IoT domains, such as transportation, energy, agriculture \cite{yoshida2016iot}, etc. However, this area is more interesting from a machine learning perspective as the heterogeneous data coming from different domains lead to big data, which can result in high-quality output when analyzed using DL models. Smart city benefits from advances in other domain to achieve efficient resource management for the whole city. For example, to improve public transportation infrastructure and offer new improved services, getting analytics and patterns out of public transportation behaviors is of interest for local authorities.

Toshiba has recently developed a DL testbed jointly with Dell Technologies, and used this testbed in a Smart Community Center in Kawasaki, Japan, to evaluate the data collected in the Center \cite{toshiba10_17_2016}.
The aim of running the testbed is to measure the effectiveness of using DL architectures in IoT ecosystems, and identify the best practices for service improvement including increasing machines' availability, optimizing monitoring sensors, and lowering maintenance expenses. The big data that feeds the testbed were gathered from building management, air conditioning and building security.

One important issue for smart city is predicting crowd movements patterns, and their use in public transportation. Song \textit{et al.} \cite{song2016deeptransport} developed a system based on DNN models to achieve this goal on a city level. Their system is built upon a four-layer LSTM neural network to learn from human mobility data (GPS data) joint with their transportation transition modes (e.g., stay, walk, bicycle, car, train). They treated the prediction of people's mobility and transportation mode as two separated tasks instead of joining all these features together. Consequently, their learning system is based on a multi-task deep LSTM architecture to jointly learn from the two sets of features. The choice of LSTM was driven by the spatio-temporal nature of human mobility patterns. The authors assert that their approach based on multi-task deep LSTM achieves better performance compared to both shallow LSTMs having only one single LSTM layer as well as deep LSTMs without multi-tasking. 

Liang \textit{et al.} \cite{liang2016mercury} presented a real-time crowd density prediction system in transportation stations that leverages the mobile phone users' telecommunication data known as caller detail record (CDR). CDR data are gathered when a user takes a telecommunication action (i.e., call, SMS, MMS, and Internet access) on the phone, which usually includes data about user ID, time, location, and the telecommunication action of the user. They built their system based on an RNN model for metro stations, and reported more accurate predictions compared to nonlinear autoregressive neural network models. 

Waste management and garbage classification is another related task for smart cities. A straightforward method to perform this automation is through vision-based classifications using deep CNNs as it has been done in \cite{mittal2016spotgarbage}. Monitoring air quality and predicting the pollution is another direction for city management. Li \textit{et al.} \cite{li2016deep} developed a DL-based air quality prediction model using a stacked AE for unsupervised feature extraction and a logistic regression model for regression of the final predictions.

Amato \textit{et al.} in \cite{amato2016deep} developed a decentralized system to identify the occupied and empty spots in parking lots using smart cameras and deep CNNs. The authors deployed a small architecture of a CNN on smart cameras, which are equipped with Raspberry Pi 2 model. These embedded devices in smart cameras can thus run the CNN on each device to classify images of individual parking spaces as occupied or empty. The cameras then send only the classification output to a central server. Valipour \textit{et al.} \cite{valipour2016parking} also developed a system for detecting parking spots using CNN, which has shown improved results compared to SVM baselines. Table \ref{tbl:smartCityApps} summarizes the aforementioned attempts. \\

\begin{table}[]
\renewcommand{\arraystretch}{1.2}
\centering
\caption{Typical IoT-Based Services in Smart City}
\label{tbl:smartCityApps}
\begin{tabular}{|l|l|l|l|}
\hline 
\multicolumn{1}{|c|}{\textbf{Service}}                                                                 & \multicolumn{1}{c|}{\textbf{Reference}} & \multicolumn{1}{c|}{\textbf{Input data}}                              & \multicolumn{1}{c|}{\begin{tabular}[c]{@{}l@{}}\textbf{DL} \\\textbf{model} \end{tabular}}   \\ \hline \hline
\multirow{2}{*}{\begin{tabular}[c]{@{}l@{}}Crowd density/\\ transportation \\ prediction\end{tabular}} & \cite{song2016deeptransport}                                        & GPS/ transition mode                                                  & LSTM                                          \\ \cline{2-4} 
                                                                                                       &  \cite{liang2016mercury}                                       & \begin{tabular}[c]{@{}l@{}}Telecommunication \\ data/CDR\end{tabular} & RNN                                           \\ \hline
\begin{tabular}[c]{@{}l@{}}Waste \\ management\end{tabular}                                            & \cite{mittal2016spotgarbage}                                        & Garbage images                                                        & CNN                                           \\ \hline
\begin{tabular}[c]{@{}l@{}}Parking lot \\ management\end{tabular}                                      &  \cite{amato2016deep,valipour2016parking}                                       & \begin{tabular}[c]{@{}l@{}}Images of parking \\ spaces\end{tabular}   & CNN                                           \\ \hline
\end{tabular}
\end{table}

\subsubsection{Energy} $\quad$\\
The two way communication between energy consumers and the smart grid is a source of IoT big data. In this context, smart meters are in the role of data generation and acquisition for the fine grained level of energy consumption measurement. Energy providers are interested to learn the local energy consumption patterns, predict the needs, and make appropriate decisions based on real-time analytics. Mocanu \textit{et al.} in \cite{mocanu2016big} have developed a kind of RBM to identify and predict the buildings' energy flexibility in real-time. Energy flexibility is about modifying a household's electricity consumption while minimizing the impact on the occupants and operations. In the mentioned work, time-of-use and consumption of individual appliances are predicted to achieve flexible energy control. The advantage of this model beyond showing good performance and accuracy is that flexibility identification can be performed with flexibility prediction concurrently. In \cite{mocanu2016deep}, two variations of RBMs are used to forecast energy consumption for short term intervals in residential houses. The model includes a Conditional RBM (CRBM) and a Factored Conditional RBM (FCRBM). Their results indicate that FCRBM performs better that CRBM, RNN and ANN. Moreover, by extending the forecasting horizon, FCRBM and CRBM show more accurate predictions than the RBM and ANN. 

On the smart grid side, forecasting the power from solar, wind, or other types of natural sustainable sources of energy is an active research field. DL is increasingly used in many applications in this domain. For example, in \cite{gensler2016deep}, Gensler \textit{et al.} investigate the performance of several DL models, such as DBNs, AEs, and LSTMs, as well as MLP for predicting the solar power of 21 photovoltaic plants. For solar power prediction, a main element of the input is a numeric value for weather forecasting in a given time horizon. From their evaluation, the combination of AEs and LSTMs (Auto-LSTM) has been shown to produce the best results compared to other models, followed by DBN. The reason for obtaining a good prediction score by Auto-LSTM is that they are able to extract features from raw data, which is not the case for ANN and MLP. In \cite{hada2016deep}, an online forecasting system based on LSTM is proposed to predict the solar flare power 24 hours ahead. \\

\subsubsection{Intelligent Transportation Systems} $\quad$\\
Data from Intelligent Transportation Systems (ITS) is another source of big data that is becoming ubiquitous every day. Ma \textit{et al.} \cite{ma2015large} presented a system of transportation network analysis based on DL. They have employed RBM and RNN architectures as their models in a parallel computing environment, and GPS data from participating taxies as the input of the models. The accuracy of their system to predict traffic congestion evolution over one hour of aggregated data is reported to be as high as $88\%$ which was computed within less than 6 minutes. \cite{tian2015predicting} also reported the investigation on short-term traffic flow prediction. They used LSTM as their learning model and reported better accuracy for LSTM compared to other methods including SVM, simple feed forward neural networks, and stacked AEs. For different intervals ($15$, $30$, $45$, and $60$ min) LSTM showed the lowest mean absolute percentage error (MAPE) rate. However, for short intervals of 15 minutes, the error rate of SVM is slightly higher than the LSTM model. This result can be interpreted by the fact that the small number of data points in short intervals does not make stronger discrimination boundaries for the classification task in the LSTM model compared to the SVM model. In another study \cite{kang2016intrusion}, ITS data are exposed to an intrusion detection system based on DNN to improve the security of in-vehicular network communications. 

ITS also motivate the development of methods for traffic sign detection and recognition. Applications such as autonomous driving, driver assistance systems, and mobile mapping need such sort of mechanisms to provide reliable services. Cireşan \textit{et al.} \cite{cirecsan2012multi} presented a traffic sign recognition system based on DNNs of convolutional and max-pooling layers. They introduced a multi-column DNN architecture that includes several columns of separate DNNs, and reported increased accuracy with this approach. The input is preprocessed by several different preprocessors, and a random number of columns receives the preprocessed input to proceed with training. The final prediction output is the average of all the DNNs' outputs. Their results show that this proposed method, achieving a recognition rate of $99.46\%$, has been able to recognize traffic signs better than the humans on the task with $0.62\%$ more accuracy. 

In order to be applicable in real scenarios, these analytics need to be performed in real-time. Lim \textit{et al.} in \cite{lim2017real} proposed a real-time traffic sign detection based on CNN that has been integrated with a general purpose GPU. They reported F1 measure of at least 0.89 in their results with data having illumination changes. To have a faster inference engine, they used a CNN with two convolutional layers.  

Furthermore, self-driving cars use DNNs in performing many tasks, such as detecting pedestrians, traffic signs, obstacles, etc. There are several startups that use DL in their self-driving cars to perform different tasks when driving in the streets \cite{ackerman2017driveai}.  \\

\subsubsection{Healthcare and Wellbeing} $\quad$\\
IoT combined with DL has been also employed in providing healthcare and wellbeing solutions for individuals and communities. For instance, developing solutions based on mobile apps to accurately measure dietary intakes is a track of research that can help control the health and wellbeing of individuals. Liu \textit{et al.} in \cite{liu2017new} and \cite{liu2016deepfood} developed a system to recognize food images and their relevant information, such as types and portion sizes. Their image recognition algorithm is based on CNNs that achieved competitive results compared to the baseline systems.

DL for classification and analysis of medical images is a hot topic in the healthcare domain. For example, Pereira \textit{et al.}~\cite{pereira2016convolutional} used the idea of recognizing handwritten images by CNNs to help identifying Parkinson's disease in its early stages. Their model learns features from the signals of a smart pen that uses sensors to measure handwritten dynamics during the individual’s exam. Muhammad \textit{et al.} \cite{muhammad2017smart} propose a voice pathology detection system using IoT and cloud frameworks, in which patients' voice signals are captured through sensor devices and are sent to a cloud server for analytics. They used an extreme learning machine trained by voice signals to diagnose the pathology. In \cite{wang2017detecting}, DL was employed for detection of cardiovascular diseases from mammograms. In their study, Wang \textit{et al.} used breast arterial calcification (BAC) revealed in mammograms as a sign of coronary artery disease. They developed a CNN with twelve layers to identify the existence of BAC in a patient. Their results show that the accuracy of their DL model is as good as the human experts. Although this work has been done offline, it shows the potential of developing or extending mammogram devices in IoT contexts for online and early detection of such diseases. 

Feng \textit{et al.} \cite{feng2014deep} report the use of RBMs and DBNs for fall detection in a home care environment. Normal postures in such environment are standing, sitting, bending, and lying. Lying on the floor longer than a threshold is considered as a fallen posture. Their evaluation shows that RBM outperforms DBN in terms of classification accuracy. The lack of large datasets and performing offline detection are the restrictions of their method.

Researchers also used time series medical data in conjunction with RNN based models for early diagnosis and prediction of diseases. Lipton \textit{et al.} \cite{lipton2015learning}  investigated the performance of LSTM networks to analyze and recognize patterns in multivariate time series of medical measurements in intensive care units (ICUs). The input data in their system consist of sensor data of vital signs as well as lab test results. Their performance results show that an LSTM model trained on raw time-series data outperforms a MLP network. A survey of DL in health informatics is provided in~\cite{ravi2017deep}.\\

\subsubsection{Agriculture} $\quad$\\
Producing healthy crops and developing efficient ways of growing plants is a requirement for a healthy society and sustainable environment. Disease recognition in plants using DNNs is a direction that has shown to be a viable solution. In a study that is reported by Sladojevic \textit{et al.} \cite{sladojevic2016deep}, the authors developed a plant disease recognition system based on the classification of leave images. They have used a deep convolutional network model implemented using the Caffe framework. In this model, diseased leaves in $13$ categories can be identified from the healthy ones with an accuracy of about $96\%$. Such recognition model can be exploited as a smart mobile applications for farmers to identify the fruit, vegetable, or plant disease based on their leaf images captured by their mobile devices. It can also allow them to select remedies or pesticides in conjunction with complementary data.

DL also has been used in remote sensing for land and crop detection and classification \cite{kussul2017deep} \cite{kuwata2015estimating}  \cite{scott2017training}. The direction established in these works enabled the automated monitoring and management of the agricultural lands in large scales. In most of such works, deep convolutional networks are used to learn from images of the land or crops. In \cite{kussul2017deep}, it is reported that using CNN has yielded an accuracy of $85\%$ in detecting major crops, including wheat, sunflower, soybeans, and maize, while outperforming other approaches such as MLP and random forest (RF).

Furthermore, DL has been reported to be utilized for prediction and detection tasks for automatic farming. For example, \cite{steen2016using} has used a DL model based on deep CNNs for obstacle detection in agricultural fields, which enables autonomous machines to operate safely in them. The proposed system was able to detect a standardized object with an accuracy between $90.8\%$ to $99.9\%$, based on the field (e.g., row crops or grass mowing).

Moreover, fruit detection and finding out the stage of fruit (raw or ripe) is critical for automated harvesting. In \cite{sa2016deepfruits}, Sa \textit{et al.} used a variation of CNN, called Region-based CNN, for image analysis of fruits. The input image of the system comes in two modes: one containing RGB colors and the other is near-infrared. The information of these images are combined in the model and has achieved detection improvement compared to pixel-based training models.\\

\subsubsection{Education} $\quad$\\
IoT and DL contribute to the efficiency of education systems, from kindergarten to higher education. Mobile devices can gather learners' data and deep analytical methods can be used for prediction and interpretation of learners progress and achievements. Augmented reality technology combined with wearables and mobile devices are also potential applications for DL methods in this area to make students motivated, lessons and studies to be interesting, and make educational learning methods to be efficient\cite{ibanez2014experimenting,kwok2015vision}. Moreover, DL can be used as a personalized recommendation module \cite{wang2015collaborative} to recommend more relevant content to the educator. The applications of DL in other domains, such as natural language translation and text summarization, would be of help for smart education when it comes to online learning on mobile devices.

Furthermore, the advent of Massive Open Online Courses (MOOCs) and their popularity among the students has led to generating a huge amount of data from the learners' behavior in such courses. MOOCs analysis can help identify struggling students in early sessions of a course, and provide sufficient support and attention from instructors to those students to achieve a better performance. Yang \textit{et al.} \cite{yang2017behavior} proposed a method for predicting student grades in MOOCs. They use clickstream data collected from lecture videos when students are watching the video and interacting with it. Clickstream data are fed to a time series RNN that learns from both prior performance and clickstream data. In addition, Piech \textit{et al.} applied RNN and LSTM networks to model the prediction of educator answers to exercises and quizzes, based on their past activities and interactions in MOOCs \cite{piech2015deep}. Results showed improvement over Bayesian Knowledge Tracing (BKT) methods, which employ a Hidden Markov Model (HMM) for updating probabilities of single concepts. Mehmood \textit{et al.} \cite{mehmood2017utilearn} also used DNNs for a personalized ubiquitous e-teaching and e-learning framework, based on IoT technologies, aiming for the development and delivery of educational content in smart cities. Their proposed framework is built on top of an IoT infrastructure (e.g., smart phone sensors, smart watch sensors, virtual reality technologies) connecting the users in order to optimize the teaching and learning processes. They used DNN for human activity recognition to deliver adaptive educational content to the students.

Classroom occupancy monitoring is another application that has been investigated by Conti \textit{et al.} \cite{conti2014brain}. In this work, the authors propose two methods for head detection and density estimation, both based on CNN architecture for counting students in a classroom. The algorithms have been deployed on off-the-shelf embedded mobile ARM platform. Their algorithm receives the images that are taken from the cameras in three classrooms with a rate of three pictures every $10$ minutes. They report that the root-mean-square (RMS) error of their algorithms is at most $8.55$.\\

\subsubsection{Industry} $\quad$\\
For the industry sector, IoT and cyber-physical systems (CPS) are the core elements to advance manufacturing technologies toward smart manufacturing (a.k.a Industry $4.0$). Providing high-accuracy intelligent systems is critical in such applications, as it directly leads to increased efficiency and productivity in assembly/product lines, as well as decreased maintenance expenses and operation costs. Therefore, DL can play a key role in this field. Indeed, a wide range of applications in industry (such as visual inspection of product lines, object detection and tracking, controlling robots, fault diagnosis, etc.) can benefit from introduction of DL models. 

In \cite{luckow2016deep}, visual inspection is investigated using CNN architectures including AlexNet and GoogLeNet over different platforms (Caffe, Tensorflow, and Torch). In this work, several images of produced vehicles in the assembly line along with their annotation are submitted to a DL system. It has been found that the best performance is achieved using Tensorflow with accuracy of $94\%$. Moreover, Tensorflow was the fastest framework in terms of training time, where the model reached its peak accuracy in a shorter time, followed by Torch and then Caffe.

Shao \textit{et al.} \cite{shao2017enhancement} used DNNs for feature extraction in a fault diagnosis (also referred as fault detection and classification (FDC)) system for rotating devices. Models using denoising auto-encoder (DAE) and contractive auto-encoder (CAE) were developed. The learned features from these models were both refined using a method called locality preserving projection (LPP), and fed to a softmax classifier for fault diagnosis. The input to the system is vibration data of a rotating device. In their system, seven operating conditions were considered, including normal operation, rubbing fault, four degrees of unbalance faults and compound faults (rub and unbalance). Given the vibration data, the diagnosis system identifies whether the device is in normal condition or in one of the fault conditions. Based on their experiments for fault diagnosis of rotor and locomotive bearing devices, the proposed approach is reported to outperform CNN and shallow learning methods.  

In another study reported by Lee \cite{lee2017framework}, a DBN model was proposed in conjunction with an IoT deployment and cloud platform to support fault detection of defect types in cars' headlight modules in a vehicle manufacturer setting. Their results confirmed the superior performance of the DBN model over two baseline methods, using SVM and radial basis function (RBF), in terms of error rate in test dataset. However, the reported error rate for their training dataset in the DBN model is comparable to that of the SVM model. 

For the problem of fault detection and classification (FDC) in noisy settings, \cite{lee2017deep} employed stacked denoising AEs (SdA) to both reduce the noise of sensory data caused by mechanical and electrical disturbances, and perform fault classification. Their system was applied for fault detection in wafer samples of a photolithography process. Results show that SdA leads to $14\%$ more accuracy in noisy situations compared to several baseline methods including K-Nearest Neighbors and SVM. Yan \textit{et al.} \cite{yan2015accurate} have also used SdA joint with extreme learning machines for anomaly detection in the behavior of gas turbine combustion system. Based on their results, the use of learned features by SdA leads to a better classification accuracy compared to the use of hand crafted features in their system.\\

\subsubsection{Government} $\quad$\\
Governments can gain great potential advantages through enhanced and intelligent connectivity that comes from the convergence of IoT and DL. Indeed, a wide variety of tasks that pertains to the governments or city authorities require precise analysis and prediction. For instance, the recognition and prediction of  natural disasters (landslide, hurricane, forest fires, etc.) and environmental monitoring is of high importance for governments to take appropriate actions. Optical remote sensing images that are fed to a deep AEs network and softmax classifiers were proposed by Liu \textit{et al.}~\cite{liu2016geological} to predict geological landslides. An accuracy of $97.4\%$ was reported for the proposed method, thus outperforming SVM and ANN models. In another study \cite{wang2017earthquake}, an LSTM network is used for the prediction of earthquakes. They used the historical data from US Geological Survey website for training. Their system was shown to achieve an accuracy of $63\%$ and $74\%$ with 1-D and 2-D input data, respectively. In another study by Liu \textit{et al.} \cite{liu2016application}, a CNN architecture is used for detection of extreme climate events, such as tropical cyclones, atmospheric rivers and weather fronts. Training data in their system included image patterns of climate events. The authors developed their system in Neon framework and achieved accuracy of $89\%$-$99\%$.

In addition, damage detection in the infrastructures of the cities, such as roads, water pipelines, etc., is another area where IoT and DL can provide benefits to governments. In \cite{maeda2016lightweight}, the problem of road damage detection was addressed using DNNs that gets its data through crowd-sourcing, which can be enabled by IoT devices. Citizens can report the damage through a mobile app to a platform. However, these citizens have no expert knowledge to accurately assess the status of road damage, which may lead to uncertain and/or wrong assessments. To eliminate these instances, the app can determine the status of the road damage by analyzing the image of the scene. The analysis is performed by a deep CNN that is trained by citizen reports as well as road manager inspection results. Since the training phase is out of the capability of mobile phones, the DL model is created on a server and trained everyday. An Android application can then download the latest model from the server upon each launch, and identify the status of road damages reported by images. Evaluations showed a damage classification accuracy of $81.4\%$ in $1$ second of analysis on the mobile devices.\\

\subsubsection{Sport and Entertainment} $\quad$\\
Sports analytics have been evolving rapidly during the recent years and plays an important role to bring a competitive advantage for a team or player. Professional sport teams nowadays have dedicated departments or employees for their analytics \cite{steinberg2015changing}. Analytics and predictions in this field can be used to track the players' behavior, performance, score capturing, etc. DL is new to this area and only few works have used DNNs in different sports. 

In \cite{liu2017deep}, a DL method has been proposed for making an intelligent basketball arena. This system makes use of SVM to choose the best camera for real-time broadcasting from among the available cameras around the court. They also fed basketball energy images\footnote{Basketball energy image is the spatial-temporal tracks of basketballs in the hotspot area. Hotspot area includes basketball backboard, hoop, and basket.} to a CNN to capture the shoot and scoring clips from the non-scoring ones, hence providing accurate online score reporting and interesting highlight clips. This system was shown to achieve an accuracy of $94.59\%$ in capturing score clips with $45$ ms of processing time for each frame.
 
In another work by Wang \textit{et al.} \cite{wang2016classifying}, an RNN has been used for classification of offensive basketball plays in NBA games. The authors used video clips of the games from SportVU\footnote{http://go.stats.com/sportvu} dataset. This dataset provides videos of the rate of $25$ frames per second to detect players' unique ID, their location on the court, and the position of the ball. Their model is shown to achieve accuracy of $66\%$ and $80\%$ for top-1 and top-3 accuracies, respectively. Similarly, \cite{shah2016applying} used an RNN with LSTM units over the same dataset to predict the success rates of three-point shots, and reported better classification accuracy compared to gradient boosted machine (GBM) and generalized linear model (GLM).

Kautz \textit{et al.} \cite{kautz2017activity} investigated players' activity recognition in volleyball. Wearable sensor data and CNN were employed to achieve this task, and a classification accuracy of $83.2\%$ to identify players activities was observed. 

Group activity recognition is another interesting direction for sport teams. Ibrahim \textit{et al.} \cite{ibrahim2016hierarchical} investigated this option in a volleyball team using a hierarchical LSTM model. In this work, a single LSTM model was built to derive the activities of each player, and a top-level LSTM model was designed to aggregate the individual models to identify the overall behavior of the team. A CNN model was utilized to extract features from video frames, and feed them to the individual LSTM models. Compared to several baseline models, the proposed hierarchical model obtained better classification results.\\

\subsubsection{Retail}$\quad$\\
Due to the proliferation of mobile devices, online shopping has increased greatly. A recent shift toward product image retrieval through visual search techniques was noticed \cite{bell2015learning}. CNNs have been used for this visual search of clothes and fashion market, to find items in online shops that are identical or similar to what you have seen in a movie \cite{xiao2016exact} or in the street \cite{hadi2015buy}.

Moreover, shopping for visually impaired people needs to be made convenient. A combination of IoT technologies, including smart carts, integrated with DL methods can be a solution to this problem. In \cite{advani2017multitask}, a visual grocery assistance system that includes smart glasses, gloves, and shopping carts was designed to help visually impaired people in shopping. This system also used a CNN to detect items in the aisles. 

Moreover, check-out counters in retail stores are usually the bottlenecks where people queue up to pay their shoppings. The development of smart carts can enable real-time self check-out and enhancing such system with prediction capabilities can offer an item that a customer may need based on his/her past shopping.

Furthermore, recommending items to shoppers is a popular application of IoT for retails that uses different technologies, like BLE signals or visual cameras. The latter approach can be done through identifying the shop items or shoppers actions (e.g., reach to a shelf, retract from a shelf, etc.) \cite{singh2016multi} and providing a list of related items for the detected action.

To analyze the customer interest in merchandise, Liu \textit{et al.} \cite{liu2016joint} proposed a customer pose and orientation estimation system based on a DNN consisting of a CNN and RNN. The input data comes from surveillance cameras. The CNN network is used to extract image features. The image features and the last predicted orientation features are then fed to an RNN to get the output pose and orientation.\\

\subsubsection{Smart IoT Infrastructure}$\quad$\\
IoT environments consist of a large number of sensors, actuators, media and many other devices that generate big M2M and network traffic data. Therefore, the management, monitoring, and coordination of these devices are subject to processing such big data, with advanced machine learning techniques, to identify bottlenecks, improve the performance, and guarantee the quality of service.

One popular task for infrastructure management would be anomaly detection. For example, spectrum anomaly detection in wireless communications using AE was proposed by Feng \textit{et al.} in \cite{feng2017anomaly}. In this work, an AE model was developed to detect the anomaly that may happen due to sudden change in signal-to-noise ratio of the communications channel. The model is trained on features based on the time–frequency diagram of input signals. Their result showed that a deeper AE performs better than the conventional shallow networks. Lopez-Martin \textit{et al.} \cite{lopez2017conditional} and Shone \textit{et al.} \cite{shone2017deep} have used conditional VAE and deep AEs, respectively, for network intrusion detection. In the conditional VAE, the labels of the samples are used in addition to the latent variables as extra inputs to the decoder network.

The tiny traces of IoT traffic may not lead to congestion at the backbone. However, the need to access the channel simultaneously by a large number of IoT devices can lead to contention during the channel access phase. The contention in channel access turns to a severe problem when the access delays are increased \cite{hasan2013random}. Therefore, load balancing is a viable solution that can be performed by DL models to predict the traffic metrics and propose alternate routes. Kim \textit{et al.} \cite{kim2017load} used DBNs to perform load balancing in IoT. Their DL model is trained on a large amount of user data and network loads. Interference identification can also be handled by DNNs as demonstrated by Schmidt \textit{et al.} \cite{schmidt2017wireless}, where a wireless interference identification systems based on CNN was proposed. Ahad \textit{et al.} \cite{ahad2016neural} provided a survey focused on the application of neural networks to wireless networks. They reviewed related literature on quality of service and quality of experience, load balancing, improved security, etc. 

Since the emerging 5th Generation (5G) cellular networks constitute one of the main pillars of IoT infrastructure, it is necessary to utilize cutting-edge technologies to enhance the different aspects of cellular networks, including radio resource management, mobility management, service provisioning management, self-organization, and to find an efficient and accurate solution for complicated configuration problems \cite{li2017intelligent}. As part of these efforts, using crowd-sourced cellular networks data (e.g., signal strength) can help to come up with reliable solutions. For example, such big data can be used to create more precise coverage maps for cellular networks to improve the performance of the network \cite{fida2017zipweave}, as was performed by the OpenSignal mobile application \cite{opensignal2018open}.

\subsection{Lessons Learned} 
In this section, we have identified five classes of IoT services as the foundational services that can be used in a wide range of IoT applications. We discussed how DL has been used to achieve these services. Moreover, we went through a wide range of IoT domains to find out how they exploit DL to deliver an intelligent service. Table \ref{tbl:usage_foundational_IoT} shows the works that utilized foundational services in IoT domains. 

Many IoT domains and applications have greatly benefited from image recognition. The interest is expected to grow faster as the high-resolution cameras embedded in smart phones will result in easier generation of image and video data. The usage of other fundamental applications, especially physiological and psychological detections as well as localization, can be seen in different fields. However, the utilization of security and privacy services is shown to be limited. This is the gap in developing intelligent IoT applications, where the potential activities of hackers and attackers are ignored. Also, voice recognition with DL has not been used widely in IoT applications belonging to several domains, such as smart homes, education, ITS, and industry. There are works that use voice recognition with traditional machine learning approaches. Voice recognition has shown remarkable advancement with DL. One reason for the few appearance of this technique in IoT applications is the lack of comprehensive training datasets for each domain, as there is a need for large training datasets to train voice recognition DNNs. 

Foundational services need fast data analytics to be efficient in their context. Despite several works in this direction, IoT fast data analytics based on DL has many spaces for development of algorithms and architectures.

\begin{table*}
\renewcommand{\arraystretch}{1.2}
\centering
\caption{Usage of Foundational Services in IoT Domains.}
\label{tbl:usage_foundational_IoT}
\begin{tabular}{l|l|l|l|l|l|l|}
\cline{3-7}
\multicolumn{2}{l|}{\multirow{2}{*}{}}                            & \multicolumn{5}{c|}{\textbf{IoT Foundational Services}}                                                                                                         \\ \cline{3-7} 
\multicolumn{2}{l|}{}                                             & \textbf{Image Recognition} & \textbf{Voice Recognition} & \begin{tabular}[c]{@{}l@{}}\textbf{Physiological} \textbf{\&} \\ \textbf{Psychological}\\ \textbf{Detection}\end{tabular} & \textbf{Localization} & \begin{tabular}[c]{@{}l@{}}\textbf{Security} \textbf{\&} \\\textbf{Privacy} \end{tabular}  \\ \hline 
\multicolumn{1}{|l|}{\multirow{11}{*}{\rotatebox[origin=c]{90}{\textbf{IoT Domains}}}} & Smart Home  &                   &                   &                                                                                      &              &     \\ \cline{2-7} 
\multicolumn{1}{|l|}{}                              & Smart City  & \cite{mittal2016spotgarbage,amato2016deep,valipour2016parking
}                  &                   &                                                                                       &  \cite{song2016deeptransport,liang2016mercury}            &     \\ \cline{2-7} 
\multicolumn{1}{|l|}{}                              & Energy      &                   &                   &                                                                                       &              & \cite{he2017real}    \\ \cline{2-7} 
\multicolumn{1}{|l|}{}                              & ITS         &  \cite{cirecsan2012multi,lim2017real}                 &                   &                                                                                       &              & \cite{kang2016intrusion}    \\ \cline{2-7} 
\multicolumn{1}{|l|}{}                              & Healthcare  & \cite{liu2017new,liu2016deepfood,pereira2016convolutional,wang2017detecting}                  &  \cite{muhammad2017smart}                 &  \cite{feng2014deep}                                                                                      &              &     \\ \cline{2-7} 
\multicolumn{1}{|l|}{}                              & Agriculture & \cite{sladojevic2016deep, kussul2017deep,kuwata2015estimating,scott2017training,steen2016using,sa2016deepfruits}                  &                   &                                                                                       &              &     \\ \cline{2-7} 
\multicolumn{1}{|l|}{}                              & Education   &  \cite{conti2014brain}                 &                   &  \cite{mehmood2017utilearn}                                                                                     &              &     \\ \cline{2-7} 
\multicolumn{1}{|l|}{}                              & Industry    &  \cite{luckow2016deep,lee2017framework}                 &                   &                                                                                       &  \cite{yang2016rfid}            &     \\ \cline{2-7} 
\multicolumn{1}{|l|}{}                              & Government  & \cite{liu2016geological,maeda2016lightweight,liu2016application}                  &                   &                                                                                       &              &     \\ \cline{2-7} 
\multicolumn{1}{|l|}{}                              & Sport       & \cite{liu2017deep,wang2016classifying,shah2016applying}                  &                   & \cite{kautz2017activity,ibrahim2016hierarchical}                                                                                       & \cite{lu2017multimodal}             &     \\ \cline{2-7} 
\multicolumn{1}{|l|}{}                              & Retail      &  \cite{bell2015learning,xiao2016exact,hadi2015buy,advani2017multitask}                 &                   &  \cite{liu2016joint}                                                                                     & \cite{singh2016multi}             &     \\
\hline
\end{tabular}
\end{table*}

\begin{figure}
	\begin{center}		
		\includegraphics[width=.5\textwidth]{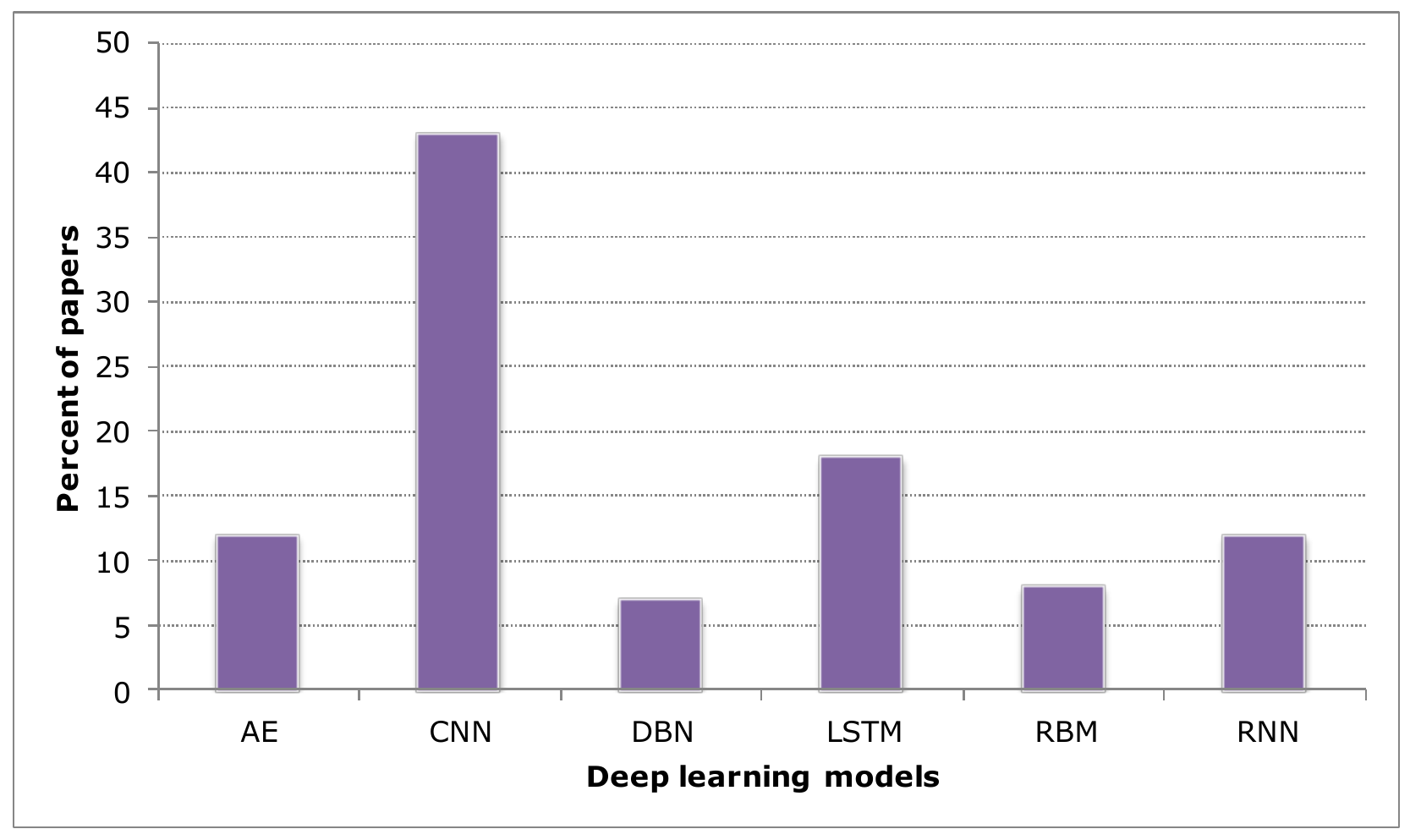}		
	\end{center}
	\caption{The percentage of surveyed papers that have used DL models}\label{fig:fig-DL_models_stats}	
\end{figure}

Table \ref{tbl:IoT_domains_DL_models} summarizes the research in each domain, and their DL model. Figure~\ref{fig:fig-DL_models_stats} also depicts the frequency of different models that have been used in the different research works. About $43\%$ of the papers have used CNN in building their proposed systems while DBN are less used compared to other models (about $7\%$). RNNs and LSTMs together, as time-series models, have been used in $30\%$ of the works. The table also emphasizes the great impact of works related to image recognition on IoT applications. Moreover, one third of the IoT applications are related to time-series or serial data, in which employing RNNs is a helpful approach.\\ 

\subsubsection{Complexity vs. Performance}$\quad$\\

Canziani \textit{et al.} \cite{canziani2016analysis} analyzed several state-of-the-art DNN models to find out the relationship between their accuracy, memory usage, operations count, inference time, and power consumption. They found out that the accuracy and inference time show a hyperbolic relationship such that a minor increase in accuracy leads to a long computational time. They also illustrated that the number of operations in a network model have a linear relationship with the inference time. Their results also indicated that imposing an energy constraint would limit the maximum achievable accuracy. Regarding the memory footprint and batch size, the results showed that the maximum memory usage is constant during the initial memory allocation of the model and then linearly increases with the batch size. Given that the neurons are the main building blocks of a model that performs the operations, the number of operations is proportional to the number of neurons. So, the complexity can be expressed as the number of neurons in the network, such that increasing the number of neurons directly impacts the run-time.  

However, there is not a clear relationship between the accuracy and number of layers (i.e., depth) or number of neurons. There are reports indicating degradation of accuracy after increasing the number of layers beyond some point. For example, Zhang \textit{et al.} \cite{zhang2016deep} assert that the number of hidden layers and neurons have a direct effect on the accuracy of the localization system. Increasing the layers initially leads to better results, but at some point when the network is made deeper, the results start degrading. Their best result was obtained with a network of three hidden layers. On the other hand, the depth of representations has been shown to be most beneficial for many image recognition tasks \cite{he2016deep}. The high accuracy of vision-based tasks, in part, is due to introducing deeper networks with larger number of parameters (e.g., over $1000$ layers as presented in \cite{he2016deep, he2016identity}). There are many hyper-parameters to optimize (e.g., epoch count, loss function, activation function, optimization function, learning rate, etc.) that complicate the process of developing good and accurate DL models. Table~\ref{tbl:dnn_sizes} summarizes the characteristics of DNN models in several applications. In the table, the test time is for one sample unless specified otherwise.  

\subsubsection{Pitfalls and Criticisms}$\quad$\\
DL models were demonstrated to be as a great step toward creating powerful AI systems, but they are not a single solution for all problems. DL techniques are known as black boxes that show high predictability but low interpretability. While powerful prediction capability is most desired from the scientific perspective, the interpretability and explicability of the models are preferred from a business perspective \cite{bourguignat2014interpret}. In \cite{chollet2018deep}, Chollet argued that problems requiring reasoning, long-term planning, and algorithmic-like data manipulation, cannot be solved by deep learning models. This is because of the nature of DL techniques that only transform one vector space into another, no matter how much data you feed them. 

Moreover, there are criticisms on the performance of DNN models, suggesting that the traditional models may achieve comparable results or even outperform deep models \cite{hellendoorn2017deep}. According to Chatfield \textit{et al.} \cite{chatfield2014return}, the dimensionality of the convolutional layers in CNNs can be shrunk without adversely affecting the performance. They also discussed that the shallow techniques can reach a performance analogous to that of deep CNN models if the former models use the data augmentation techniques that are commonly applied to CNN-based methods. Ba \textit{et al.} \cite{ba2014deep} performed several empirical tests asserting that shallow FNNs can learn the complex functions and achieve accuracies that were previously possible only by deep models. In their work on optical communication systems, Eriksson \textit{et al.} \cite{eriksson2017applying} showed that using pseudo random bit sequences or short repeated sequences can lead to overestimating the signal-to-noise ratio.

Generally, DL models are sensitive to the structure, and size of the data. Compared to shallow models, they work better when there is a large body of training data with a wide range of attributes. Otherwise, shallow models typically lead to better results.

\begin{table*}
\renewcommand{\arraystretch}{1.2}
\centering
\caption{The common use of different DNN models in IoT domains.}
\label{tbl:IoT_domains_DL_models}
\begin{tabular}{|l|c|c|c|c|c|c|}
\hline
\multirow{2}{*}{\textbf{Domain}} & \multicolumn{6}{c|}{\textbf{Usage of DNNs}}          \\ \cline{2-7} 
                        & \textbf{AE} & \textbf{CNN} & \textbf{DBN} & \textbf{LSTM} & \textbf{RBM} & \textbf{RNN}  \\ \hline \hline
Image Recognition & \cite{liu2016geological} & \begin{tabular}[c]{@{}l@{}} \cite{mittal2016spotgarbage,liu2017new,sladojevic2016deep,amato2016deep,valipour2016parking}\\ \cite{cirecsan2012multi,lim2017real,pereira2016convolutional,wang2017detecting}\\ \cite{kussul2017deep,steen2016using,conti2014brain}\\ \cite{liu2017deep,ibrahim2016hierarchical,xiao2016exact} \end{tabular} & \cite{lee2017framework} & \cite{shah2016applying} &  & \cite{wang2016classifying,shah2016applying}\\ \hline
\begin{tabular}[c]{@{}l@{}}Physiological \& \\ Psychological Detection\end{tabular} & \cite{fragkiadaki2015recurrent,kahou2016emonets} & \begin{tabular}[c]{@{}l@{}} \cite{toshev2014deeppose,ordonez2016deep,li2016deepRfidActivity}\\ \cite{pigou2015beyond,kahou2016emonets,neverova2016learning} \end{tabular} & \cite{kahou2016emonets} & \cite{ordonez2016deep,tao2016multi} &  & \cite{pigou2015beyond,fragkiadaki2015recurrent,neverova2016learning} \\ \hline
Localization & \cite{gu2015semi,zhang2016deep} & \cite{liu2017fusion, becker2017indoor} & & \cite{lu2017multimodal} & \cite{wang2015deepfi} & \\ \hline  
Privacy and Security &  & \cite{shokri2015privacy} & \cite{he2017real,yuan2014droid} &  &  & \\ \hline
Smart home              &     &  \cite{manic2016intelligent}    &     & \cite{manic2016intelligent}   &                 &  \\ \hline
Smart city              &     & \cite{mittal2016spotgarbage,amato2016deep, valipour2016parking}     &     & \cite{song2016deeptransport}   &                  & \cite{liang2016mercury} \\ \hline
Energy                  & \cite{gensler2016deep}     &      & \cite{gensler2016deep}    & \cite{gensler2016deep} \cite{hada2016deep}  &  \cite{mocanu2016big} \cite{mocanu2016deep}                & \cite{mocanu2016deep} \\ \hline
ITS             &     &  \cite{cirecsan2012multi,lim2017real}    & \cite{kang2016intrusion}     & \cite{tian2015predicting}   & \cite{ma2015large}                 & \cite{ma2015large} \\ \hline
Healthcare              &  & \cite{liu2016deepfood,liu2017new,pereira2016convolutional,wang2017detecting}          &  \cite{feng2014deep}   & \cite{lipton2015learning}   &    \cite{feng2014deep}              &  \\ \hline
Agriculture             &     & \cite{sladojevic2016deep,kussul2017deep,kuwata2015estimating,scott2017training,steen2016using,sa2016deepfruits}     &      &    &                  &  \\ \hline
Education             &     & \cite{conti2014brain}     &      &  \cite{piech2015deep}  &                  & \cite{yang2017behavior,piech2015deep} \\ \hline
Industry              & \cite{shao2017enhancement,lee2017deep,yan2015accurate}    & \cite{luckow2016deep}     & \cite{lee2017framework}    &    &                  &  \\ \hline
Government              & \cite{liu2016geological}    & \cite{maeda2016lightweight,liu2016application}    &     & \cite{wang2017earthquake}   &                  &  \\ \hline
Sport              &     & \cite{liu2017deep,kautz2017activity,ibrahim2016hierarchical}     &     &  \cite{shah2016applying,ibrahim2016hierarchical}  &                  & \cite{wang2016classifying} \\ \hline
Retail              &     &  \cite{bell2015learning,xiao2016exact,hadi2015buy,advani2017multitask}   &     &    &                  & \cite{singh2016multi}  \\ \hline
IoT Infrastructure              & \cite{feng2017anomaly,lopez2017conditional,shone2017deep}     &  \cite{schmidt2017wireless}    & \cite{kim2017load}    &    &                  &   \\ \hline

\end{tabular}
\end{table*}

\begin{table*}[]
\renewcommand{\arraystretch}{1.2}
\centering
\caption{DNN Sizes and Complexities in Different Applications. (NA: Not Available, C: Convolutional Layer, F: Fully Connected layer, FF: Feedforward layer, SG: Summation (collapse) layer, P: Pooling layer, LRN: local response normalization layer, SM: Softmax layer, L: LSTM layer, R: RNN layer, RBM: RBM hidden layer)}
\label{tbl:dnn_sizes}
\begin{tabular}{|l|l|l|l|l|l|l|}
\hline
\textbf{Work}     & \textbf{Application}  & \textbf{Type of DNN} & \textbf{Depth}                                                       & \textbf{Layers Sizes}                                                                                                 & \textbf{Training Time}                                                               & \textbf{Test Time} \\ \hline \hline
\cite{ma2015large}  & Transportation analysis  & RNN+RBM & 2                                                             & \begin{tabular}[c]{@{}l@{}} R(100)-RBM(150) \end{tabular}                                                                                        &  NA                                                                         & 354 (s), whole test set \\ \hline
\cite{wang2015deepfi} & Localization & \begin{tabular}[c]{@{}l@{}}RBM\\ DBN\end{tabular}                  & \begin{tabular}[c]{@{}l@{}} 4\\4 \end{tabular} & \begin{tabular}[c]{@{}l@{}}500-300-150-50\\ 300-150-100-50\end{tabular}                                     & NA                                                                          & NA        \\ \hline
\cite{gu2015semi} & Localization & \begin{tabular}[c]{@{}l@{}}SdA\\ DBN\\ ML-ELM\\ SDELM\end{tabular} & \begin{tabular}[c]{@{}l@{}}4\\ 3\\ 5\\ 5\end{tabular} & \begin{tabular}[c]{@{}l@{}}26-200-200-71\\ 26-300-71\\ 26-300-300-1500-71\\ 26-300-300-1500-71\end{tabular} & \begin{tabular}[c]{@{}l@{}}451 (s)\\ 110 (s)\\ 14 (s)\\ 24 (s)\end{tabular} & NA        \\ \hline
\cite{zhang2016deep} & Localization & SdA  &  5                                                            & 163-200-200-200-91                                                                                          & NA                                                                          & 0.25 (s)  \\ \hline
\cite{toshev2014deeppose} & Pose detection & CNN   & 12                                                            & \begin{tabular}[c]{@{}l@{}}C(55$\times$55$\times$96)-LRN-P-\\C(27$\times$27$\times$256)-LRN-P-\\C(13$\times$13$\times$384)-\\C(13$\times$13$\times$384)-\\C(13$\times$13$\times$256)-P-\\F(4096)-F(4096)-SM \end{tabular}                                                                                        & NA                                                                          & 0.1 (s) \\ \hline
\cite{ordonez2016deep} & Human activity detection  & CNN+LSTM    &   7                                                         & \begin{tabular}[c]{@{}l@{}} C(384)-C(20544)-C(20544)-\\C(20544)-L(942592)-\\L(33280)-SM \end{tabular}                                                                                        & 340 (min)                                                                          & 7 (s), whole test set \\ \hline
\cite{tao2016multi} & Human activity detection  & LSTM  &  5                                                             & \begin{tabular}[c]{@{}l@{}} L(4)-FF(6)-L(10)-SG-SM \end{tabular}                                                                                        &  NA                                                                         & 2.7 (ms) \\ \hline
\cite{he2017real}  & FDI detection  & DBN  &  4                                                           & \begin{tabular}[c]{@{}l@{}} 50-50-50-50 \end{tabular}                                                                                        &  NA                                                                         & 1.01 (ms) \\ \hline
\cite{yuan2014droid}  & Malware detection  & DBN  & 3                                                            & \begin{tabular}[c]{@{}l@{}} 150-150-150 \end{tabular}                                                                                        &  NA                                                                         & NA \\ \hline
\cite{valipour2016parking}  & Parking space  & CNN & 8                                                             & \begin{tabular}[c]{@{}l@{}} C(64$\times$11$\times$11)-C(256$\times$5$\times$5)-\\C(256$\times$3$\times$3)-C(256$\times$3$\times$3)-\\C(256$\times$3$\times$3)-F(4096)-F(2)-SM \end{tabular}                                                                                        &  NA                                                                         & 0.22 (s) \\ \hline
\cite{lim2017real}  & Traffic sign detection  & CNN & 6                                                             & \begin{tabular}[c]{@{}l@{}} C(36$\times$36$\times$20)-P-\\C(14$\times$14$\times$50)-P-\\FC(250)-SM \end{tabular}                                                                                        &  NA                                                                         & \begin{tabular}[c]{@{}l@{}} 29.6 (ms) on GPU\\4264 (ms) on CPU \end{tabular} \\ \hline
\cite{liu2016deepfood}  & food recognition  & CNN & 22                                                             & \begin{tabular}[c]{@{}l@{}} Used GoogLeNet \cite{szegedy2015going} \end{tabular}                                                                                        &  NA                                                                         & \begin{tabular}[c]{@{}l@{}} 50 (s) \end{tabular} \\ \hline
\cite{kussul2017deep}  & Crop recognition  & CNN & 6                                                             & \begin{tabular}[c]{@{}l@{}} C(96$\times$7$\times$7)-P-\\C(96$\times$4$\times$4)-P-F(96)-F(96) \end{tabular}                                                                                        &  12 (h)                                                                         & NA \\ \hline
\cite{conti2014brain}  & Classroom Occupancy  & CNN & 5                                                             & \begin{tabular}[c]{@{}l@{}} C(6$\times$28$\times$28)-P-\\C(16$\times$10$\times$10)-P-F(120) \end{tabular}                                                                                        &  2.5 (h)                                                                         & 2 (s) (4 thread) \\ \hline
\cite{shao2017enhancement}  & Fault diagnosis  & AE & 4                                                             & \begin{tabular}[c]{@{}l@{}} 300-300-300-300 \end{tabular}                                                                                        &  NA                                                                         & 91 (s) \\ \hline
\cite{maeda2016lightweight}  & Road damage detection  & CNN & 8                                                             & \begin{tabular}[c]{@{}l@{}} Used AlexNet \cite{krizhevsky2012imagenet}  \end{tabular}                                                                                        &  NA                                                                         & 1.08 (s) \\ \hline
\cite{wang2016classifying}  & Classifying offensive plays  & RNN & 3                                                             & \begin{tabular}[c]{@{}l@{}}  10-10-11 \end{tabular}                                                                                        &  NA                                                                         & 10 (ms) \\ \hline
\end{tabular}
\end{table*}

\section{DL on IoT Devices}\label{sec:DL_on_devices}

Prior to the era of IoT, most research on DL targeted the improvement of its models and algorithms to efficiently operate when the scale of the problem grows to the big data, by trying to deploy efficient models on cloud platforms. The emergence of IoT has then opened up a totally different direction when the scale of the problems shrank down to resource-constrained devices and to the need for real-time analytics. 

Smart objects need to support some sort of light-weight intelligence. Due to DL's successful results in speech and video applications, which are among the fundamental services and common uses of IoT, adapting its models and approaches for deployment on resource-constrained devices became a very crucial point of study. So far, DL methods can hardly be used in IoT and resource-constrained devices for training purposes since DL models require a large portion of resources, such as the processors, battery energy, and memory. In some cases, the available resources are even not sufficient for running a pre-trained DL algorithm for inference tasks \cite{lane2015early}. Luckily, it has been recently shown that many parameters that are stored in DNNs may be redundant \cite{denil2013predicting}. It is also sometimes unnecessary to use a large number of hidden layers to get a high accuracy \cite{ba2014deep}. Consequently, efficiently removing these parameters and/or layers will considerably reduce the complexity of these DNNs without significant degradation of the output \cite{denil2013predicting,ba2014deep} and make them IoT-friendly. In the remaining of this section, we will discuss methods and technologies to achieve this results, and illustrate their applications in different domains.

\subsection{Methods and Technologies}
DL models may consist of millions or even billions of parameters which need sophisticated computing and large storage resources. In this section, we discuss several state-of-the-art approaches that bring DL models to IoT embedded and resource constrained devices.\\

\subsubsection{Network Compression}$\quad$\\
One way of adopting DNNs to resource-constrained devices is through the use of network compression, in which a dense network is converted to a sparse network. This approach helps in reducing the storage and computational requirements of DNNs when they are used for classification or other sorts of inference on IoT devices. The main limitation of this approach is that they are not general enough to support all kinds of networks. It is only applicable to specific network models that can exhibit such sparsity.

Another interesting study to adopt compressed DL models on IoT devices is the one performed by Lane \textit{et al.}~\cite{lane2015early}. In this study, the authors measure different factors that embedded, mobile, and wearable devices can bear for running DL algorithms. These factors included measurements of the running time, energy consumption, and memory footprint. The study focused on investigating the behavior of CNNs and DNNs in three hardware platforms that are used in IoT, mobile, and wearable applications, namely \textit{Snapdragon 800} used in some models of smart phones and tablets, \textit{Intel Edison} used in wearable and form-factor sensitive IoT, and \textit{Nvidia Tegra K1} employed in smart phones as well as IoT-enabled vehicles. Torch has been used for developing and training DNNs, and AlexNet \cite{krizhevsky2012imagenet} was the dominant model used in these platforms. Their measurement of energy usage indicated that all the platforms, including Intel Edison (which is the weakest one), were able to run the compressed models. In terms of execution time for CNNs, it has been shown that the later convolutional layers tend to consume less time as their dimensions decrease. 

Moreover, it is known that feed-forward layers are much faster than the convolutional layers in CNNs. Consequently, a good approach for improving CNN models on resource-constrained devices is to replace convolutional layers with feed-forward layers whenever possible. In addition, choosing the employed activation function in DNNs can have a great effect on time efficiency. For instance, several tests have shown that ReLU functions are more time-efficient followed by Tanh, and then Sigmoid. However, the overall runtime reduction of such selection is not significant (less than $7\%$) compared to the execution time of layers (at least $25\%$). In terms of memory usage, CNNs use less storage than DNNs due to the fewer stored parameters in convolutional layers compared to their counterpart in DNNs.    

\begin{figure}[]
	\begin{center}		
		\includegraphics[width=.5\textwidth]{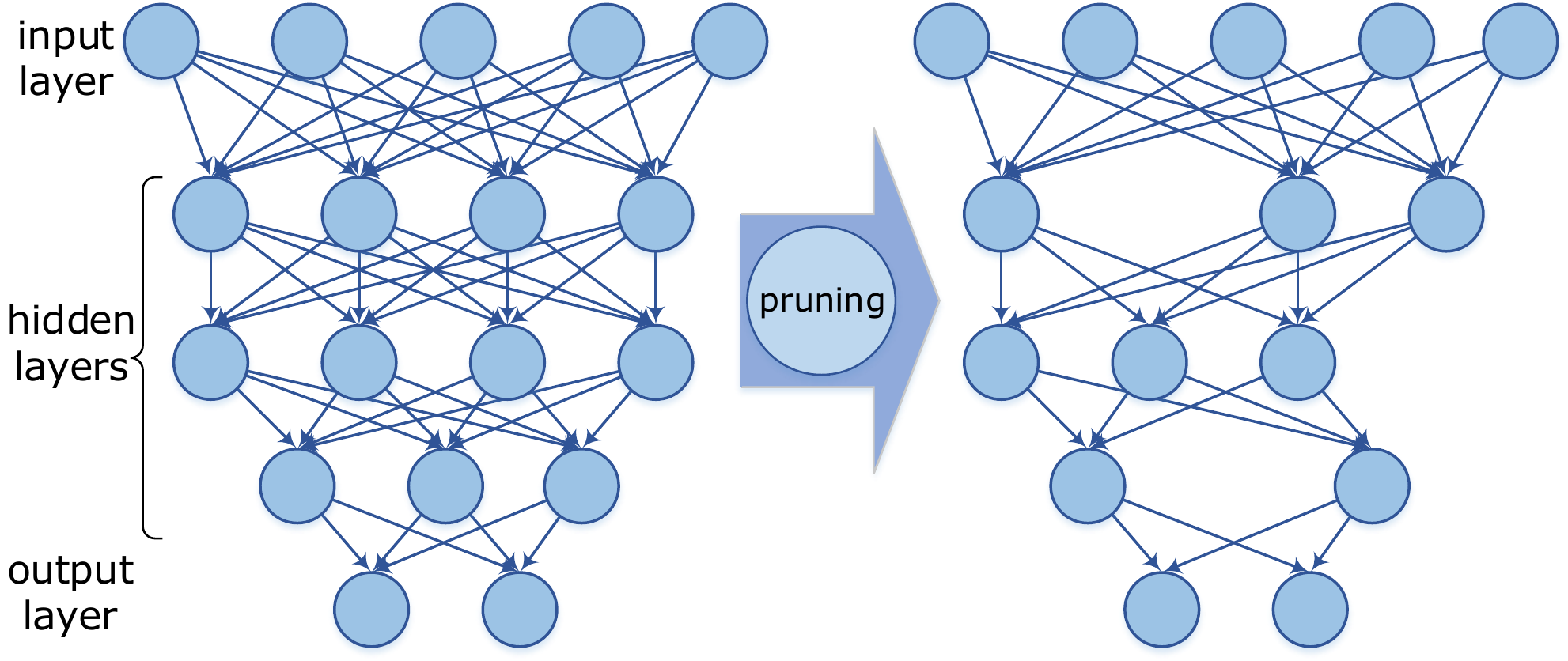}		
	\end{center}
	
	\caption{The overall concept of pruning a DNN.}\label{fig:pruning}	
\end{figure}

As previously stated, reducing the number of employed parameters in DNNs, by pruning redundant and less-important ones, is another important approach to make DNNs implementable on resource-constrained devices. One of the first works on this approach is Optimal Brain Damage \cite{lecun1989optimal} in $1989$. At the core of this method, the second order derivative of the parameters are used to compute the importance of the parameters and prune unimportant parameters from the network. The method presented in \cite{han2015learning} also works based on pruning redundant and unnecessary connections and neurons as well as using weight sharing mechanisms. Weight sharing replaces each weight with an $n$ bit index from a shared table that has $2^n$ possible values. The steps to prune the network as describe by Han \textit{et al.} in \cite{han2015learning} consist of:
\begin{itemize}
\item Train the network to find the connections with high weights.
\item Prune the unimportant connections that have a weight less than a threshold.
\item After pruning, there may remain some neurons with no input nor output connections. The pruning process identifies these neurons and removes them as well as all their remaining connections.
\item Retrain the network to fine-tune the weight of the updated model. The weights should be transferred from the previous trained steps instead of initializing them, otherwise the performance will degrade to some extent.
\end{itemize}
The authors evaluated this approach on four models related to vision, namely AlexNet, VGG-16, LeNet-300-100, and LeNet-5. The models were compressed at least 9 times for AlexNet and 13 times at VGG-16, while the accuracy of the models were almost preserved. One limitation of this approach is that it cannot be used for other types of DNN models. Moreover, the resulting compressed networks are not efficient enough on all hardware platforms and CPU architectures, and thus need new kinds of accelerators that can handle dynamic activation sparsity and weight sharing. Figure~\ref{fig:pruning} illustrates the concept of pruning a DNN.  

In \cite{han2016eie}, an inference engine, called EIE, was designed with a special hardware architecture and SRAMs instead of DRAMs, and was shown to work well with compressed network models. In this architecture, customized sparse matrix vector multiplication and weight sharing are handled without losing the efficiency of the network. The engine consists of a scalable array of processing elements (PEs), each of which keeping a part of the network in an SRAM and performing its corresponding computations. Since most of the energy that is used by neural networks is consumed for accessing the memory, the energy usage is reported to be 120 times fewer with this designed accelerator than the energy consumption of the corresponding original network.

\begin{table*}
\centering
\caption{Methods and Technologies to Bring DL on IoT Devices}
\label{tbl:ondevice_DL_methods}
\begin{tabular}{|l|l|l|l|}
\hline
\multicolumn{1}{|c|}{\textbf{Method / Technology}} & \multicolumn{1}{c|}{\textbf{Reference}} & \multicolumn{1}{c|}{\textbf{Pros}}                                                                                              & \multicolumn{1}{c|}{\textbf{Cons}}                                                                                                                           \\ \hline \hline
Network Compression          &  \cite{han2015learning} \cite{chen2015compressing} \cite{courbariaux2016binarynet}                              & \tabitem Reduce storage and computation                                                                                         & \begin{tabular}[c]{@{}l@{}}\tabitem Not general for all DL models \\ \tabitem Need specific hardware \\ \tabitem The pruning process bring \\overload to training\end{tabular} \\ \hline
Approximate Computing        &   \cite{venkataramani2014axnn, moons2016energy}                             & \begin{tabular}[c]{@{}l@{}} \tabitem Makes fast DL models \\ \tabitem Save energy \end{tabular}                                                                                      & \tabitem Not suitable for precise systems                                                                                                                    \\ \hline
Accelerators                 & \begin{tabular}[c]{@{}l@{}} \cite{price2017scalable} \cite{han2016eie}  \cite{ramasubramanian2014spindle}\\ \cite{chen2017eyeriss} \cite{chen2014diannao} \cite{lane2016deepx} \end{tabular}                            & \begin{tabular}[c]{@{}l@{}}\tabitem Integrates DL model with the hardware \\ \tabitem Efficient computations\end{tabular}               & \tabitem \begin{tabular}[c]{@{}l@{}}Does not work with the traditional \\hardware platforms \end{tabular}                                                                                                   \\ \hline
Tinymote with DL             &  \cite{bang201714}                             & \begin{tabular}[c]{@{}l@{}}\tabitem Good for time-critical IoT apps \\ \tabitem Energy-efficient\\ \tabitem Provides more security and privacy for data\end{tabular} & \begin{tabular}[c]{@{}l@{}} \tabitem Special-purpose networks \end{tabular}                                                                                                                                                   \\ \hline
\end{tabular}
\end{table*}

In HashedNets \cite{chen2015compressing}, the neural network connection weights are randomly grouped into hash buckets using a hash function. All connections that fall into a same bucket are represented by a single parameter. Backpropagation is used to fine-tune the parameters during the training. Testing results show that the accuracy of this hash-based compressed model outperforms all other compression baseline methods. 

The work in \cite{courbariaux2016binarynet} by Courbariaux \textit{et al.} proposed to binarize network weights and neurons at both the inference phase and the entire training phase, in order to reduce the memory footprint and accesses. The network can also perform most of the arithmetic operations through bit-wise operations, leading to a decreased power consumption. MNIST, CIFAR-10 and Street View House Numbers (SVHN) \cite{netzer2011svhn} datasets were tested over Torch7 and Theano frameworks using this approach, and results were found to be promising.\\

\subsubsection{Approximate Computing}$\quad$\\
Approximate computing is another approach to both implement machine learning tools on IoT devices and contribute to the energy saving of their hosting devices~\cite{venkataramani2014axnn, moons2016energy}. The validity of this approach arises from the fact that, in many IoT applications, machine learning outputs (e.g., predictions) need not to be exact, but rather to be in an acceptable range providing the desired quality. Indeed, these approaches need to define quality thresholds that the output must not pass. Integrating DL models with approximate computing can lead to more efficient DL models for resource-constrained devices. Venkataramani \textit{et al.}~\cite{venkataramani2014axnn} proposed the extension of approximate computing to neural networks, and converted a neural network to an approximate neural network. In their approach, the authors extend backpropagation to identify the neurons with the least effect on output accuracy. Then, the approximate NN is formed by substituting less important neurons in the original network with their approximate counterparts. Making approximate neurons is performed by an approximate design technique called precision scaling. Instead of using a typical fixed number of bits (16-bit or 32-bit format) to present computations, various number of bits ($4$ - $10$ bits) are used in this technique. After forming the approximate network, the precisions of the inputs and weights of neurons are adjusted to come up with an optimal trade-off between accuracy and energy. There are also other attempts that have reported applying approximate computing with precision scaling on CNNs~\cite{moons2016energy} and DBNs~\cite{xu2017approxdbn}. However, the current practice requires that the process of training the model and converting it to approximate DL takes place on a resource rich platform and then the converted model is deployed on a resource-constrained device.\\

\subsubsection{Accelerators}$\quad$

Designing specific hardware and circuits is another active research direction aiming to optimize the energy efficiency \cite{chen2017eyeriss} and memory footprint \cite{chen2014diannao} of DL models in IoT devices. The focus of such research works is on the inference time of DL models, since the training procedure of complex models is time and and energy intensive. In \cite{venkataramani2016efficient}, several approaches for improving the intelligence of IoT devices are identified including designing accelerators for DNNs, and using Post-CMOS technologies such as spintronics that employs electron spinning mechanism~\cite{awschalom2007challenges}. This latter technology suggests a direction toward the development of hybrid devices that can store data, perform computations and communications within the same material technology. 

The works in \cite{ramasubramanian2014spindle} and \cite{chen2017eyeriss} have reported an investigation of developing accelerators for DNNs and CNNs, respectively. Beyond the hardware accelerators, the work in \cite{lane2016deepx} proposed the use of a software accelerator for the inference phase of DL models on mobile devices. It employs two resource control algorithms at run-time, one that compresses layers and the other that decomposes deep architecture models across available processors. This software accelerator can be a complementary solution for the hardware accelerator designs. \\ 

\subsubsection{Tinymotes}$\quad$\\
In addition to all prior solutions, developing tiny size processors (micromotes) with strong DL capabilities is on the rise \cite{bang201714} \cite{bourzac2017speck}. Designed within the range of one cubic millimeter, such processors can be operated by batteries, consuming only about $300$ microwatts while performing on-board analysis and prediction using deep network accelerators. By this technology, many time-critical IoT applications can perform decision-making on the device instead of sending data to high performance computers and waiting for their response. For applications where data security and privacy are the main concerns, this integration of hardware and DL alleviates these concerns to some extent, as no or only limited data needs to be sent to the cloud for analysis. Moons \textit{et al.}~\cite{moons20160} also developed a tiny processor for CNNs (total active area of $1.2\times 2$ mm$^2$) that is power efficient (power consumption is from $25$ to $288$ mW).

\subsection{Applications}

There are existing mobile apps that employ pre-trained DNNs to perform their analytic and prediction tasks, such as using a CNN to identify garbage in images \cite{mittal2016spotgarbage}. However, resource consumption on these apps is still very high. Indeed, \cite{mittal2016spotgarbage} reports about $5.6$ seconds for returning a prediction response, while consuming $83$\% of the CPU and $67$ MB of memory. Howard \textit{et al.} \cite{howard2017mobilenets} proposed MobileNets  architecture for use in mobile and embedded vision applications. By restructuring a complex model through factorizing a standard convolution layer into a depthwise convolution and a $1 \times 1$ convolution, they were able to reach a smaller and computationally efficient models for GoogleNet and VGG16. They also demonstrated several use cases of their model including object detection, image classification, and identifying face attributes.

Amato \textit{et al.} \cite{amato2016deep} run a CNN on Raspberry Pi boards that were incorporated in smart cameras to find out the empty parking slots. Ravi \textit{et al.} in \cite{ravi2016deep} have reported the development of a fitness app for mobile devices that uses DL for classifying human activities. The DL model is trained on a standard machine and then transferred to the mobile platform for activity recognition. However, the input of the DL model is mixed with several engineered features to improve the accuracy. As the authors describe, the small number of layers in the DNN models dedicated for a resource-constrained device is a potential reason for them achieving a poor performance. In addition, the performance would not be satisfactorily if the training data is not well representing the entire ecosystem.  

Nguyen \textit{et al.} \cite{nguyen2016cognitive} proposed a conceptual software-hardware framework to support IoT smart applications. Their framework consists of a cognitive engine and a smart connectivity component. The cognitive engine, which provides cognitive functionality to smart objects, utilizes both DL algorithms and game-theoretic decision analytics. To be suitable for IoT, these algorithms must be deployed on low-power application-specific processors. The smart connectivity component integrates with cognitive radio transceivers and baseband processors to cater flexible and reliable connections to IoT smart objects. 

\subsection{Lessons Learned}
 In this section, the need to move toward supporting DL on IoT embedded and resource constrained devices were discussed. The adversary characteristics of IoT devices and DL techniques make this direction more challenging since IoT devices can rarely host DL models even to only perform predictions due to their resource constraints. To tackle these challenges, several methods were introduced in the recent literature including: 
\begin{itemize}
\item DNN compression
\item Approximate computing for DL 
\item Accelerators 
\item Tinymotes with DL.
\end{itemize}

These approaches focus on the inference functionality of available or pre-trained DL models. So, training DL models on resource-constrained and embedded devices is still an open challenge. Shifting the training process to IoT devices is desired for scalable and distributed deployment of IoT devices. For example, having hundreds of smart security cameras deployed in a community for face-based authentication, the training process for each camera can be done on site. Network compression involves identifying unimportant connections and neurons in a DNN through several rounds of training. While this is a promising approach for getting close to real-time analytics on IoT devices, more investigations need to be performed to handle several challenges such as: 
\begin{itemize}
\item It is not clear whether network compression approaches are suitable for data streaming, especially when the DL model is dynamic and may evolve over time.
\item The compression methods for time-series architectures, such as RNN and LSTM, have not been well investigated, and there is a gap to see if the existing compression methods are applicable to these DL architectures.
\item There is a need to specify the trade-off between the rate of compression and accuracy of a DNN, as more compression leads to degraded accuracy.
\end{itemize}

More recently, approximate computing approaches have also been utilized in making DL models simpler and more energy-efficient, in order to operate them on resource constrained devices. Similar to network compression techniques, these methods also take advantage of insignificant neurons. However, instead of manipulating the network structure, they preserve the structure but change the computation representations through bit-length reduction. For that reason, they seem applicable to a variety of DL architectures and can even cover the dynamic evolution of network models during run-time. Keeping a balance between accuracy and energy usage is their common goal. Nonetheless, more works are needed to find out the superiority of one of these approaches for embedding DL models in IoT devices.

Moreover, we discussed the emergence of special and small form-factor hardware that is designed to efficiently run DL models on embedded and resource constrained devices. These architectures can be utilized in wearable, mobile, and IoT devices, due to their reduced resource demands and their applicability to time-sensitive IoT applications. However, their generality to support any kind of DNN as well as their interoperability and compatibility with existing hardware platforms remain as clear challenges. 

Table \ref{tbl:ondevice_DL_methods} summarizes the methods and technologies utilized in the recent literature to host DL analytics on IoT devices along with their pros and cons.

We also reviewed some applications that have implemented DL on resource constrained devices. Due to the aforementioned challenges, there are not many well developed applications in this category. However, by resolving these challenges and barriers, we will see the rise of many IoT applications where their core DL model is embedded into the sensors, actuators, and IoT smart objects.

\section{Fog and Cloud-centric DL for IoT}\label{sec:cloud_centric_DL}

Cloud computing is considered a promising solution for IoT big data analytics. However, it may not be ideal for IoT data with security, legal/policy restrictions (e.g., data should not be transferred into cloud centers that are hosted outside of national territory), or time constraints. On the other hand, the high-level abstraction of data for some analytics purposes should be acquired by aggregating several sources of IoT data; hence, it is insufficient to deploy analytic solutions on individual IoT nodes in these cases.

Instead of being only on the cloud, the idea of bringing computing and analytics closer to the end-users/devices has been recently proposed under the name of fog computing. Relying on fog-based analytics, we can benefit from the advantages of cloud computing while reducing/avoiding its drawbacks, such as network latency and security risks. It has been shown that, by hosting data analytics on fog computing nodes, the overall performance can be improved due to the avoidance of transmitting large amounts of raw  data to  distant cloud nodes \cite{tang2017incorporating}. It is also possible to perform real-time analytics to some extent since the  fog is hosted locally close to the source of data. Smart application gateways are the core elements in this new fog technology, performing some of the tasks currently done by cloud computing such as data aggregation, classification, integration, and interpretation, thus facilitating the use of IoT local computing resources.

The work in \cite{al2015toward} proposed an intelligent IoT gateway that supports mechanisms by which the end users can control the application protocols in order to optimize the performance. The intelligent gateway primarily supports the inter-operation of different types of both IoT and resource-rich devices, causing them to be treated similarly. In the proposed intelligent gateway, a lightweight analytic tool is embedded to increase the performance at the application level. Equipping IoT gateways and edge nodes with efficient DL algorithms can localize many complex analytical tasks that are currently performed on the cloud. Table \ref{tbl:DL_products_fog} summarizes several products that have incorporated DL in their intelligent core, and can serve IoT domains in the fog or cloud. 

In the following subsection, we review several state-of-the-art enabling technologies that facilitate deep learning on the fog and cloud platforms.

\subsection{Enabling Technologies and Platforms}

Despite introducing DL analytics on fog infrastructure, cloud computing remains the only viable solution for analytics in many IoT applications that cannot be handled by fog computing. For example, complex tasks such as video analysis require large and complex models with a lot of computing resources. Thus, designing scalable and high performance cloud-centric DNN models and algorithms, which can perform analytics on massive IoT data, is still an important research area. Coates \textit{et al.} \cite{catanzaro2013deep} proposed a large-scale system, based on a cluster of GPU servers, which can perform the training of neural networks with $1$ billion parameters on $3$ machines in few days. The system can be also scaled to train networks with $11$ billion parameters on 16 machines.

Project Adam \cite{chilimbi2014project} is another attempt to develop a scalable and efficient DL model. The system is based on distributed DL, where the computation and communication of the whole system are optimized for high scalability and efficiency. The evaluation of this system using a cluster of $120$ machines shows that training a large DNN with $2$ billion connection achieves two times higher accuracy compared to a baseline system, while using $30$ times fewer machines.

Google's Tensor Processing Unit (TPU) \cite{jouppi2017datacenter} is a specialized co-processor for DNNs in Google's data centers. It was designed in $2015$ with the aim to accelerate the inference phase of DNNs that are written by TensorFlow framework. From $95\%$ of DNN representatives in their data centers, CNNs only constitute about $5\%$ of the workload, while MLPs and LSTMs cover the other $90\%$. Performance evaluation showed that TPU outperforms its contemporary GPUs or CPUs on average, by achieving 15 to $30$ times faster operation execution, while consuming $30$ to $80$ times fewer energy per TeraOps/second.

\begin{figure}
	\begin{center}		
		\includegraphics[width=.35\textwidth]{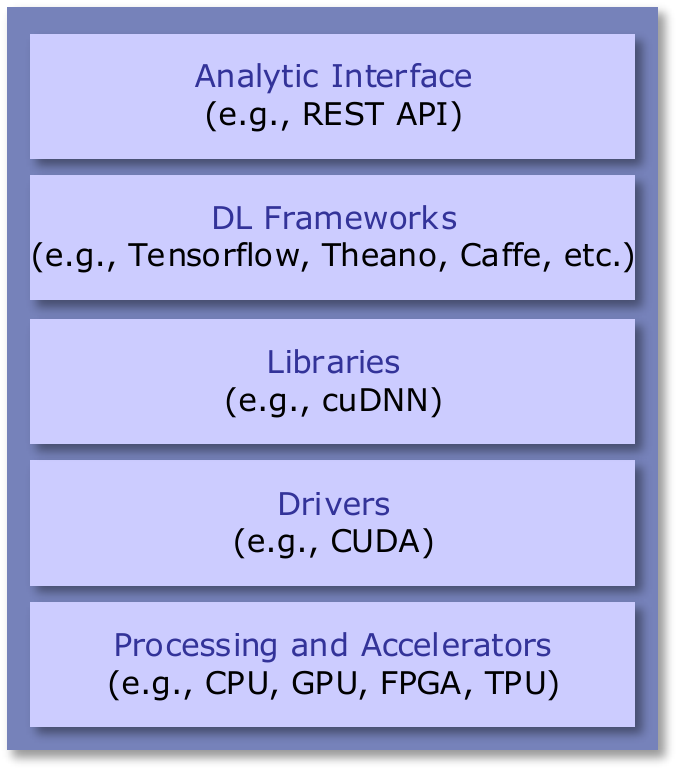}		
	\end{center}
	\caption{A general stack of DL models as a service in the cloud platforms.}\label{fig:DLStack}	
\end{figure}

Beyond the infrastructural advancements to host scalable DL models on cloud platforms, there is a need for mechanisms and approaches to make DL models accessible through APIs, in order to be easily integrated into IoT applications. This aspect has not been investigated much, and only a few products are available, such as Amazon's AWS DL AMIs\footnote{https://aws.amazon.com/amazon-ai/amis/}, Google cloud ML\footnote{https://cloud.google.com/products/machine-learning/}, and IBM Watson\footnote{https://www.ibm.com/watson/}. This creates opportunities for cloud providers to offer ``\textit{DL models as a service}'' as a new sub-category of Software as a Service (SaaS). However, this imposes several challenges for cloud providers, since DL tasks are computationally intensive and may starve other cloud services. Moreover, due to the data thirstiness of DL models, data transfers to the cloud may become a bottleneck. In order to deliver DL analytics on the cloud, Figure~\ref{fig:DLStack} presents a general stack for DL models as a service. Different providers may use their customized intelligence stack \cite{daniel2016lessons} \cite{hemsoth2015gpu}. 

\subsection{Challenges}

When DL analytics come to fog nodes, several challenges need to be addressed, including: 
\begin{itemize}
\item DL service discovery: Fog nodes are densely distributed in geographical regions, and each node may have specific analytics capabilities (e.g., one node runs CNN models for image detection, another node runs RNNs for time-series data prediction, etc.). So, the devices need to identify the sources of appropriate analytic providers through some sort of extended service discovery protocols for DL analytics. 
\item  DL model and task distribution: Partitioning the execution of DL models and tasks among the fog nodes, and optimally distributing of the data stream among the available nodes are critical for time-sensitive applications~\cite{simmhan2016big}. Aggregating the final results from the computing nodes and returning the action with the least latency are the other side of the coin.
\item Design factors: Since fog computing environments are in their infancy and are expected to evolve, it is worthwhile to investigate how the design factors of the fog environment (e.g., architectures, resource management, etc.) and the deployment of DL models in this environment can impact the quality of analytic services. Alternatively, it would be also interesting to see how far these design factors can be tailored/extended to improve the operation and quality of DL analytics. 
\item Mobile edge: Through the ubiquity of mobile edge computing environments and their contribution to the IoT analytics, it is important to consider the dynamicity of such environments for designing edge-assisted DL analytics since mobile devices may join and leave the system. Also, the energy management of mobile edge devices should be accurate when analytic tasks are delegated to them.
\end{itemize}

A few attempts reported the integration of DL on fog nodes in the IoT ecosystems. For example, a proof of concept for deploying CNN models on fog nodes for machine health prognosis was proposed by Qaisar \textit{et al.}~\cite{qaisar2017fog}. In their work, a thorough search among fog nodes is done to find free nodes to delegate analytic tasks to. Also, Li \textit{et al.}~\cite{li2016deepcham} proposed a system that leverages the collaboration of mobile and edge devices running CNN models for object recognition.\\

\begin{table}
\centering
\caption{Some Products that used Deep Learning and serving IoT Domains on the Fog or Cloud.}
\label{tbl:DL_products_fog}
\begin{tabular}{|l|l|l|l|}
\hline
\textbf{Product}          & \textbf{Description}                    & \textbf{Application}           & \textbf{Platform} \\ \hline \hline
Amazon Alexa    & \begin{tabular}[c]{@{}l@{}}Intelligent \\personal \\assistant (IPA) \end{tabular} & Smart home            & Fog      \\ \hline
Microsoft Cortana          & IPA & Smart Car, XBox       & Fog      \\ \hline
Google Assistant & IPA & \begin{tabular}[c]{@{}l@{}}Smart Car, \\Smart home\end{tabular} & Fog      \\ \hline
IBM Watson       & \begin{tabular}[c]{@{}l@{}}Cognitive \\framework \end{tabular}           & IoT domains           & Cloud    \\ \hline
\end{tabular}
\end{table}

\subsection{Lessons Learned}
In this section, we highlighted the role of cloud and fog computing and their enabling technologies, platforms and challenges to deliver DL analytics to IoT applications. The great success of cloud computing in support of DL is backed by the advancement and employment of optimized processors for neural networks as well as scalable distributed DL algorithms. Deploying DL models on fog platforms for IoT applications, such as smart homes and smart grids, would draw the attention of the end users due to the ease of accessibility and fast response time. Nevertheless, cloud-based DL analytics would be of great importance for long-term and complex data analytics that exceed the capabilities of fog computing. Some smart city applications, government sector, and nation-wide IoT deployments need to utilize cloud-based DL infrastructures.

Currently, the integration of DL analytics into IoT applications is limited to RESTful APIs, which are based on the HTTP protocol. While there exist several other application protocols that are extensively used in IoT applications, such as Message Queue Telemetry Transport (MQTT), Constrained Application Protocol (CoAP), Extensible Messaging and Presence Protocol (XMPP), and Advanced Message Queuing Protocol (AMQP), the integration of these protocols with the DL analytic interfaces calls for enhancing their compatibility with the aforementioned protocols to eliminate the need for message conversion proxies, which imposes extra overhead on the analytics response time.

We identified several challenges related to the deployment and usage of DL models in support of analytics on fog nodes. DL service discovery is a necessary requirement due to the dense deployment of fog nodes, which makes brute-force search for available services an inefficient approach. Currently used service discovery protocols in IoT applications, such as multicast DNS (mDNS) or DNS Service Discovery (DNS-SD)~\cite{al2015internet}, need to be extended to support DL service discovery (e.g., declare the type of analytics, DL model, input shape, etc.). Efficient distribution of DL models and tasks, and distribution of data streams on fog nodes and the aggregation of the results are other requirements that need to be addressed. 

\begin{table*}[]
		\centering
		\caption{Common Data sets for Deep Learning in IoT.}
		\label{tbl:datasets}
		\begin{tabular}{|l|l|l|l|l|}
\hline
			
			{\textbf{Dataset Name}} & 
			{\textbf{Domain}} & 
			{\textbf{Provider}} & 
			{\textbf{Notes}} & 
			{\textbf{Address/Link}}\\ 
			\hline \hline

			CGIAR dataset                                                                                 & \begin{tabular}[c]{@{}l@{}}Agriculture, \\Climate \end{tabular}  & CCAFS                                                                           & \begin{tabular}[c]{@{}l@{}}High-resolution climate \\ datasets for a variety \\ of fields including agricultural\end{tabular}                                                                             & http://www.ccafs-climate.org/                                                                                                                                    \\ \hline
			\begin{tabular}[c]{@{}l@{}}Educational \\Process \\ Mining\end{tabular}                         & Education              & \begin{tabular}[c]{@{}l@{}}University \\ of Genova\end{tabular}                 & \begin{tabular}[c]{@{}l@{}}Recordings of 115 subjects' \\ activities through a logging \\ application while learning \\ with an educational simulator\end{tabular}                                        & \begin{tabular}[c]{@{}l@{}}http://archive.ics.uci.edu/ml/\\ datasets/Educational+Process+\\ Mining+\%28EPM\%29\%3A+\\ A+Learning+Analytics+Data+Set\end{tabular} \\ \hline
			\begin{tabular}[c]{@{}l@{}}Commercial \\Building \\ Energy Dataset\end{tabular}                 & \begin{tabular}[c]{@{}l@{}}Energy, \\Smart Building \end{tabular} & IIITD                                                                           & \begin{tabular}[c]{@{}l@{}}Energy related data set \\from a  commercial building \\where data is sampled \\more than once a minute.\end{tabular}                                                        & http://combed.github.io/                                                                                                                                         \\ \hline
			\begin{tabular}[c]{@{}l@{}}Individual \\household\\  electric power \\ consumption\end{tabular} & \begin{tabular}[c]{@{}l@{}}Energy, \\Smart home \end{tabular}     & \begin{tabular}[c]{@{}l@{}}EDF R\&D, \\ Clamart, \\France\end{tabular}            & \begin{tabular}[c]{@{}l@{}}One-minute sampling rate \\ over a period of almost \\4 years\end{tabular}                                                                                                       & \begin{tabular}[c]{@{}l@{}}http://archive.ics.uci.edu/ml/\\ datasets/Individual+household+\\ electric+power+consumption\end{tabular}                             \\ \hline
			AMPds dataset                                                                                 & \begin{tabular}[c]{@{}l@{}}Energy, \\Smart home \end{tabular}    &  S. Makonin                                                                               & \begin{tabular}[c]{@{}l@{}}AMPds contains electricity, \\water, and natural gas \\measurements at one minute \\intervals for 2 years of \\monitoring\end{tabular}                                        & http://ampds.org/                                                                                                                                                \\ \hline
			\begin{tabular}[c]{@{}l@{}}UK Domestic \\ Appliance-Level \\ Electricity\end{tabular}         & \begin{tabular}[c]{@{}l@{}}Energy, \\Smart Home \end{tabular}    & \begin{tabular}[c]{@{}l@{}}Kelly and \\  Knottenbelt\end{tabular}   & \begin{tabular}[c]{@{}l@{}}Power demand from five \\houses. In each house both \\the whole-house  mains \\power demand as well as\\  power demand from individual\\  appliances are recorded.\end{tabular} & \begin{tabular}[c]{@{}l@{}}http://www.doc.ic.ac.uk/\\ $\sim$dk3810/data/\end{tabular}                                                                            \\ \hline
			\begin{tabular}[c]{@{}l@{}}PhysioBank\\ databases\end{tabular}                                                                          & Healthcare             & PhysioNet                                                                       & \begin{tabular}[c]{@{}l@{}}Archive of over 80 \\physiological datasets.  \end{tabular}                                                                                                                                                                           & \begin{tabular}[c]{@{}l@{}}https://physionet.org/physiobank\\ /database/\end{tabular}                                                                            \\ \hline
			\begin{tabular}[c]{@{}l@{}}Saarbruecken \\Voice Database \end{tabular}                                                                                       & Healthcare               & \begin{tabular}[c]{@{}l@{}} Universität \\des \\Saarlandes \end{tabular}    & \begin{tabular}[c]{@{}l@{}} A collection of voice \\recordings from more than \\2000 persons for pathological \\voice detection. \end{tabular}                                                    &  \begin{tabular}[c]{@{}l@{}}http://www.stimmdatenbank.\\coli.uni-saarland.de/\\help\_en.php4 \end{tabular}                                             \\ \hline
			T-LESS                                                                                        & Industry               & \begin{tabular}[c]{@{}l@{}}CMP at \\Czech \\ Technical \\University\end{tabular}    & \begin{tabular}[c]{@{}l@{}}An RGB-D dataset and \\evaluation methodology for \\detection and 6D pose \\estimation of texture-less \\objects\end{tabular}                                                    & http://cmp.felk.cvut.cz/t-less/                                                                                                                                  \\ \hline
			\begin{tabular}[c]{@{}l@{}}CityPulse Dataset\\  Collection\end{tabular}                       & Smart City             & \begin{tabular}[c]{@{}l@{}}CityPulse \\EU FP7 \\project\end{tabular}             & \begin{tabular}[c]{@{}l@{}}Road Traffic Data, Pollution \\Data, Weather, Parking\end{tabular}                                                                                                            & \begin{tabular}[c]{@{}l@{}}http://iot.ee.surrey.ac.uk:8080\\ /datasets.html\end{tabular}                                                                         \\ \hline
			\begin{tabular}[c]{@{}l@{}}Open Data \\Institute - node \\Trento\end{tabular}                  & Smart City             & \begin{tabular}[c]{@{}l@{}}Telecom \\Italia \end{tabular}                                                                  & \begin{tabular}[c]{@{}l@{}}Weather, Air quality, \\Electricity,  \\Telecommunication\end{tabular}                                                                                                           & http://theodi.fbk.eu/openbigdata/                                                                                                                                \\ \hline
			Málaga datasets                                                                               & Smart City             & \begin{tabular}[c]{@{}l@{}}City of \\Malaga \end{tabular}                                                                 & \begin{tabular}[c]{@{}l@{}}A broad range of categories \\such as energy, ITS, \\weather, Industry, Sport, etc.\end{tabular}                                                                             & \begin{tabular}[c]{@{}l@{}}http://datosabiertos.malaga.eu\\ /dataset\end{tabular}                                                                                \\ \hline
			\begin{tabular}[c]{@{}l@{}}Gas sensors for \\ home activity \\ monitoring\end{tabular}        & Smart home             & \begin{tabular}[c]{@{}l@{}}Univ. of \\ California \\San Diego\end{tabular}   &  \begin{tabular}[c]{@{}l@{}} Recordings of 8 gas sensors \\under three conditions \\including background, wine \\and banana presentations. \end{tabular}                                                                                                                                                                                                        & \begin{tabular}[c]{@{}l@{}}http://archive.ics.uci.edu/ml\\ /datasets/Gas+sensors+for+\\ home+activity+monitoring\end{tabular}                                    \\ \hline
        
        \begin{tabular}[c]{@{}l@{}}CASAS datasets \\for  activities of\\ daily living\end{tabular}        & Smart home             & \begin{tabular}[c]{@{}l@{}}Washington \\State \\ University\end{tabular}   &  \begin{tabular}[c]{@{}l@{}} Several public datasets related \\to Activities of Daily Living \\(ADL) performance in a two-\\story home, an apartment, \\and an office settings. \end{tabular}                                                                                                                                                                                                        & \begin{tabular}[c]{@{}l@{}}http://ailab.wsu.edu/casas/\\datasets.html\end{tabular}                                    \\ \hline
        \begin{tabular}[c]{@{}l@{}}ARAS Human \\Activity Dataset\end{tabular}        & Smart home             & \begin{tabular}[c]{@{}l@{}}Bogazici \\University\end{tabular}   &  \begin{tabular}[c]{@{}l@{}} Human activity recognition \\datasets collected from two \\real houses with multiple \\residents during two months. \end{tabular}   & \begin{tabular}[c]{@{}l@{}}https://www.cmpe.boun.edu.tr\\/aras/\end{tabular}                                    \\ \hline
        
        MERLSense Data                                                                                       & \begin{tabular}[c]{@{}l@{}}Smart home, \\building\end{tabular}                  & \begin{tabular}[c]{@{}l@{}} Mitsubishi \\Electric \\Research \\Labs \end{tabular}                                                                       & \begin{tabular}[c]{@{}l@{}}Motion sensor data of \\residual traces from a \\network of over 200 sensors \\for two years, containing \\over 50 million records. \end{tabular} & http://www.merl.com/wmd \\ \hline
        SportVU                                                                                       & Sport                  & Stats LLC                                                                       & \begin{tabular}[c]{@{}l@{}}Video of basketball and \\soccer games captured from \\6 cameras.  \end{tabular}                                                                                                                                                                   & http://go.stats.com/sportvu                                                                                                                                      \\ \hline
        RealDisp                                                                                       & Sport                  & O. Banos                                                                       & \begin{tabular}[c]{@{}l@{}}Includes a wide range of \\physical activities (warm up, \\cool down and fitness \\exercises).  \end{tabular}                                                                                                                                                                   & \begin{tabular}[c]{@{}l@{}}http://orestibanos.com/\\datasets.htm\end{tabular}                                                                                                                                     \\ \hline
            \end{tabular}
	\end{table*}

	\begin{table*}[]
		\centering
		\begin{tabular}{|l|l|l|l|l|}
			\multicolumn{5}{c}{TABLE \ref{tbl:datasets} -- Continued from previous page.}
			\\ \hline
			
			\textbf{Dataset Name} & 
			\textbf{Domain} & 
			\textbf{Provider} & 
			\textbf{Notes} & 
			\textbf{Address/Link}\\ 
			\hline \hline

			\begin{tabular}[c]{@{}l@{}}Taxi Service \\ Trajectory\end{tabular}                            & Transportation         & \begin{tabular}[c]{@{}l@{}}Prediction \\Challenge,\\  ECML \\PKDD 2015\end{tabular} & \begin{tabular}[c]{@{}l@{}}Trajectories performed by \\all the 442 taxis \\running in the city \\of Porto, in Portugal.\end{tabular}                                                                      & \begin{tabular}[c]{@{}l@{}}http://www.geolink.pt/\\ ecmlpkdd2015-challenge/\\ dataset.html\end{tabular}                                                          \\ \hline
			\begin{tabular}[c]{@{}l@{}}GeoLife GPS\\  Trajectories\end{tabular}                           & Transportation         & Microsoft                                                                       & \begin{tabular}[c]{@{}l@{}}A GPS trajectory by a \\sequence of time-stamped \\points\end{tabular}                                                                                                          & \begin{tabular}[c]{@{}l@{}}https://www.microsoft.com\\ /en-us/download/details.aspx?\\ id=52367\end{tabular}                                                     \\ \hline
			\begin{tabular}[c]{@{}l@{}}T-Drive trajectory\\  data\end{tabular}                            & Transportation         & Microsoft                                                                       & \begin{tabular}[c]{@{}l@{}}Contains a one-week \\trajectories of 10,357 taxis\end{tabular}                                                                                                               & \begin{tabular}[c]{@{}l@{}}https://www.microsoft.com/\\ en-us/research/publication/\\ t-drive-trajectory-data-sample/\end{tabular}                               \\ \hline
			\begin{tabular}[c]{@{}l@{}}Chicago Bus \\Traces  data\end{tabular}                            & Transportation         & \begin{tabular}[c]{@{}l@{}}M. Doering\end{tabular}                                                                       & \begin{tabular}[c]{@{}l@{}}Bus traces from the \\Chicago Transport
				Authority \\for 18 days with a rate \\between 20 and 40 seconds.\end{tabular}                                                                                                               & \begin{tabular}[c]{@{}l@{}}http://www.ibr.cs.tu-bs.de/\\users/mdoering/bustraces/\end{tabular} \\ \hline
			\begin{tabular}[c]{@{}l@{}}Uber trip \\data\end{tabular}                            & Transportation         & \begin{tabular}[c]{@{}l@{}}FiveThirty-\\Eight\end{tabular}                                                                       & \begin{tabular}[c]{@{}l@{}}About 20 million Uber \\pickups in New York City \\during 12 months.\end{tabular}                                                                                                               & \begin{tabular}[c]{@{}l@{}}https://github.com/fivethirtyeight/\\uber-tlc-foil-response\end{tabular} \\ \hline
			\begin{tabular}[c]{@{}l@{}}Traffic Sign \\Recognition\end{tabular}                            & Transportation         & \begin{tabular}[c]{@{}l@{}}K. Lim\end{tabular}                                                                       & \begin{tabular}[c]{@{}l@{}}Three datasets: Korean \\daytime, Korean nighttime, \\and German daytime \\traffic signs based on \\Vienna traffic rules.\end{tabular}                                                                                                               & \begin{tabular}[c]{@{}l@{}}https://figshare.com/articles\\/Traffic\_Sign\_Recognition\_\\Testsets/4597795
				
			\end{tabular} 
			\\ \hline
			\begin{tabular}[c]{@{}l@{}}DDD17\end{tabular}                            & Transportation         & \begin{tabular}[c]{@{}l@{}}J. Binas\end{tabular}                                                                       & \begin{tabular}[c]{@{}l@{}}End-To-End DAVIS \\Driving Dataset.\end{tabular}                                                                                                               & \begin{tabular}[c]{@{}l@{}}http://sensors.ini.uzh.ch/\\databases.html
				
			\end{tabular} 
			\\ \hline
		\end{tabular}
\end{table*}

\section{IoT Challenges for Deep Learning, and Future Directions}\label{sec:challenges}

In this section we first review several challenges that are important from the machine learning point of view to implement and develop IoT analytics. Then we point out  research directions that can fill the existing gaps for IoT analytics based on DL approaches.

\subsection{Challenges}

\subsubsection{Lack of Large IoT Dataset}$\quad$\\
The lack of availability of large real-world datasets for IoT applications is a main hurdle for incorporating DL models in IoT, as more data is needed for DL to achieve more accuracy. Moreover, more data prevents the overfitting of the models. This shortage is a barrier for deployment and acceptance of IoT analytics based on DL since the empirical validation and evaluation of the system should be shown promising in the natural world. Access to the copyrighted datasets or privacy considerations are another burdens that are more common in domains with human data such as healthcare and education. Also, a portfolio of appropriate datasets would be of a lot of help for developers and researchers. A general list of useful datasets has been compiled in Wikipedia \cite{wikipedia2017datasetList}. For the convenience of researchers in machine learning applications in IoT, table \ref{tbl:datasets} presents a collection of common datasets suitable to use for DL.\\

\subsubsection{Preprocessing}$\quad$\\
Preparing raw data in an appropriate representation to be fed in DL models is another challenge for IoT applications. Most DL approaches need some sort of preprocessing to yield good results. For example, image processing techniques by CNNs work better when the input data at the pixel level are normalized, scaled into a specific range, or transformed into a standard representation \cite{krizhevsky2012imagenet}\cite{mnih2013playing}. For IoT applications, preprocessing is more complex since the system deals with data from different sources that may have various formats and distributions while showing missing data.\\

\subsubsection{Secure and Privacy Preserving Deep Learning}$\quad$\\
Ensuring data security and privacy is a main concern in many IoT applications, since IoT big data will be transferred through the Internet for analytics, and can be thus observed around the world. While anonymization is used in many applications, these techniques can be hacked and re-identified as anonymized data. Moreover, DL training models are also subject to malicious attacks, such as False Data Injection or adversarial sample inputs, by which many functional or non-functional (e.g., availability, reliability, validity, trustworthiness, etc.) requirements of the IoT systems may be in jeopardy. Indeed, DL models learn the features from the raw data, and can therefore learn from any invalid data feed to it. In this case, DL models must be enhanced with some mechanism to discover abnormal or invalid data. A data monitoring DL model accompanying the main model should work in such scenarios. Papernot \textit{et al.} \cite{papernot2016limitations} have investigated the vulnerability of DNNs in adversarial settings where an adversary tries to provide some inputs that lead to an incorrect output classification and hence corrupting the integrity of the classification. Developing further techniques to defend and prevent the effect of this sort of attacks on DL models is necessary for reliable IoT applications.\\

\subsubsection{Challenges of 6V's}$\quad$\\
Despite the recent advancement in DL for big data, there are still significant challenges that need to be addressed to mature this technology. Each characteristic of IoT big data imposes a challenge for DL techniques. In the following we highlight these challenges.

The massive volume of data poses a great challenge for DL, especially for time and structure complexity. The voluminous number of input data, their broad number of attributes, and their high degree of classification result in a very complex DL model and affect running time performance.  Running DL on distributed frameworks or clusters of CPUs with parallel processing is a viable solution that has been developed \cite{chen2014big-deep}. The high volume of IoT big data also brings another challenge, namely the noisy and unlabeled data. Even though DL is very good at tolerating noisy data and learning from unlabeled data, it is not clear to what extent DL models can be accurate in the presence of such abnormal data.

The variety of IoT data formats that come from various sources pops up the challenge of managing conflicts between different data sources. In case of no conflict in data sources, DL has the ability to effectively work on heterogeneous data. 

The high velocity of IoT big data, i.e., the high rate of data generation, also brings the challenge of high speed processing and analysis of data. Online learning is a solution for high velocity and has been proposed for DNNs. However, more research is needed to augment DL with online learning and sequential learning techniques.

The veracity of IoT big data also presents challenges for DL analytics. The IoT big data analytics will not be useful if the input data is not coming from a trustworthy source. Data validation and trustworthiness should be checked at each level of big data analytics, especially when we are dealing with online streams of input data to an analytic engine \cite{najafabadi2015deep}.

Moreover, the variability of IoT big data (variation in the data flow rates) rises challenges for online analytics. In case of immense streams of data, DL techniques, and in particular the online ones, handle them. Data sampling techniques would be beneficial in these scenarios.

Finally, a main challenge for business managers to adopt big data is that it is not clear for them how to use big data analytics to get value out of it and improve their business\cite{lavalle2011big}. Beyond that, the analytic engine may produce abstractions that are not important for the stakeholders, or are not clear enough for them. 
\\

\subsubsection{Deep Learning for IoT Devices}$\quad$\\
Developing DL on IoT devices poses a new challenge for IoT device designers, to consider the requirements of handling DNNs in resource-constrained devices. These requirements are expected to grow as the datasets sizes are growing every day, and new algorithms arise to be part of the solutions for DL in IoT.\\ 

\subsubsection{Deep Learning Limitations}$\quad$\\
Despite showing impressive results in many applications, DL models still have several limitations. Nguyen \textit{et al.} \cite{nguyen2015deep} reported about the false confidence of DDN for predicting images that are unrecognizable by humans. By producing fooling examples that are totally unrecognizable by humans, the DNN classifies them as familiar objects.

The other limitation is the focus of DL models on classification, while many IoT applications (e.g., electricity load forecasting, temperature forecasting) need a kind of regression at their analytics core. Few works tried to enrich DNNs with regression capabilities, such as the work in \cite{qiu2014ensemble} proposing the ensemble of DBN and Support Vector Regression (SVR) for regression tasks. However, more investigation is required to clear many aspects of regression with DL. 

\subsection{Future Directions}

\subsubsection{IoT Mobile Data}$\quad$\\
One remarkable part of IoT data comes from mobile devices. Investigating efficient ways to utilize mobile big data in conjunction with DL approaches is a way to come up with better services for IoT domains, especially in smart city scenarios. In \cite{alsheikh2016mobile}, the capabilities of DL models in mobile big data analytics were investigated using a distributed learning framework that executes an iterative MapReduce task on several parallel Spark workers.\\

\subsubsection{Integrating Contextual Information}$\quad$\\ 
The environment's situation cannot be understood by the IoT sensor data alone. Therefore, IoT data needs to be fused with other sources of data, namely context information that complement the understanding of the environment \cite{perera2014context}. This integration can also help for fast data analytics and quick reasoning due to the bounded search space for the reasoning engine. For example, a smart camera with capability of face pose recognition can perform its job in various contexts such as security gates in smart homes or government buildings, or in smart cars for driving assistance. In all these situations, complementary contextual information (e.g., time within the day, daily habits, etc.) helps the system to reason about the best action that can be done based on the detected pose of the person.  \\

\subsubsection{Online Resource Provisioning for IoT Analytics}$\quad$\\ 
The deployment of fast DL based data analytics on the fog and cloud would require online provisioning of fog or cloud resources to host the stream of data. Due to the streaming nature of IoT data, knowing the volume of data sequence in advance is not feasible. In this regard, we need a new class of algorithms that work based on the current stream of data and do not rely on the prior knowledge of the data stream. A DL mechanism and an online auctioning algorithm are proposed in \cite{borkowski2016predicting} and  \cite{gharaibeh2017online}, respectively, to support online provisioning of fog and cloud resources for IoT applications.\\

\begin{figure}
	\begin{center}		
		\includegraphics[width=.45\textwidth]{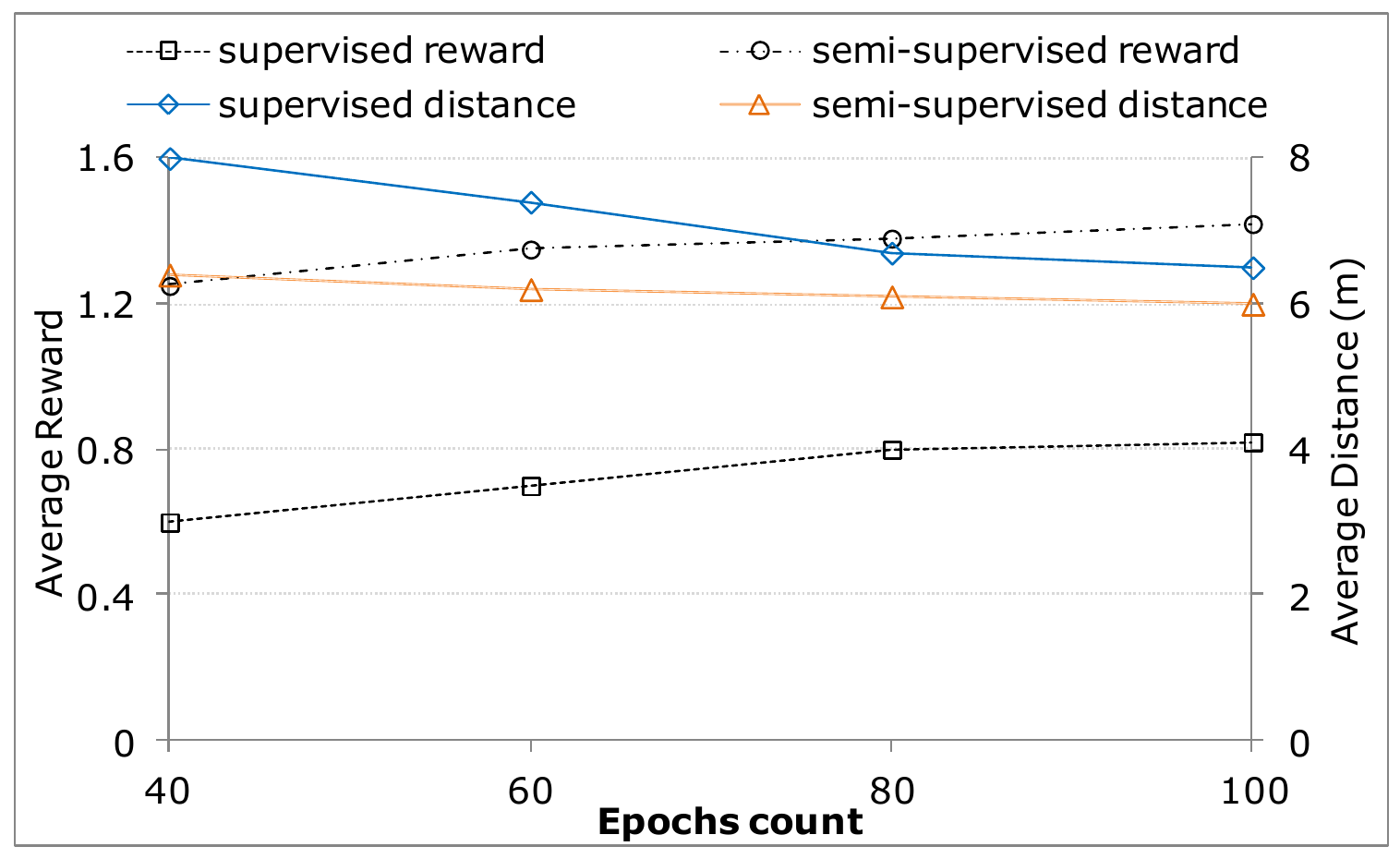}		
	\end{center}
	\caption{Deep reinforcement learning with only labeled data (supervised) vs. with labeled and unlabeled data (semisupervised). At each epoch, semi-supervised model outperforms the supervised model both in terms of total received rewards and closeness to the target.}\label{fig:fig-DRL_result}	
\end{figure}

\subsubsection{Semi-supervised Analytic Frameworks}$\quad$\\ 
Most of the analytic algorithms are supervised, thus needing a large amount of training labeled data that is either not available or comes at a great expense to prepare. Based on IDC's report \cite{gantz2012digital}, it is estimated that by $2012$ only about $3\%$ of all data in the digital universe has been annotated, which implies the poor source of training datasets for DL. A combination of advanced machine learning algorithms designed for semi-supervised settings fits well for smart cities systems, where a small training dataset can be used while the learning agent improves its accuracy using a large amount of unlabeled data~\cite{mohammadi2018enabling}. Figure \ref{fig:fig-DRL_result} illustrates the role of semi-supervised learning in improving the output accuracy for deep reinforcement learning \cite{mohammadi2017semi} in indoor localization experiments. In their experiments, only $15\%$ of data was labeled but the results were strengthened by utilizing unlabeled data in the algorithm.\\

\subsubsection{Dependable and Reliable IoT Analytics}$\quad$\\
As we rely more on CPS and IoT in large scales, the need for mechanisms to ensure the safety of the system against malicious attacks as well as failures become more crucial~\cite{nsf2017cps}. DL approaches can be applied in these directions by analyzing the huge amount of log traces of CPS and IoT systems, in order to identify and predict weak points of the system where attacks may occur or the functionality is defected. This will help the system to prevent or recover from faults and consequently increase the level of dependability of CPS and IoT systems.\\

\subsubsection{Self-organizing Communication Networks}$\quad$\\
With a huge number of IoT devices, the configuration and maintenance of their underlying physical M2M communications and networking become harder. Although the large body of network nodes and their relation is a challenge for traditional machine learning approaches, it opens the opportunity for DL architectures to prove their competency in this area by providing a range of self-services such as self-configuration, self-optimization, self-healing, and self-load balancing. Valente \textit{et al.} \cite{valente2017survey} have provided a survey of traditional machine learning approaches for self-organizing cellular networks.\\

\subsubsection{Emerging IoT Applications}$\quad$\\
\textbf{Unmanned aerial vehicles:} The usage of Unmanned aerial vehicles (UAVs) is a promising application that can improve service delivery in hard-reaching regions or in critical situations. UAVs have been also used for many image analysis in real-time such as surveillance tasks, search-and-rescue operations, and infrastructure inspection~\cite{lee2017real}. These devices face several challenges for their adoption, including routing, energy saving, avoiding private regions, and obstacle avoidance \cite{tai2016deep} etc. DL can be of great impact in this domain for the prediction and decision-making of tasks to get the best out of UAVs. Moreover, UAVs can be seen as on-the-fly analytics platforms that potentially can provide temporarily fog computing analytic services as well as distributed analytics.\\

\noindent\textbf{Virtual/Augmented Reality:} Virtual/augmented reality is another application area that can benefit from both IoT and DL. The latter can be used in this field to provide services such as object tracking \cite{akgul2016applying}, activity recognition, image classification, and object recognition \cite{sutanto20173d} to name a few. Augmented reality can greatly affect several domains such as education, museums, smart connected cars, etc. \\

\noindent\textbf{Mobile Robotics:} 
Mobile robots are being used in many commercial and industrial settings for moving materials or performing tasks in hazardous environments. There are many research efforts that have benefited from DL to develop an intelligent control software for mobile robots \cite{tai2016deep_robotics} \cite{goeddel2016learning}. Being able to perceive the environment through different kinds of sensors, such as LIDARs and cameras, have made them a top topic to assess the performance of CNN techniques for a variety of vision-based tasks. A strict requirement for mobile robots is that DL models should be able to provide real-time responsiveness.

\section{Conclusion}\label{sec:Conclusion}

DL and IoT have drawn the attention of researchers and commercial verticals in recent years, as these two technology trends have proven to make a positive effect on our lives, cities, and world. IoT and DL constitute a chain of data producer-consumer, in which IoT generates raw data that is analyzed by DL models and DL models produce high-level abstraction and insight that is fed to the IoT systems for fine-tuning and improvement of services. 

In this survey, we reviewed the characteristics of IoT data and its challenges for DL methods. In specific, we highlighted IoT fast and streaming data as well as IoT big data as the two main categories of IoT data generation and their requirements for analytics. We also presented several main architectures of DL that is used in the context of IoT applications followed by several open source frameworks for development of DL architectures. Reviewing different applications in various sectors of IoT that have utilized DL was another part of this survey in which we identified five foundational services along with eleven application domains. By distinguishing foundational services, as well as IoT vertical applications, and reviewing their DL approaches and use cases, the authors provided a basis for other researchers to understand the principle components of IoT smart services and apply the relevant techniques to their problems. The new paradigm of implementing DL on IoT devices was surveyed and several approaches to achieve it were introduced. DL based on fog and cloud infrastructures to support IoT applications was another part of this survey. We also identified the challenges and future research direction in the path of DL for IoT applications. 
\newpage

\section*{List Of Acronyms}

\begin{table}[H]
\begin{tabular}{ll}
5G & 5th Generation (Cellular networks)\\
AE & Auto-encoder \\
AI & Artificial Intelligence \\
AMQP & Advanced Message Queuing Protocol\\
ANN & Artificial Neural Network \\
BAC & Breast Arterial Calcification \\
BLE & Bluetooth Low Energy \\
BPTT & Backpropagation Through Time \\
CAE & Contractive Auto-encoder\\
CDR & Caller Detail Record \\
CIFAR & Canadian Institute for Advanced Research \\
CNN &  Convolutional Neural Network \\
CoAP & Constrained Application Protocol\\
CPS & Cyber-physical System \\
CRBM & Conditional Restricted Boltzmann Machine \\
DAE & Denoising Auto-encoder \\
DBN & Deep Belief Network \\
DL &  Deep Learning \\
DNN & Deep Neural Network \\
DNS-SD & DNS Service Discovery \\
DRL & Deep Reinforcement Learning \\
ELM & Extreme Learning Machine \\
FDC & Fault Detection and Classification \\
FDI & False Data Injection \\
FNN & Feedforward Neural Network \\
GAN & Generative Adversarial Network \\
GBM & Gradient Boosted Machine \\
GLM & Generalized Linear Model \\
GPU & Graphics Processing Unit \\
HMM & Hidden Markov Model \\
HVAC & Heating, Ventilation and Air Conditioning \\
INS & Inertia Navigation System \\
IoT & Internet of Things \\
IPA & Intelligent Personal Assistant \\
ITS & Intelligent Transportation System\\
LSTM & Long Short-term Memory \\
M2M & Machine-to-Machine \\
MAPE & Mean Absolute Percentage Error \\
mDNS & multicast DNS \\
ML & Machine Learning \\
MLP & Multi-layer Perceptron \\
MNIST & Modified National Institute of Standards and Technology \\
MOOC & Massive Open Online Courses \\
MQTT & Message Queue Telemetry Transport\\
RBN & Restricted Boltzmann Machine \\
ReLU & Rectified Linear Units \\
RL & Reinforcement Learning \\
RNN & Recurrent Neural Network \\
SaaS & Software as a Service\\
SdA & Stacked denoising Autoencoder \\
SGD & Stochastic Gradient Descent \\
SVM & Support Vector Machine \\
SVR & Support Vector Regression\\
TPU & Tensor Processing Unit \\
UAV & Unmanned Aerial Vehicle \\
VAE & Variational Auto-encoder\\
VGG & Visual Geometry Group \\
VLC & Visual Light Communication \\
WSN & Wireless Sensor Network\\
XMPP & Extensible Messaging and Presence Protocol \\
\end{tabular}
\end{table}

\bibliographystyle{IEEEtran}

\bibliography{references}

\begin{IEEEbiography}[{\includegraphics[width=1in,
height=1.25in,clip,keepaspectratio]{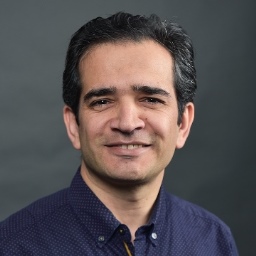}}]{Mehdi Mohammadi} (S'14) received his B.S. degree in Computer Engineering from Kharazmi University, Tehran, Iran in 2003 and his M.S. degree in Computer Engineering (Software) from Sheikhbahaee University, Isfahan, Iran in 2010. He received his Ph.D. degree in Computer Science from Western Michigan University (WMU), Kalamazoo, MI, USA. His research interests include Internet of Things, IoT data analytics, Machine Learning, and Cloud Computing. He served as reviewer for multiple journals including IEEE Internet of Things Journal, IEEE Communications Magazine, IEEE Communications Letters, IEEE Transactions on Emerging Topics in Computational Intelligence,  Wiley's Security and Wireless Communication Networks Journal and Wiley's Wireless Communications and Mobile Computing Journal. He was the recipient of six travel grants from the National Science Foundation (NSF). 
\end{IEEEbiography}

\begin{IEEEbiography}[{\includegraphics[width=1in,
height=1.25in,clip,keepaspectratio]{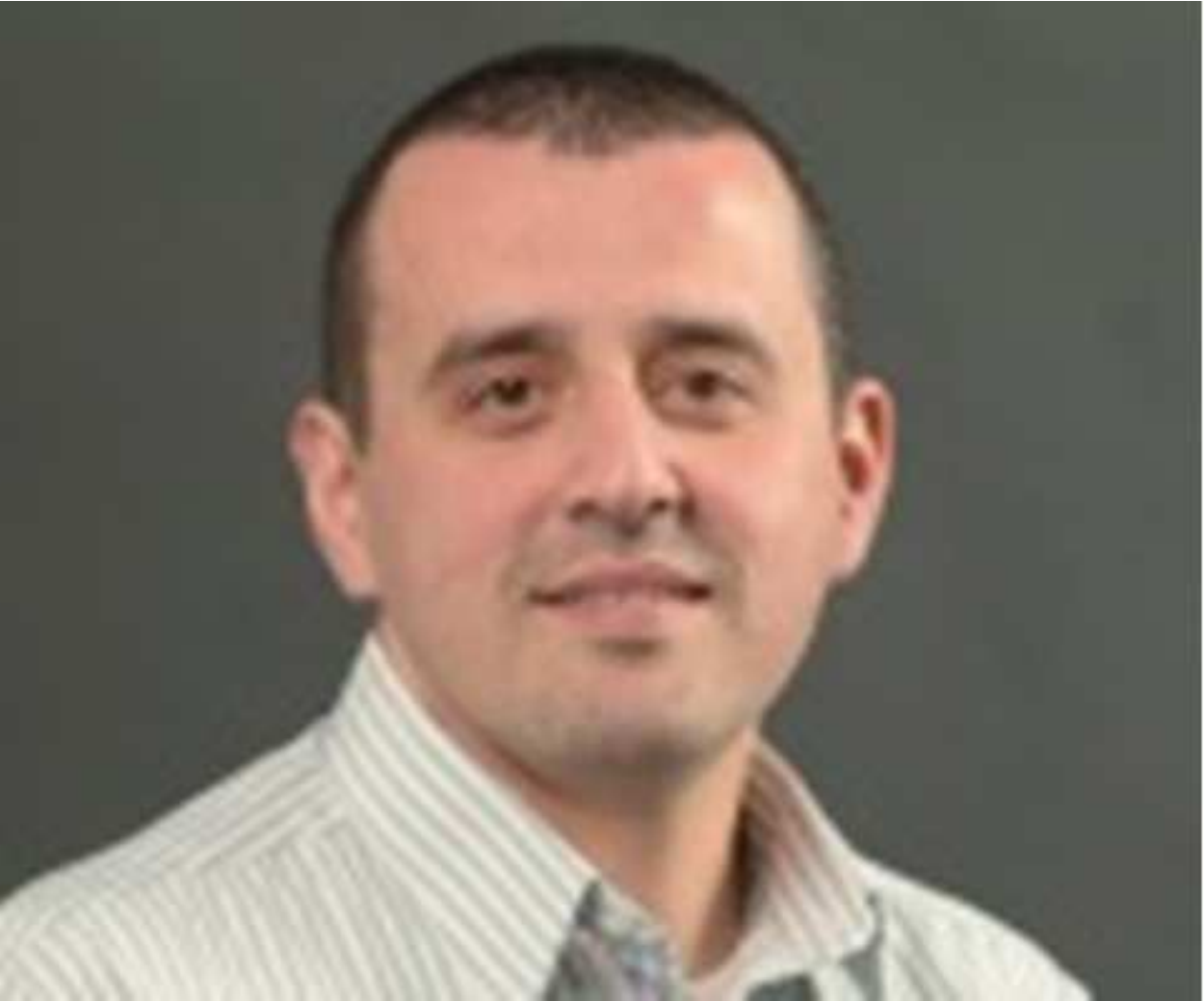}}]{Ala Al-Fuqaha} (S'00-M'04-SM'09) received his M.S. from the University of Missouri-Columbia and Ph.D. from the University of Missouri-Kansas City. Currently, he is a Professor and director of NEST Research Lab at the Computer Science Department of Western Michigan University. His research interests include the use of machine learning in general and deep learning in particular in support of the data-driven and self-driven management of large-scale deployments of IoT and smart city infrastructure and services, Wireless Vehicular Networks (VANETs), cooperation and spectrum access etiquettes in cognitive radio networks, and management and planning of software defined networks (SDN). He is a senior member of the IEEE and an ABET Program Evaluator (PEV). He served on editorial boards and technical program committees of multiple international journals and conferences.
\end{IEEEbiography}

\begin{IEEEbiography}[{\includegraphics[width=1in,height=1.25in,clip,keepaspectratio]{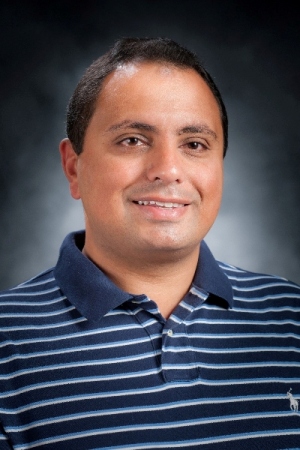}}]{Sameh Sorour} 
 (S'98, M'11, SM'16) is an Assistant Professor at the Department of Electrical and Computer Engineering, University of Idaho. He received his B.Sc. and M.Sc. degrees in Electrical Engineering from Alexandria University, Egypt, in 2002 and 2006, respectively. In 2011, he obtained his Ph.D. degree in Electrical and Computer Engineering from University of Toronto, Canada. After two postdoctoral fellowships at University of Toronto and King Abduallah University of Science and Technology (KAUST), he joined King Fahd University of Petroleum and Minerals (KFUPM) in 2013 before moving to University of Idaho in 2016. His research interests lie in the broad area of advanced communications/networking/computing/learning technologies for smart cities applications, including cyber physical systems, internet of things (IoT) and IoT-enabled systems, cloud and fog networking, network coding, device-to-device networking, autonomous driving and autonomous systems, intelligent transportation systems, and mathematical modelling and optimization for smart systems.

\end{IEEEbiography}

\begin{IEEEbiography}[{\includegraphics[width=1in,
height=1.25in,clip,keepaspectratio]{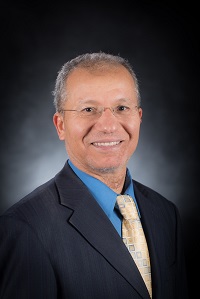}}]{Mohsen Guizani} (S'85, M'89, SM'99, F'09) received the B.S. (with distinction) and M.S. degrees in electrical engineering, the M.S. and Ph.D. degrees in computer engineering from Syracuse University, Syracuse, NY, USA, in 1984, 1986, 1987, and 1990, respectively. He is currently a Professor and the ECE Department Chair at the University of Idaho, USA. Previously, he served as the Associate Vice President of Graduate Studies, Qatar University, Chair of the Computer Science Department, Western Michigan University, and Chair of the Computer Science Department, University of West Florida. He also served in academic positions at the University of Missouri-Kansas City, University of Colorado-Boulder, Syracuse University, and Kuwait University. His research interests include wireless communications and mobile computing, computer networks, mobile cloud computing, security, and smart grid. He currently serves on the editorial boards of several international technical journals and the Founder and the Editor-in-Chief of Wireless Communications and Mobile Computing journal (Wiley). He is the author of nine books and more than 450 publications in refereed journals and conferences. He guest edited a number of special issues in IEEE journals and magazines. He also served as a member, Chair, and General Chair of a number of international conferences. He received the teaching award multiple times from different institutions as well as the best Research Award from three institutions. He was the Chair of the IEEE Communications Society Wireless Technical Committee and the Chair of the TAOS Technical Committee. He served as the IEEE Computer Society Distinguished Speaker from 2003 to 2005. He is a Fellow of IEEE and a Senior Member of ACM.

\end{IEEEbiography}

\end{document}